\documentclass[trackchanges,twocolumn,raggedbottom]{aastex701}

\usepackage{float}
\usepackage{makecell}

\DeclareRobustCommand{\okina}{%
  \raisebox{\dimexpr\fontcharht\font`A-\height}{%
    \scalebox{0.8}{`}%
  }%
}

\begin{document}

\title{SN~2025aico: An Interesting Case Of \\ $^{56}$Ni Mixing, Ejecta Asymmetries, and Dust Formation in a Type IIb Supernova}
%%-------------------------------------------------------------------------
\author[orcid=0009-0009-8560-6952]{Grace Davis}
\affiliation{Institute for Astronomy, University of Hawai\okina i at M\=anoa, 2680 Woodlawn Dr., Honolulu, HI 96822, USA}
\affiliation{Department of Astronomy and Astrophysics, The Pennsylvania State University, 251 Pollock Road, University Park, PA 16802, USA}
\email[show]{gmd5786@psu.edu}

\author[orcid=0000-0001-7186-105X]{Kyle Medler}
\affiliation{Institute for Astronomy, University of Hawai\okina i at M\=anoa, 2680 Woodlawn Dr., Honolulu, HI 96822, USA}
\email[show]{kmedler@hawaii.edu}

\author[orcid=0000-0002-5221-7557]{Chris Ashall}
\affiliation{Institute for Astronomy, University of Hawai\okina i at M\=anoa, 2680 Woodlawn Dr., Honolulu, HI 96822, USA}
\email{cashall@hawaii.edu}

\author[orcid=0000-0003-3953-9532]{Willem~B.~Hoogendam}
\affiliation{Institute for Astronomy, University of Hawai\okina i at M\=anoa, 2680 Woodlawn Dr., Honolulu, HI 96822, USA}
\email{willemh@hawaii.edu}

\author[orcid=0000-0002-7305-8321]{Cameron Pfeffer}
\affiliation{Institute for Astronomy, University of Hawai\okina i at M\=anoa, 2680 Woodlawn Dr., Honolulu, HI 96822, USA}
\email{cpfeffer@hawaii.edu}

\author[0000-0002-6230-0151]{David~O.~Jones}
\affiliation{Institute for Astronomy, University of Hawai\okina i, 640 N. A\okina ohoku Pl., Hilo, HI 96720, USA}
\email{dojones@hawaii.edu}

\author[orcid=0000-0001-6142-6556]{Shunsaku Horiuchi}
  \email{horiuchi@phys.sci.isct.ac.jp}
 \affiliation{ Department of Physics, Institute of Science Tokyo, 2-12-1 Ookayama, Meguro-ku, Tokyo 152-8551, Japan}

\author[orcid=0000-0001-5888-2542]{Tyco Mera}
  \email{tycomera@gmail.com}
 \affiliation{Institute for Astronomy, University of Hawai\okina i at M\=anoa, 2680 Woodlawn Dr., Honolulu, HI 96822, USA}

\author[orcid=0000-0003-4631-1149]{Benjamin Shappee}
    \email{shappee@hawaii.edu}
\affiliation{Institute for Astronomy, University of Hawai\okina i at M\=anoa, 2680 Woodlawn Dr., Honolulu, HI 96822, USA}

\author[orcid=0000-0002-4557-6682]{Charlotte Ward}
\affiliation{Department of Astronomy and Astrophysics, The Pennsylvania State University, 251 Pollock Road, University Park, PA 16802, USA}
\email{cvw5890@psu.edu}

\author[orcid=0009-0000-2871-9330]{Grace Showerman}
    \email{gshowe@hawaii.edu}
\affiliation{Institute for Astronomy, University of Hawai\okina i at M\=anoa, 2680 Woodlawn Dr., Honolulu, HI 96822, USA}
%%-------------------------------------------------------------------------
\begin{abstract}
Stripped-envelope supernovae provide a window into how massive stars lose their layers, mix radioactive material, produce dust, and explode asymmetrically. We present optical, near-infrared (NIR), and mid-infrared (MIR) observations of SN~2025aico, a Type~IIb supernova in LEDA~35384. Our spectroscopic sequence, spanning $+1$ to $+167$\,d after explosion, follows its evolution from a photospheric phase exhibiting both hydrogen and helium features to prominent helium emission and, ultimately, nebular-phase ejecta. Using \ion{He}{1} $1.083$ and $2.0581\,\mu$m transitions, we investigated the kinematics and geometry of the helium-rich material. Both transitions exhibit a three-phase, non-monotonic velocity evolution: an initial rapid decline, a subsequent increase, and an eventual plateau. We interpret this behavior as evidence for limited outward mixing of $^{56}$Ni, such that radioactive energy deposition reaches the outer helium-rich ejecta progressively as the ejecta expand, producing the non-thermal electrons responsible for helium excitation. After $\sim100$\,d, both NIR \ion{He}{1} transitions develop double-peaked emission profiles. Similar structure in the oxygen emission indicates ejecta asymmetry. Comparison with other supernovae suggests a tentative connection between this structure and explosion energy, potentially linked to the delay between core collapse and explosion. Serendipitous \textit{JWST} observations at $+124.4$\,d reveal an infrared excess. Modeling favors warm ($\sim800$--$1500$K) carbon dust newly formed in the ejecta, together with cooler carbon or silicate dust likely associated with pre-existing circumstellar material. SN~2025aico demonstrates how continuous optical-to-MIR observations can connect progenitor evolution, explosion physics, ejecta geometry, and dust production.
\end{abstract}
%%-------------------------------------------------------------------------
%%-------------------------------------------------------------------------
\section{Introduction}
Core-collapse supernovae (CCSNe) mark the violent endpoint of stars with initial masses $M \gtrsim 8~M_\odot$, occurring when nuclear burning can no longer support the stellar core against gravitational collapse \citep{Woosley2002,Smartt2009,Janka_2012, Tsujimoto_2022}. They are among the most energetic phenomena in the Universe and play a key role in the nucleosynthesis and chemical enrichment of galaxies \citep{Woosley1995, Nomoto2013, Neopane2024Long}. The diversity of CCSNe subtypes reflects the wide range of progenitor systems and the history of mass-loss that preceded the collapse of their progenitors \citep{gilkis2025}. Among these, the hydrogen-rich Type~II Supernovae (SNe II) and the hydrogen-poor/helium-rich Type~Ib Supernovae (SNe Ib) represent the two ends of the stripping of the hydrogen envelope, with Type~IIb supernovae (SNe IIb) occupying a transitional class between them \citep{filippenko_1997, modjaz2019, 2024ApJ...974..316D,jerkstrand2025}. Recent measurements of the local CCSN population find that stripped-envelope SNe constitute a substantial fraction of CCSNe, highlighting the importance of understanding the progenitor systems and stripping mechanisms that produce these events \citep{Pessi2025}.

SNe IIb are characterized by a spectral evolution that initially resembles SNe II, with prominent hydrogen features during the early photospheric phase, before transitioning to helium-dominated spectra at later epochs \citep{filippenko_1997, Turatto_2003}. Within SNe IIb there is diversity in the amount of hydrogen present in the outer envelope, resulting in the duration and strength of these hydrogen features varying considerably \citep{Gilkis2022,long2022,jerkstrand2025}. Inferred residual hydrogen masses span $\sim0.016$--$0.233~M_\odot$ among well-studied stripped-envelope SNe, including SN~1993J \citep{Hoflich1993}, SN~2008ax \citep{Pastorello_2008}, SN~2011dh \citep{Maund_2011}, SN~2013df \citep{Ciabattari2013}, iPTF13bvn \citep{Cao2013}, SN~2016gkg \citep{Kilpatrick_2016}, SN~2019yvr \citep{Ferrari2024}, and SN~2020acat \citep{Ergon2024}. A residual hydrogen mass of approximately $0.033~M_\odot$ has been proposed as the boundary between SNe~IIb and hydrogen-poor SNe~Ib; variations in the remaining envelope mass also contribute to the broad range of inferred SN~IIb progenitor radii \citep{Gilkis2022,Hachinger2012,long2022, 2024ApJ...974..316D}. The light curve morphology of these events is affected by the progenitor radius: compact progenitors ($R \lesssim 100~R_\odot$ ; \citealt{Gangopadhyay2023, Zhao_2026}) produce single-peaked light curves dominated by radioactive $^{56}$Ni decay, while more extended progenitors ($R \gtrsim 100$--$200~R_\odot$; \citealt{Woolsey1994, Kumar2013}) display double-peaked light-curves with an initial peak powered by the cooling of shock-heated envelope material followed by a second, radioactively powered maximum \citep{Chevalier2010,barmentloo2024}.

The low hydrogen masses inferred for SNe~IIb result from substantial envelope stripping during the progenitor's lifetime. Two principal mechanisms have been proposed: mass transfer to a binary companion \citep{Podsiadlowski1993,Stancliffe2009,Claeys2011,Benvenuto2013,Yoon2017} and line-driven winds from a single star \citep{Groh2013,long2022}. Because the efficiency of line-driven winds increases with metallicity \citep{Woosley_2006}, single-star stripping is expected to operate primarily at near-solar or super-solar metallicity and generally requires higher progenitor masses than those inferred for most well-studied SNe~IIb \citep{Vink2001,Sravan2019}. Wolf--Rayet stars, whose strong winds can remove most or all of the hydrogen envelope, are therefore more commonly associated with SNe~Ib/c than with SNe~IIb \citep{Groh2013}.

In the binary channel, interaction with a companion removes most, but not necessarily all, of the hydrogen envelope, naturally producing an SN~IIb progenitor. Support for this scenario comes from pre-explosion detections of the progenitors of SN~1993J, SN~2011dh, SN~2013df, and SN~2016gkg, which are consistent with extended yellow or blue supergiants rather than compact Wolf--Rayet stars \citep{Hoflich1993,Maund_2011,Ciabattari2013,Kilpatrick_2016}. The binary scenario is further supported by the high binary fraction among massive stars \citep{Sana2012} and the comparatively low ejecta masses inferred for SNe~IIb, which are more readily explained by binary-stripped progenitors than by massive single stars \citep{Lyman_2016}. SNe~IIb also preferentially occur in metal-poor environments, implying low-metallicity progenitors whose line-driven winds are generally too weak to remove most of the hydrogen envelope without binary interaction \citep{Vink2001, Fang2019}. Binary interaction is therefore expected to dominate the SN~IIb population, although single-star wind stripping may contribute a minority of events, particularly at the high-mass and high-metallicity end of the progenitor population \citep{Sravan2019,Zapartas2021}.

Until recently the majority of Type~IIb events have been observed primarily in the optical regime, limiting our view to the outer ejecta at relatively late times. The extension to near-infrared (NIR) wavelengths has opened a complementary window into the physics of explosions and progenitor properties \citep{Taubenberger_2011,Ergon_2014,Bianco_2014,Davis_2021, Shahbandeh2023,khakpash2024, Medler2025,yamanaka2026,hwangbo2026}. Spectral lines in this wavelength regime are generally less susceptible to blending and probe lower optical depths, revealing the inner ejecta at earlier epochs than is possible in the optical \citep{Dessart_2015,Hsiao_2018,Davis2019}. In particular, while optical \ion{He}{1} lines are often blended with neighboring transitions, the NIR \ion{He}{1} $1.083\,\mu$m and $2.0581\,\mu$m lines are comparatively isolated, making them among the cleanest diagnostics of the helium distribution and velocity structure within the ejecta \citep{Maurer2010,Davis_2021,Medler2025}. These transitions remain prominent from the photospheric phase through the nebular phases and therefore provide powerful probes of the excitation, kinematics, and geometry of the helium-rich ejecta \citep{Medler2025}.

The geometry of stripped-envelope supernova (SE-SNe) explosions remains poorly understood. Increasing evidence from nebular spectroscopy and spectropolarimetry indicates that asymmetry is common among core-collapse supernovae, including SNe~IIb \citep{Maund_2007,Wang_2008}, with departures from spherical symmetry extending across multiple compositional layers \citep{Tanaka_2012,Finn_2016,stevance2019}. These asymmetries are thought to arise during the core-collapse explosion itself through large-scale hydrodynamic instabilities, such as neutrino-driven convection and the standing accretion shock instability (SASI), which produce anisotropic shock propagation and asymmetric mixing of freshly synthesized $^{56}$Ni into the helium-rich layers \citep{Kifonidis_2003,Janka_2017, giudici2025}. Because the isolated NIR \ion{He}{1} $1.083\,\mu$m and $2.0581\,\mu$m transitions remain strong well into the nebular phase, they provide a particularly powerful means of probing the three-dimensional distribution of helium-rich ejecta and constraining the geometry of the explosion \citep{Maeda2008, Taubenberger_2011}.

\begin{figure*}
    \centering
    \includegraphics[width=\linewidth]{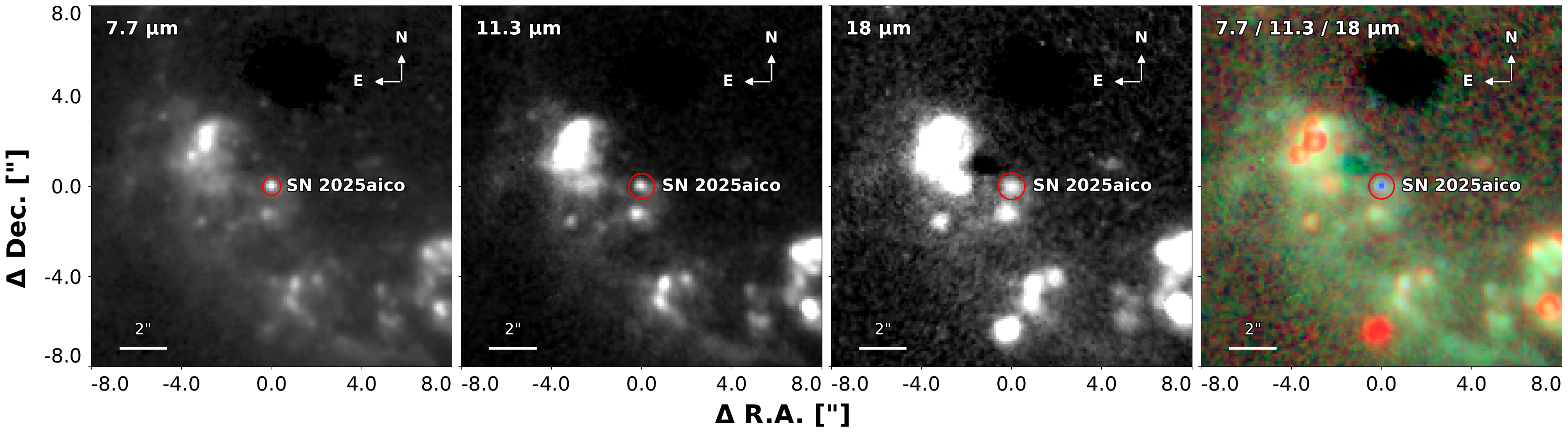}
    \caption{JWST/MIRI imaging of the field around SN~2025aico in the F770W ($7.7\,\mu$m), F1130W ($11.3\,\mu$m), and F1800W ($18.0\,\mu$m) filters, plus an RGB composite (R = F1800W, G = F1130W, B = F770W). Each panel shows an approximately $16''\times16''$ field of view centered on the SN.}
    \label{fig:miri_cutout}
\end{figure*}

Thermal dust emission can appear as an infrared excess above the underlying SN continuum, revealing dust that may remain undetectable at optical wavelengths \citep{Gall_2014,Wesson2015}. Ground-based NIR observations primarily probe dust near its sublimation temperature ($T\gtrsim2000$~K), whereas much of the emission from lower-temperature dust occurs at longer wavelengths that are difficult to access from the ground \citep{Fox2011, Rho2018,Park2025,2025ApJ...993..191M}. Early-time observations of CCSNe typically infer newly formed dust masses of $\lesssim10^{-3}$--$10^{-2},M_\odot$, substantially below the $\sim0.1$--$1,M_\odot$ per event predicted by theoretical models and estimated to be necessary for CCSNe to account for the dust content of high-redshift galaxies \citep{Gall_2014,Wesson2015}. In contrast, observations of older CCSNe and supernova remnants have revealed substantially larger dust reservoirs: SN~1987A evolved from $\sim10^{-3}M_\odot$ at $\sim615$~d to $\sim0.6$--$0.8 M_\odot$ after roughly two decades \citep{Matsuura_2011, Wesson2015}, while estimates for Cas~A reach $\sim0.4$--$0.6 M_\odot$ \citep{De_Looze_2016}. Similar large dust masses have been inferred in other CCSN remnants, demonstrating that individual CCSNe are capable of producing several tenths of a solar mass of dust \citep{Gomez_2012,Owen_2015,De_Looze_2019}. However, whether CCSNe ultimately provide a substantial fraction of the cosmic dust budget remains uncertain because the timescale over which these large dust masses develop and the fraction of grains that survive processing by the reverse shock and are injected into the ISM remain poorly constrained \citep{Gall_2014,Rho2018}. The mid-infrared wavelength coverage provided by the \textit{James Webb Space Telescope} (\textit{JWST}), particularly with its Mid-Infrared Instrument (MIRI), enables a more complete characterization of dust emission in CCSNe and provides important constraints on their potential contribution to dust production in the early Universe \citep{Hosseinzadeh_2023,Shahbandeh2023,2025ApJ...993..191M,Davis2026}.

% The combination of nebularoptical and infrared observations offers a powerful means of investigating both the geometry and composition of SE-SNe ejecta. Non-standard line profiles provide insight into the three-dimensional distribution of the explosion products, while infrared observations probe the presence of newly formed or pre-existing dust that may otherwise remain hidden at optical wavelengths. Obtaining both types of observations for the same object therefore provides a unique opportunity to study the connection between explosion geometry, progenitor structure, and dust formation in Type~IIb SNe.

Here we present optical, NIR, and JWST MIR observations of the nearby Type~IIb supernova SN~2025aico. SN~2025aico was discovered on 2025 December 24 at $04^{\rm h}50^{\rm m}58.848^{\rm s}$ UT (MJD~61033.20) at coordinates (J2000) $\alpha=11^{\rm h}29^{\rm m}16^{\rm s}.362$, $\delta=+20\arcdeg35\arcmin08\farcs41$, in the nearby galaxy LEDA~35384 ($z=0.004737\pm0.000117$; \citealt{zhao2026}). The event exhibited an early shock-cooling phase lasting fewer than four days \citep{Andrews2025}. Early optical photometry and spectroscopy of SN~2025aico have been modeled by \citet{zhao2026}, who combined a shock-cooling component with a radioactively powered diffusion model to constrain the explosion epoch, rise time, and ejecta properties. 

Our observations extend this early-time picture with extensive optical and NIR spectroscopic coverage from the first few days after explosion through the nebular phase. Combined with the proximity of SN~2025aico, this dataset provides a rare opportunity to investigate the kinematic and geometric evolution of the helium-rich ejecta using the isolated NIR \ion{He}{1} $1.083\,\mu$m and $2.0581\,\mu$m transitions. In addition, serendipitous late-time \textit{JWST}/MIRI observations extend the wavelength coverage into the MIR, enabling us to examine the spectral energy distribution and search for evidence of warm dust emission. The combination of early-time spectroscopy, nebular-phase NIR observations, and late-time MIR imaging provides a unique opportunity to investigate the connection between progenitor structure, explosion geometry, and dust formation in H-rich SE-SNe. 

This paper is organized as follows. In Section~\ref{sec:observations_and_data}, we describe the observations, data reduction, and flux calibration of the optical, NIR, and \textit{JWST} datasets. Section~\ref{sec:emissionfeatures} presents the spectral evolution of SN~2025aico and identifies the principal emission features. In Section~\ref{sec:spectroscopy}, we examine the evolution of the spectral line profiles and velocities to investigate the kinematics and geometry of the ejecta. Section~\ref{sec:comparison} compares these spectroscopic properties with those of other well-studied Type~IIb SNe. In Section~\ref{sec:progenitor}, we discuss the constraints that these observations place on the progenitor system. Section~\ref{sec:spectralenergy} presents the construction and modeling of the spectral energy distribution and examines the origin of the observed MIR excess. Finally, Section~\ref{sec:Conclusion} summarizes our main results and discusses their broader implications.

%%-------------------------------------------------------------------------
\section{Observations and Data Reduction}\label{sec:observations_and_data}

Optical and NIR spectral observations of SN~2025aico were obtained between roughly 2 days after explosion until approximately 167 days. Additional serendipitous detections of SN~2025aico in MIR imaging were obtained with \textit{JWST}/MIRI under Cycle~4 Program~7041 \citep{Berg_2025jwst}. Here, we present and analyze this data.

The NIR spectroscopy of SN~2025aico was obtained with IRTF/SpeX \citep{Rayner1998} and Keck~II/NIRES \citep{McLean_1998} as part of the Hawaii Infrared Supernova Study \citep[HISS;][]{Medler2025,Hoogendam2025a,Hoogendam2025b}, a major observational program focused on NIR observations of SNe. For details on the data reduction method of these IRTF/SpeX and Keck~II/NIRES see \citet{Medler2025}. Additional NIR spectroscopy obtained with SOAR/TripleSpec \citep{Herter2020,Hosseinzadeh2025} was retrieved from the Transient Name Server. 

Optical spectroscopy obtained by our group includes observations with Keck~I/LRIS and Keck~II/KCWI. Additional optical spectra were also obtained with MMT/Binospec \citep{Fabricant_2019,Weiner2026}, SOAR/Goodman, Gemini-North/GMOS-N \citep{Hook2004}, Apache Point Observatory (APO), and Magellan~I/IMACS (Inamori-Magellan Areal Camera and Spectrograph; \citealt{Dressler2011,Burns2021}). The Gemini-North, MMT, and Magellan~I observations formed part of the early optical spectroscopic sequence previously presented by \citet{zhao2026}, which covered approximately the first 70 days after explosion. Here, we combine this initial data with our later optical and NIR observations to construct a spectroscopic sequence extending from the shock-cooling phase into the nebular phase.

The additional publicly available optical spectra that we use in this work were retrieved from the Transient Name Server \citep{TNS2025aico} and the Weizmann Interactive Supernova Data Repository (WISeREP)\footnote{\url{https://www.wiserep.org/object/29614}} \citep{WISeREP2025aico}. These spectra were reduced by the contributing teams using their standard pipelines. Gemini-North/GMOS-N spectra were processed with the standard Gemini \textsc{IRAF} pipeline, SOAR spectra with the PASSTA pipeline, and APO and Magellan~I/IMACS spectra using standard \textsc{IRAF} procedures \citep{Hamuy2006,Folatelli2013}.

The HISS optical and NIR data were reduced and analyzed by our group. Keck~I/LRIS spectra were reduced following the methods of \citet{2025ApJ...993..191M}. Keck~II/NIRES spectra were reduced with \texttt{PypeIt} \citep{Prochaska2020} following \citet{Medler2025}, then flux calibrated and corrected for telluric absorption using the custom routines described by \citet{2025ApJ...993..191M}. Keck~II/KCWI observations were reduced using \texttt{PypeIt} in KCWI mode. The KCWI science and standard-star observations were bias corrected, flat-fielded, wavelength calibrated, and combined into three-dimensional datacubes. One-dimensional spectra were then extracted from the datacubes using point-spread-function fitting to the location of SN\,2025aico. These spectra were then flux calibrated using a standard-star observed on the same night and the SN\,2025aico observations. The blue- and red-channel KCWI spectra for each epoch were then combined to produce a single $3800 - 8800$~\AA\ spectrum for each KCWI epoch. A complete log of the observations used in this work is provided in Table~\ref{tab:A2}.

\begin{deluxetable}{ccccc}
\tablecaption{Log of spectroscopic observations of SN~2025aico spanning $+1.64$ to $+166.82$\,d post-explosion.\label{tab:A2}}
\tablehead{
  \colhead{Obs.\ Date} & \colhead{Phase} &
  \colhead{Exp.\ Time} & \colhead{Telescope/Instrument} & \colhead{Source} \\
  \colhead{[MJD]} & \colhead{[days]} & \colhead{[s]} & \colhead{} & \colhead{}
}
\startdata
\multicolumn{5}{c}{\textit{Optical spectra}} \\
\hline \\[-6pt]
61034.34 & $+1.64$   & \nodata & SOAR/Goodman        & WiseRep\tablenotemark{a} \\
61034.62 & $+1.92$   & 600     & Gemini/GMOS-N       & \citet{zhao2026} \\
61043.00 & $+10.26$  & \nodata & SOAR/Goodman           & WiseRep\tablenotemark{a} \\
61056.54 & $+23.74$  & 180     & MMT/Binospec        & \citet{zhao2026} \\
61095.00 & $+62.03$  & 450     & SOAR/Goodman        & WiseRep\tablenotemark{a} \\
61098.15 & $+65.16$  & 900     & Magellan~I/IMACS    & \citet{zhao2026} \\
61107.36 & $+74.33$  & 300     & Keck~I/LRIS         & This work. \\
61166.36 & $+133.06$ & 1200    & Keck~II/KCWI        & This work. \\
61171.42 & $+138.11$ & 300     & Keck~I/LRIS         & This work. \\
61200.27 & $+166.82$ & 1200    & Keck~II/KCWI        & This work. \\
\hline \\[-6pt]
\multicolumn{5}{c}{\textit{NIR spectra}} \\
\hline \\[-6pt]
61037.54 & $+4.83$   & 2400 & IRTF/SpeX       & This work. \\
61048.60 & $+15.83$  & 1920 & IRTF/SpeX       & This work. \\
61052.50 & $+19.72$  & 1440 & IRTF/SpeX       & This work. \\
61091.50 & $+58.54$  & 1440 & IRTF/SpeX       & This work. \\
61096.50 & $+63.52$  & 1680 & IRTF/SpeX       & This work. \\
61099.00 & $+66.00$  & 4400 & SOAR/TripleSpec & WiseRep\tablenotemark{a} \\
61135.42 & $+102.27$ & 120  & Keck~II/NIRES   & This work. \\
61158.41 & $+125.16$ & 300  & Keck~II/NIRES   & This work. \\
\enddata
\tablecomments{Phases are given in the rest frame relative to the date of explosion,
$\mathrm{MJD}_{\mathrm{exp}}=61032.69$ \citep{Andrews2025}, which was determined through light-curve modeling presented in \citet{zhao2026}. The Source column identifies the origin of each spectrum. Spectra retrieved from TNS and WISeREP were reduced by the respective contributing groups. The SOAR/GHTS exposure time was not reported on WISeREP.}
\tablenotetext{a}{\url{https://www.wiserep.org/object/29614}}
\end{deluxetable}

\subsection{JWST MIRI Photometry}\label{sec:jwst_photometry}
SN~2025aico was serendipitously observed with \textit{JWST}/MIRI at $+124.4$\,d after explosion as part of Cycle~4 Program~7041, a survey designed to image nearby star-forming galaxies \citep{Berg_2025jwst}. Although SN~2025aico was not the intended target, it fell within the MIRI field of view during observations of its host galaxy. Imaging was obtained in the \textit{F770W}, \textit{F1130W}, and \textit{F1800W} filters, which have effective wavelengths of $7.528$, $11.298$, and $17.875\,\mu$m, respectively. Figure~\ref{fig:miri_cutout} shows the SN field in each of the three filters together with an RGB composite.

The fully calibrated MIRI imaging data were downloaded from the Mikulski Archive for Space Telescopes (MAST)\footnote{\url{https://mast.stsci.edu/portal/Mashup/Clients/Mast/Portal.html}} and analyzed without additional image reprocessing. Final aperture-photometry measurements were obtained using the \texttt{space\_phot} package\footnote{\url{https://space-phot.readthedocs.io/en/latest/_modules/space_phot.html}; see also \url{https://spacetelescope.github.io/jdat_notebooks/index.html}} and circular apertures with radii of $1.0\times{\rm FWHM}$. For each filter, the local background was estimated within a circular annulus centered on the SN position. The annulus radii were selected to lie beyond the wings of the point-spread function (PSF) while avoiding contaminating flux from the host galaxy. The adopted aperture and background-annulus parameters are listed in Table~\ref{tab:jwst_photometry}.

%The sky background was estimated from annuli of $(8.92,14.64)$, $(6.00,9.00)$, and $(5.00,8.50)$ pixels for the F770W, F1130W, and F1800W filters, respectively.

\begin{table}[t]
\centering
\caption{\textit{JWST}/MIRI aperture photometry of SN~2025aico at
$+124.4$\,d.}
\label{tab:jwst_photometry}

\begin{tabular}{lccc}
\hline\hline
Filter & $F_\nu$ [$\mu$Jy] & $m_{\rm AB}$ & FWHM [px] \\
\hline
\textit{F770W}  & $32.50 \pm 4.55$ & $20.12 \pm 0.15$ & $2.445$ \\
\textit{F1130W} & $26.30 \pm 3.39$ & $20.35 \pm 0.14$ & $3.409$ \\
\textit{F1800W} & $32.10 \pm 3.84$ & $20.13 \pm 0.13$ & $5.373$ \\
\hline
\end{tabular}

\begin{minipage}{0.98\columnwidth}
\vspace{0.4em}
\footnotesize
\textit{Note.} Photometry was measured using a circular aperture with a radius equal to one FWHM. The local sky background was estimated from annuli placed outside the PSF while avoiding contaminating host-galaxy emission. Statistical flux uncertainties were determined from the
measured signal-to-noise ratios, S/N $\approx 9.1$, $8.1$, and $8.9$ for \textit{F770W}, \textit{F1130W}, and \textit{F1800W}, respectively. The variation in the aperture-corrected photometry across the tested aperture radii was treated as an additional systematic uncertainty and added in quadrature to the statistical uncertainty.
\end{minipage}

\end{table}
To evaluate the adopted background annuli and aperture corrections, we independently constructed a curve of growth for SN~2025aico in each MIRI filter using \texttt{photutils}. We performed circular aperture photometry at each aperture radius tabulated in the MIRI imaging aperture-correction reference file \texttt{jwst\_miri\_apcorr\_0014.fits}, obtained from the Calibration Reference Data System (CRDS; \citealt{2016A&C....16...41G}). We retained apertures with radii at least $0.3$ pixels smaller than the inner radius of the corresponding background annulus. At each radius, the local background was estimated using the $3\sigma$-clipped median of the pixels within the adopted annulus and subtracted from the aperture sum. For this curve-of-growth analysis, the image values were converted from $\mathrm{MJy}\,\mathrm{sr}^{-1}$ to $\mu\mathrm{Jy}$ per pixel using the pixel area recorded by the \texttt{PIXAR\_A2} header keyword. We then applied the corresponding tabulated aperture correction and converted the corrected flux densities to AB magnitudes.

We quantified the stability of the corrected photometry using the peak-to-peak span, defined as
\[
\Delta m_{\rm span}
=
\max(m_{\rm corr})-\min(m_{\rm corr}),
\]
across the tested aperture radii. The resulting spans were $0.094$, $0.040$, and $0.044$\,mag for \textit{F770W}, \textit{F1130W}, and \textit{F1800W}, respectively. To conservatively account for the observed dependence of the photometry on aperture selection, we treated the full peak-to-peak span in each filter as an aperture-dependent systematic uncertainty and added it in quadrature to the statistical magnitude uncertainty,
\[
\sigma_{m,\rm final}
=
\left(
\sigma_{m,\rm stat}^{2}
+
\Delta m_{\rm span}^{2}
\right)^{1/2}.
\]
This yields final magnitude uncertainties of $0.15$, $0.14$, and $0.13$\,mag for \textit{F770W}, \textit{F1130W}, and \textit{F1800W}, respectively.

For consistency, the aperture-dependent magnitude uncertainty was also converted to a flux-density uncertainty according to
\[
\sigma_{F,\rm ap}
=
\frac{\ln 10}{2.5}F_\nu\Delta m_{\rm span},
\]
giving $2.81$, $0.97$, and $1.30\,\mu\mathrm{Jy}$ for \textit{F770W}, \textit{F1130W}, and \textit{F1800W}, respectively. These were added in quadrature to the statistical flux uncertainties,
\[
\sigma_{F,\rm final}
=
\left(
\sigma_{F,\rm stat}^{2}
+
\sigma_{F,\rm ap}^{2}
\right)^{1/2},
\]
yielding final flux-density uncertainties of $4.55$, $3.39$, and $3.84\,\mu\mathrm{Jy}$, respectively.

The larger variation in \textit{F770W} may arise in part from its more compact PSF in pixel units compared with the longer-wavelength filters, which can make the measured flux more sensitive to subpixel centering and aperture placement. Local background structure may also contribute to the increased variation with aperture radius. Despite the larger span in \textit{F770W}, the corrected photometry remains stable to within $0.1$\,mag in all three filters and exhibits no sustained trend with aperture radius. The residual aperture dependence is therefore incorporated into the final uncertainties rather than introducing an unaccounted-for source of uncertainty in the reported photometry.

As an independent check, we repeated the curve-of-growth analysis using an isolated, unresolved source in the same mosaics at coordinates (J2000) $\alpha=11^{\rm h}29^{\rm m}17\fs5489$ and $\delta=+20\arcdeg34\arcmin48\farcs204$. For an unresolved source, the corrected flux should remain approximately independent of aperture radius after application of the appropriate encircled-energy correction. We therefore measured this source using the tabulated apertures enclosing 20--80\% of the MIRI PSF and calculated the peak-to-peak span of the corrected magnitudes. The resulting spans were $0.047$, $0.018$, and $0.081$\,mag in \textit{F770W}, \textit{F1130W}, and \textit{F1800W}, respectively. The \textit{F770W} and \textit{F1130W} measurements exhibit little dependence on aperture radius, whereas the \textit{F1800W} measurements show modest systematic brightening toward larger apertures, indicating greater sensitivity to the local background or aperture correction in that filter. Nevertheless, the corrected test-source photometry remains stable to within $0.1$\,mag in all three filters. This independent test supports the overall aperture-correction procedure and demonstrates that residual aperture-dependent variations at the $\lesssim 0.1$\,mag level can occur. Because the aperture dependence measured directly for SN~2025aico is already propagated into its final photometric uncertainties, the test-source variation is not included as an additional uncertainty term.

For the final \texttt{space\_phot} measurements, aperture corrections were calculated by linearly interpolating the values in \texttt{jwst\_miri\_apcorr\_0014.fits} to the exact $1.0\times{\rm FWHM}$ aperture radius adopted in each filter. The background-subtracted aperture sums were converted from $\mathrm{MJy}\,\mathrm{sr}^{-1}$ to Jy using the pixel solid angle, \texttt{PIXAR\_SR}, recorded in each image header, and were then multiplied by the interpolated aperture-correction factors. The aperture-dependent uncertainties derived from the curve-of-growth analysis were added in quadrature to the statistical uncertainties from the final \texttt{space\_phot} measurements. The resulting flux densities, AB magnitudes, total uncertainties, and adopted FWHM values are presented in Table~\ref{tab:jwst_photometry}.
%%-------------------------------------------------------------------------
\section{Spectral Emission Features}\label{sec:emissionfeatures}

\begin{figure*}[t!]
    \centering
    \includegraphics[width=\textwidth]{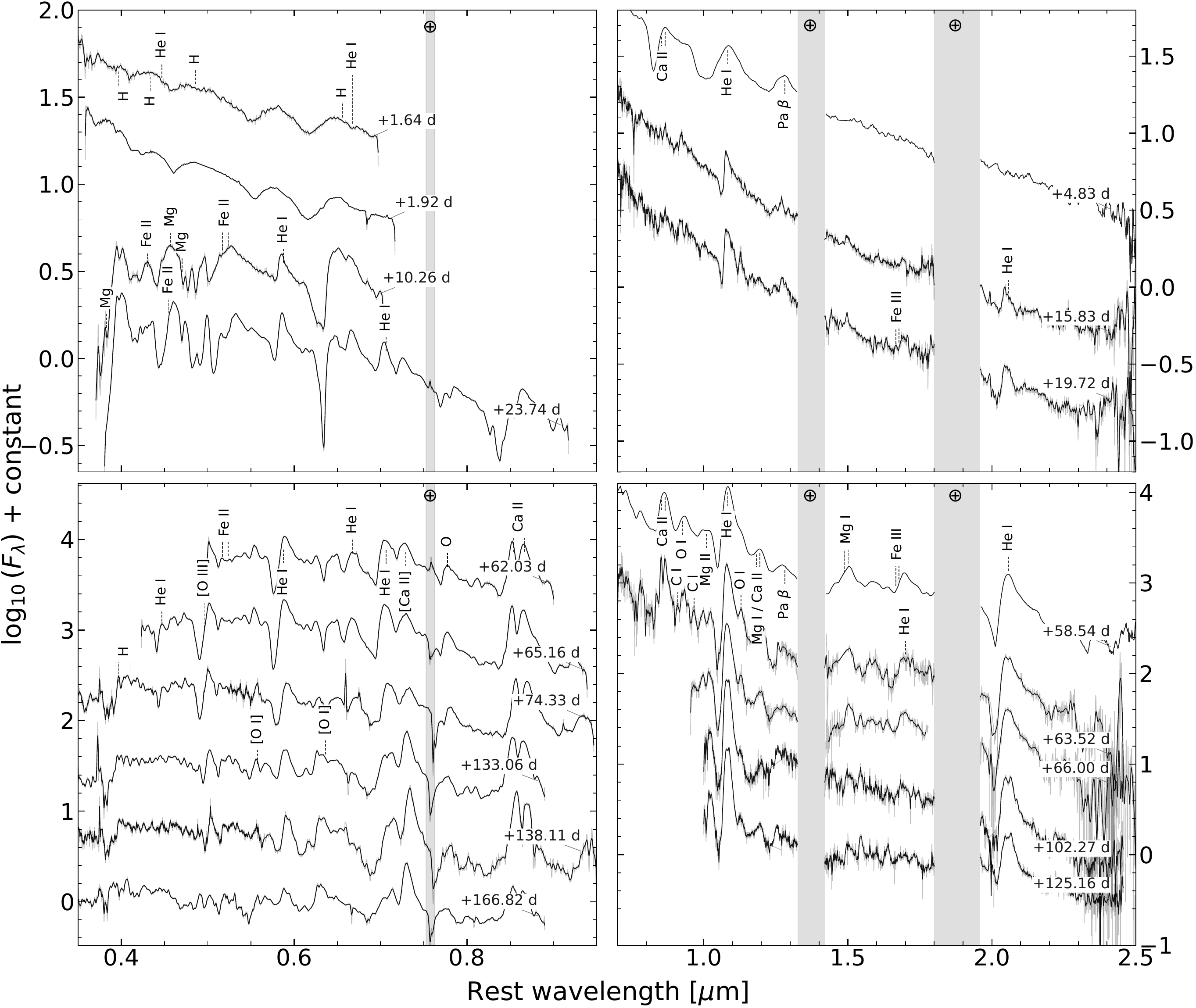}
    \caption{Optical (left) and NIR (right) spectral sequences of SN~2025aico. The optical sequence spans +1.6 to +166.8\,d post-explosion, while the NIR sequence spans +4.8 to +125.2\,d. Telluric absorption bands are indicated by the $\oplus$ symbol. Key spectral features are marked with dotted lines.}
    \label{fig:both}
\end{figure*}

The optical and NIR spectroscopic evolution of SN~2025aico from $+1.64$ to $+166.82$\,d post-explosion is shown in Figure~\ref{fig:both}. The spectra of SN~2025aico transitions from a hot photospheric phase to a nebular phase dominated by strong forbidden lines. The broad wavelength coverage and dense temporal cadence enable us to follow the evolution of hydrogen-, helium-, oxygen-, and calcium-rich material from the outer envelope into the inner ejecta, providing constraints on ejecta stratification, mixing, and explosion geometry.

The earliest optical spectra, obtained at $+1.64$ and $+1.92$\,d during the shock-cooling phase, are dominated by a blue continuum with broad H~\textsc{i} and He~\textsc{i} features, indicating a hot and rapidly expanding photosphere shortly after explosion. By $+10.26$\,d, the continuum has cooled substantially and strong Fe~\textsc{ii} features have developed throughout the $0.45$--$0.55\,\mu$m region. Prominent He~\textsc{i} and Mg~\textsc{ii} absorption features are also present, while the Ca~\textsc{ii} NIR triplet progressively strengthens and becomes one of the dominant spectral features. By approximately +62\,d, [Ca~\textsc{ii}] emission begins to emerge, while the forbidden [O~\textsc{i}] transitions become clearly apparent by +133\,d, tracing the progression toward the nebular phase as the ejecta become increasingly optically thin. Throughout the NIR time series, He~\textsc{i} $1.083\,\mu$m remains one of the strongest spectral features and is detected from the earliest NIR epoch at +4.83\,d through the latest observation at +125.16\,d. The He~\textsc{i} $2.0581\,\mu$m transition is confidently detected beginning at +15.83\,d and remains visible through +125.16\,d. In the following sections, we use the evolution of prominent hydrogen, helium, oxygen, and calcium features to probe the ejecta stratification, mixing, and geometry.

\subsection{Hydrogen}
The evolution of the hydrogen features traces the rapidly thinning outermost ejecta and provides a direct probe of the residual hydrogen envelope characteristic of SNe~IIb. Hydrogen is detected through several Balmer transitions in the optical spectra. A broad, blended absorption feature between $0.3970$ and $0.4102\,\mu$m, arising primarily from H$\epsilon$ at $0.3970\,\mu$m and H$\delta$ at $0.4102\,\mu$m, persists through $+23.74$\,d and reappears by $+74.33$\,d. H$\beta$ at $0.4861\,\mu$m is visible during the early photospheric phase, while H$\alpha$ at $0.6563\,\mu$m is strongest during the first $\sim10\,d$ and subsequently weakens substantially as the hydrogen-rich envelope becomes increasingly optically thin, before developing stronger emission again during the nebular phase at approximately +100--140\,d. In the NIR, H\,\textsc{i} Paschen~$\beta$ at $1.2818\,\mu$m first appears at $+4.83$\,d and gradually weakens with time, consistent with the evolution of the other hydrogen features.

\subsection{Helium}
Helium is detected throughout both the optical and NIR spectral sequences. In the optical, \ion{He}{1} at $0.4471\,\mu$m is observed between $+1.92$ and $+23.74$\,d before reappearing by $+65.16$\,d. The strongest optical helium transition, \ion{He}{1} at $0.5876\,\mu$m, is already present in the earliest spectrum at $+1.64$\,d, although it is initially broad and blended before becoming more clearly distinguishable as the ejecta expand and cool. \ion{He}{1} at $0.6678\,\mu$m is detected from the first epoch, while \ion{He}{1} at $0.7065\,\mu$m becomes apparent by $+23.74$\,d. In the NIR, He~\textsc{i} at $1.0830\,\mu$m is detected throughout the full observing sequence. The He~\textsc{i} transition at $2.0581\,\mu$m is first confidently detected at +15.83\,d, while the weaker He~\textsc{i} feature at $1.7002\,\mu$m emerges by +63.52\,d.

\subsection{Carbon}
Carbon is detected in the NIR through C\,\textsc{i} at $0.9093$ and $0.9658\,\mu$m, both of which emerge at $+63.52$\,d as the ejecta become increasingly optically thin, revealing the underlying C/O-rich core.

\subsection{Oxygen}
Oxygen is identified in both the optical and NIR spectra. In the optical, O\,\textsc{i} at $0.7774\,\mu$m first appears at $+62.03$\,d. At later epochs, the forbidden [O\,\textsc{i}] transitions at $0.5577$, $0.6300$, and $0.6363\,\mu$m are clearly detected by $+133.06$\,d, marking the continued transition toward the nebular phase. An emission feature near
$0.5007\,\mu$m is also tentatively identified as [O\,\textsc{iii}] $\lambda5007$ in the late-time optical spectra. In the NIR, O\,\textsc{i} at $0.9264\,\mu$m becomes visible at $+58.54$\,d, followed by O\,\textsc{i} at $1.1290\,\mu$m at $+63.52$\,d.

\subsection{Magnesium}
Magnesium is observed in both the optical and NIR spectra. In the optical, Mg\,\textsc{i} at $0.3829\,\mu$m, $0.4571\,\mu$m, and Mg\,\textsc{ii} at $0.4703\,\mu$m first appear at $+10.26$\,d. The features at $0.4571$ and $0.4703\,\mu$m remain visible through $+23.74$\,d, while the $0.3829\,\mu$m feature persists until approximately $+30$\,d. In the NIR, Mg\,\textsc{ii} at $1.0092\,\mu$m is first detected at $+58.54$\,d. Blended Mg\,\textsc{i}/Ca\,\textsc{ii} emission between $1.1828$ and $1.1950\,\mu$m and a blended Mg\,\textsc{i} feature spanning $1.4878$--$1.5033\,\mu$m are also present during the later NIR observations.

\subsection{Calcium}
Calcium is identified through the forbidden [\ion{Ca}{2}] emission lines at $0.7291$ and $0.7324\,\mu$m, which emerge by $+62.03$\,d. The permitted Ca\,\textsc{ii} NIR triplet at $0.8498$, $0.8542$, and $0.8662\,\mu$m is detected throughout the observed spectral sequence and becomes one of the dominant spectral features as the supernova evolves.

\subsection{Iron}
Iron becomes increasingly prominent as the ejecta cool. Fe\,\textsc{ii} at $0.4303\,\mu$m is observed between $+10.26$ and approximately $+30$\,d, while Fe\,\textsc{ii} at $0.4549\,\mu$m appears by $+23.74$\,d and remains visible through approximately $+30$\,d. The strong Fe\,\textsc{ii} transitions at $0.4924$, $0.5018$, $0.5169$, and $0.5235\,\mu$m together produce a broad blended absorption complex that develops by $+10.26$\,d. At late times, an emission feature near $1.64\,\mu$m emerges in the $+125$\,d NIR spectrum and is identified as the forbidden [Fe\,\textsc{ii}] $1.6440\,\mu$m transition, consistent with the ejecta becoming optically thin. Fe\,\textsc{iii} at $1.6722\,\mu$m is also present, while some contribution from \ion{He}{1} at $1.7002\,\mu$m cannot be ruled out.

%%-------------------------------------------------------------------------
\section{Spectroscopic Analysis}\label{sec:spectroscopy}
The spectroscopic sequence described above traces multiple compositional layers of the ejecta as they evolve from the photospheric to the nebular phase. We now use the velocity evolution and line-profile morphology of hydrogen, helium, oxygen, and calcium to investigate the radial stratification, radioactive mixing, and large-scale geometry of SN~2025aico.
%%----------------------
\subsection{Hydrogen Velocity Evolution}

\begin{figure*}
    \centering
    \includegraphics[width=1\linewidth]{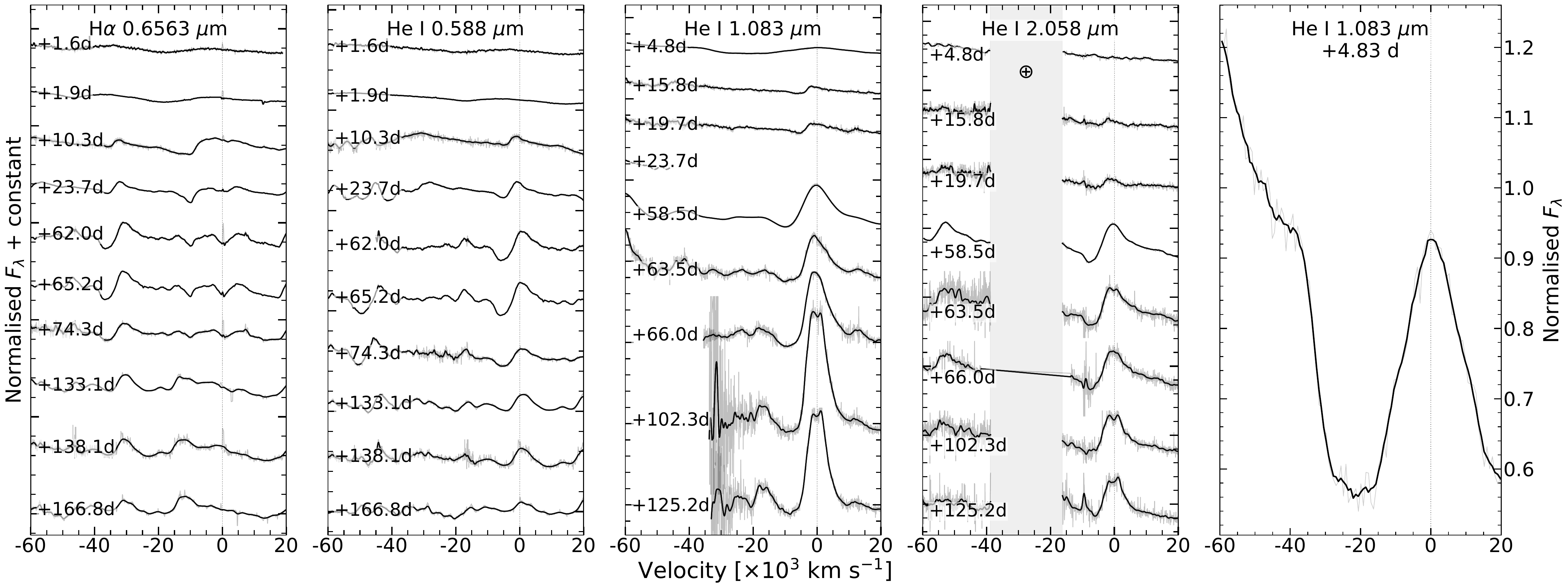}
    \caption{Evolution of the H$\alpha$ $0.6563\,\mu$m, He~\textsc{i} $0.5876\,\mu$m, He~\textsc{i} $1.083\,\mu$m, and He~\textsc{i} $2.0581\,\mu$m line profiles of SN~2025aico, plotted in velocity space relative to the rest wavelength of each transition. The optical profiles extend from the earliest post-explosion epochs through the late-time optical sequence, while the NIR He~\textsc{i} profiles extend through +125.2\,d. The H$\alpha$ profile exhibits a high-velocity absorption component during the early photospheric phase that weakens with time. The helium profiles evolve from P-Cygni absorption features to double-peaked emission profiles at late epochs.}
    \label{fig:aicoHe}
\end{figure*}
During the early photospheric phase of SNe~IIb, the line-forming region lies within the outer hydrogen-rich envelope. H$\alpha$ is strong and relatively isolated compared with many other optical transitions at these epochs, making its absorption minimum a useful probe of the velocity extent and evolution of the residual hydrogen-rich material (Figure~\ref{fig:aicoHe}).

To measure the velocity evolution of the hydrogen-rich ejecta during the optically thick phase, we fit the H$\alpha$ absorption profile at each epoch. We fit a linear local continuum to relatively featureless spectral regions blueward and redward of the line and subtract it to isolate the P-Cygni profile. We fit the absorption minimum with an inverted Gaussian. The velocity uncertainty combines the Gaussian-centroid uncertainty with the continuum-placement uncertainty estimated from 100 Monte Carlo bootstrap realizations, added in quadrature \citep{Lucy_1999}. The H$\alpha$ $0.6563\,\mu$m absorption velocity declines from $\sim14{,}000$\,km\,s$^{-1}$ at $+1.6$\,d to $\sim10{,}000$\,km\,s$^{-1}$ by approximately $+10$\,d, after which it remains nearly constant throughout the rest of the photospheric phase (Figure~\ref{fig:absorptionevolution}). In addition to the principal absorption component, H$\alpha$ exhibits a detached high-velocity component at the earliest epochs. This component weakens rapidly and is no longer apparent at later phases, consistent with the declining optical depth of the rapidly expanding outer hydrogen-rich ejecta \citep{Dessart_2012,Dessart_2022}.

\subsection{Helium Velocity Evolution}

The velocity evolution of the \ion{He}{1} lines traces changes in the line-forming region within the helium-rich ejecta. Unlike H$\alpha$, the NIR helium lines depend strongly on non-thermal excitation and ionization produced by radioactive energy deposition. The $^{56}$Ni--$^{56}$Co--$^{56}$Fe decay chain releases energy primarily through $\gamma$-rays, with positrons also produced in a fraction of $^{56}$Co decays. At the epochs considered here, $\gamma$-rays carry the majority of the non-neutrino radioactive decay energy relevant to the ejecta. Interactions of these $\gamma$-rays with the expanding ejecta, particularly through Compton scattering, produce energetic non-thermal electrons that excite and ionize helium \citep{Lucy_1991}. The resulting He~\textsc{i} line formation is therefore sensitive to both the spatial distribution of radioactive material and the transport and deposition of radioactive energy throughout the ejecta.

We focus on the relatively isolated He~\textsc{i} $1.083\,\mu$m and $2.0581\,\mu$m transitions, which remain detectable across a broad range of photospheric epochs and provide complementary constraints on the kinematics of the helium-rich ejecta \citep{Maurer2010,Davis_2021, 2025ApJ...993..191M}.

\begin{figure*}
    \centering
    \includegraphics[width=0.7\linewidth]{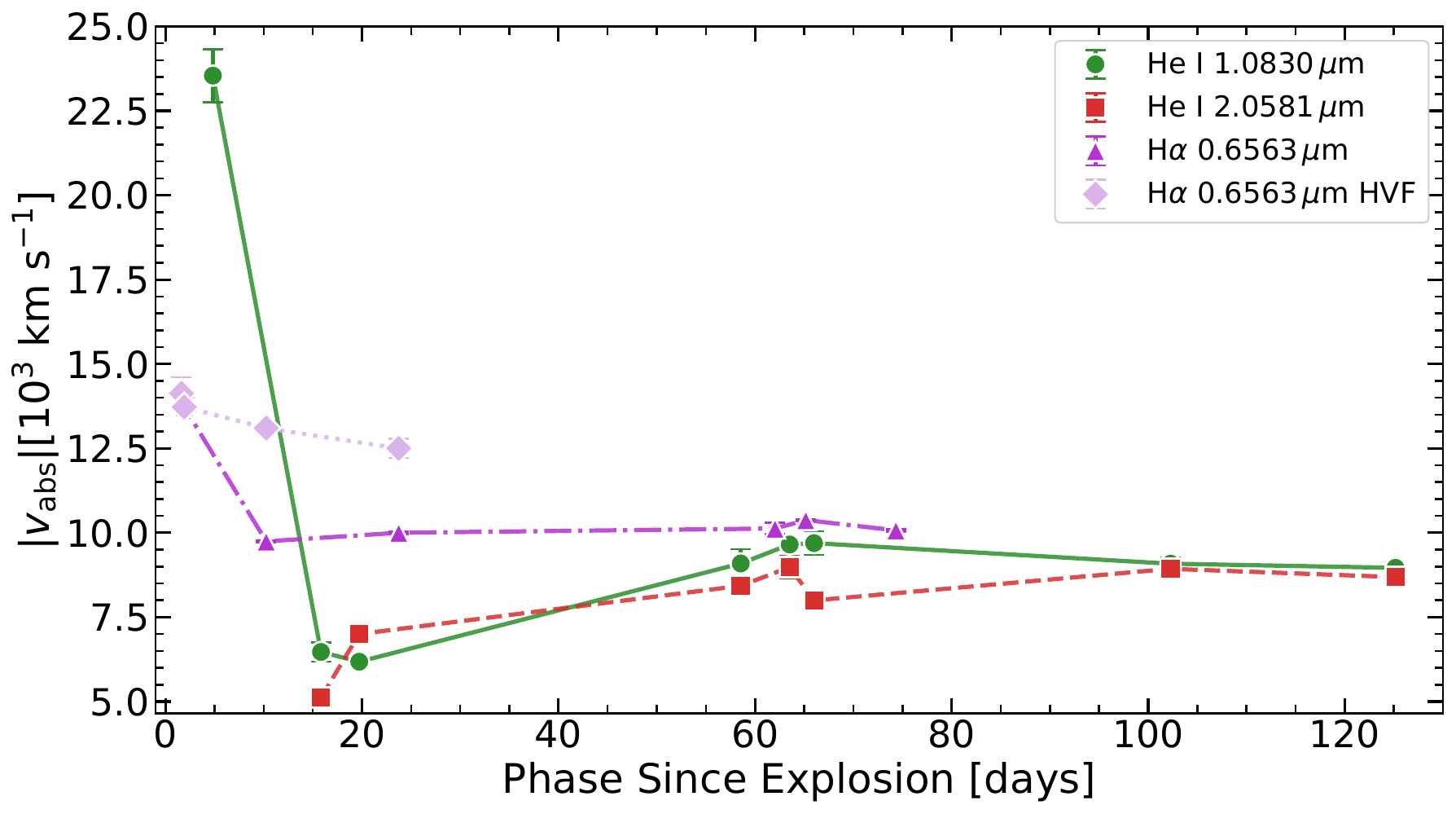}
    \caption{Absorption velocity evolution of \ion{He}{1} $1.083\,\mu$m (green squares), \ion{He}{1} $2.0581\,\mu$m (red squares), H$\alpha$ $0.6563\,\mu$m (purple triangles), and the high-velocity component of H$\alpha$ $0.6563\,\mu$m (light purple) in SN~2025aico as a function of phase since explosion.}
    \label{fig:absorptionevolution}
\end{figure*}

We measure the \ion{He}{1} $1.083\,\mu$m and $2.0581\,\mu$m absorption velocities using the same fitting method described above for H$\alpha$. The resulting evolution, shown in Figure~\ref{fig:absorptionevolution}, can be divided into three distinct phases. During the first phase, the \ion{He}{1} $1.083\,\mu$m absorption velocity declines rapidly from $\sim23{,}500~\mathrm{km~s^{-1}}$ at $+4.8$\,d to a minimum of $\sim6{,}200~\mathrm{km~s^{-1}}$ at $+19.7$\,d. At these early epochs, the helium excitation is dominated by the thermal conditions associated with the shock-heated, cooling ejecta. As the ejecta expand and cool, the effective line-forming region recedes to progressively lower velocities. Consequently, this initial decline is primarily associated with the shock-cooling evolution and is expected to be relatively insensitive to the spatial distribution of $^{56}$Ni.

During the second phase, the measured absorption velocity increases to $\sim9{,}100~\mathrm{km~s^{-1}}$. As the ejecta expand, their column density decreases and they become progressively more transparent to radioactive $\gamma$-rays, allowing the $\gamma$-rays to propagate over larger distances before depositing their energy. Compton scattering produces energetic non-thermal electrons that excite and ionize helium \citep{Lucy_1991}. The increasing spatial extent of this non-thermal excitation allows He~\textsc{i} absorption to form over a broader velocity range within the helium-rich ejecta, producing the observed increase in absorption velocity.

During the third phase, the measured absorption velocity becomes approximately constant. Within this interpretation, the plateau is consistent with energy deposition from the radioactive $\gamma$-ray channel sampling nearly the full velocity extent of the helium-rich line-forming region, such that the decreasing $\gamma$-ray optical depth no longer produces a substantial increase in the maximum velocity at which He~\textsc{i} absorption is excited. We emphasize that this interpretation refers specifically to the $\gamma$-ray component of radioactive energy deposition. Radioactive decays also produce energetic leptons, including positrons, which may remain trapped within the ejecta and deposit their energy more locally \citep{Desai_2025}.

Radiative-transfer models demonstrate that the evolution of the He~\textsc{i} line-forming region depends strongly on the spatial distribution of $^{56}$Ni \citep{Dessart_2012,Ergon_2014,Dessart_2015,Dessart_2016,Ergon_2022}. If $^{56}$Ni is mixed extensively into or near the helium-rich ejecta, non-thermal excitation can begin at earlier epochs, reducing the prominence of a distinct intermediate velocity increase. More limited outward mixing can instead produce delayed non-thermal excitation as radioactive energy deposition extends into progressively faster helium-rich layers. The presence and evolution of this intermediate phase can therefore provide a diagnostic of the extent of $^{56}$Ni mixing. In SN~2025aico, the observed second-phase velocity increase is consistent with comparatively limited outward mixing of $^{56}$Ni into the helium-rich ejecta.

Throughout the photospheric phase, the helium absorption velocities are systematically lower than those of the principal H$\alpha$ component, while the detached high-velocity H$\alpha$ component traces still faster material. Unlike the \ion{He}{1} lines, H$\alpha$ does not exhibit a comparable intermediate velocity increase. Although non-thermal processes can contribute to hydrogen excitation, He~\textsc{i} line formation is more strongly dependent on non-thermal excitation and ionization and therefore responds more sensitively to the changing distribution of radioactive energy deposition. The absence of a corresponding velocity increase in H$\alpha$ supports the interpretation that the second helium phase is associated with changing non-thermal excitation rather than a global acceleration or restructuring of the ejecta. Together, the systematically higher hydrogen velocities and lower helium velocities are consistent with stratified ejecta, in which a thin hydrogen-rich envelope surrounds deeper helium-rich material, as expected for Type~IIb progenitors that retain only a residual hydrogen layer prior to explosion \citep{Dessart_2012,Dessart_2022,jerkstrand2025}.

\subsection{Helium Structure}

\ion{He}{1} is present from our earliest NIR spectrum at $+4.8$\,d post-explosion, where the $1.083\,\mu$m feature is clearly visible and very broad, tracing the high-velocity outer helium-rich ejecta. At this epoch, the large blueshift places the \ion{He}{1} $2.0581\,\mu$m transition within a region of strong telluric absorption, preventing a confident assessment of the line. From $+15.8$ to $+125.2$\,d, both NIR \ion{He}{1} transitions are clearly detected. By $+15.8$\,d both transitions are present and exhibit single, well-defined P-Cygni absorption profiles blueward of the line center ($v=0$), consistent with line formation in an optically thick, expanding photosphere \citep{filippenko_1997}. Between $+58.5$ and $+66.0$\,d, the absorption troughs broaden and the emission from the inner $\sim500$\,km\,s$^{-1}$ of material begins to decrease, marking the onset of a more complex line profile. From $+102.3$\,d onward, both transitions develop pronounced double-peaked emission profiles that persist through the later epoch at $+125.2$\,d.
 
The double-peaked structure is somewhat more distinct in the \ion{He}{1} $2.0581\,\mu$m emission feature than in the feature associated with the \ion{He}{1} $1.083\,\mu$m line. This is likely because the $2.058\,\mu$m transition is relatively cleaner and less contaminated by nearby emission \citep{Li1995}. The emergence of this morphology coincides with the transition toward increasingly optically thin ejecta, revealing the underlying geometry of the helium-rich material \citep{sharma2024,Medler2025}. The appearance of closely matching double-peaked profiles in both isolated helium transitions, with comparable velocity separations and peaks symmetrically displaced about the rest wavelength, makes it unlikely that the observed structure arises from line blending or telluric contamination. Instead, the profiles strongly favor an intrinsic asymmetry in the helium-rich ejecta \citep{Jerkstrand_2017}. Several ejecta geometries can produce such double-peaked emission profiles, including toroidal or disk-like distributions, bipolar outflows, and large-scale clumping in the ejecta \citep{Jerkstrand_2017, jerkstrand2025}.

\subsection{Oxygen and Calcium Structure}

\begin{figure*}
    \centering
    \includegraphics[width=1\linewidth]{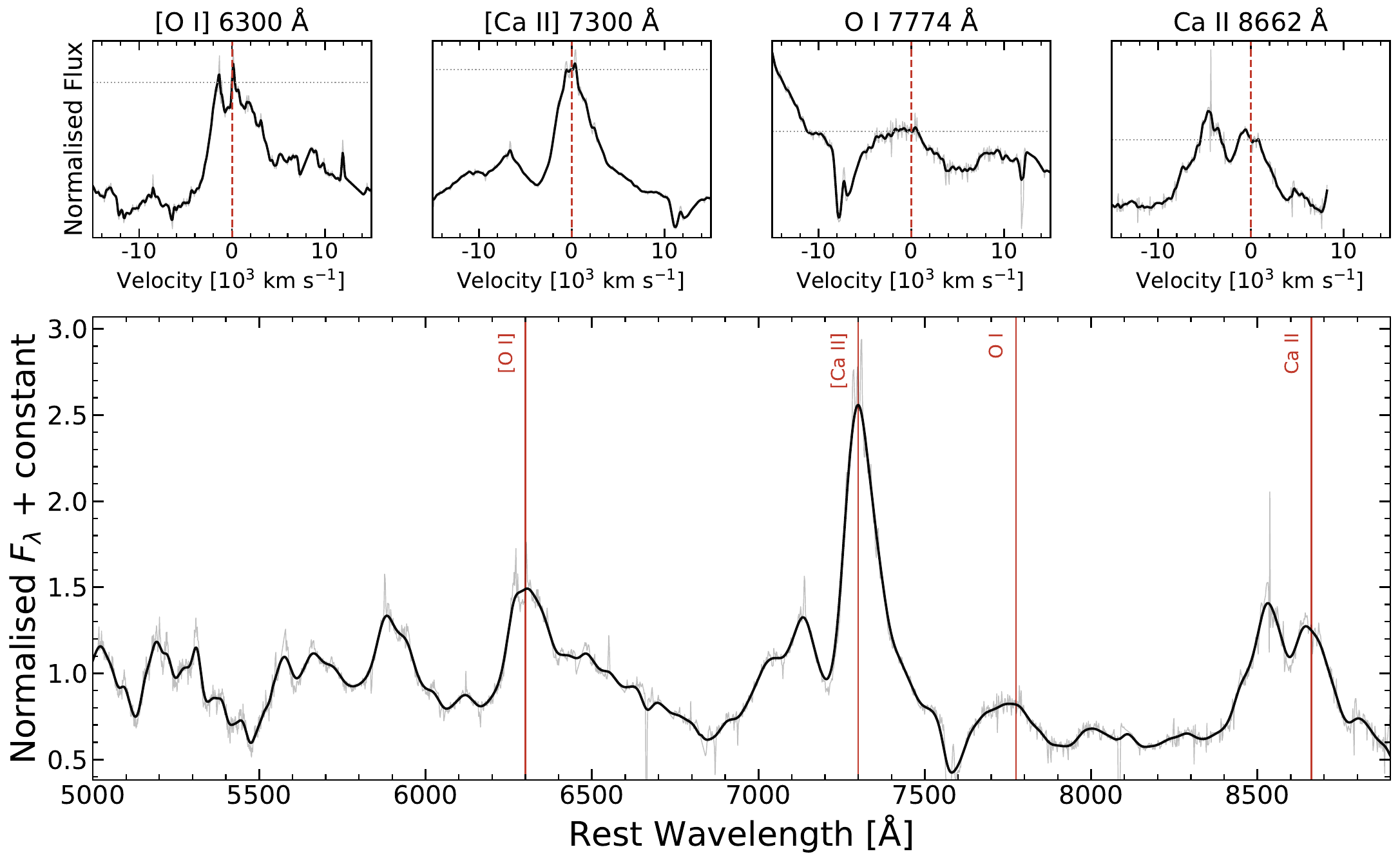}
    \caption{Late-time oxygen and calcium emission features of SN~2025aico at $+166$\,d post-explosion. The lower panel shows the optical spectrum, with the rest wavelengths of [O\,\textsc{i}] $0.6300,0.6363\,\mu$m, [Ca\,\textsc{ii}] $0.7291,0.7324\,\mu$m, O\,\textsc{i} $0.7774\,\mu$m, and Ca\,\textsc{ii} $0.8662\,\mu$m indicated by the vertical red lines. The upper panels show the corresponding line profiles in velocity space relative to the rest wavelength of each transition; these insets are shown unsmoothed to preserve the intrinsic profile shape, in contrast to the smoothed spectrum plotted in the lower panel.}
    \label{fig:O+Ca_emission}
\end{figure*}

At nebular phases, oxygen and calcium emission provides an independent view of the deeper ejecta. Figure~\ref{fig:O+Ca_emission} presents the continuum-subtracted line profiles in velocity space in the upper panels and the corresponding late-time spectrum in the lower panel. The velocity profiles probe the geometry of the oxygen- and calcium-emitting material, while the relative line strengths constrain the progenitor core. In particular, the integrated [Ca\,\textsc{ii}]/[O\,\textsc{i}] flux ratio has been widely used as a diagnostic of progenitor core and ZAMS mass because of its sensitivity to the oxygen-rich core produced during stellar evolution \citep{Fransson1989,Jerkstrand2015,Anderson2015,Jerkstrand_2017,Fang2022,Fang2023}.

At $+166$\,d post-explosion, the forbidden [O\,\textsc{i}] $0.6300$ and $0.6363\,\mu$m and [Ca\,\textsc{ii}] $0.7291$ and $0.7324\,\mu$m transitions dominate the optical spectrum of SN~2025aico, while the permitted O\,\textsc{i} $0.7774\,\mu$m and Ca\,\textsc{ii} $0.8662\,\mu$m features remain comparatively weak (Figure~\ref{fig:O+Ca_emission}). The relative strengths and velocity-space morphologies of these transitions can therefore be used to examine the distribution and asymmetry of the oxygen- and calcium-emitting material and to constrain the progenitor core \citep{Maeda2008,Jerkstrand2015,Jerkstrand_2017,Fang2019,Fang2022,Fang2023}.

The velocity-space line profiles (Figure~\ref{fig:O+Ca_emission}, upper panels) provide an additional, independent constraint on the ejecta geometry. The [O\,\textsc{i}] $0.6300\,\mu$m profile is double-peaked, with the two components on either side of zero velocity and an extended red wing reaching out to $\sim+10{,}000$\,km\,s$^{-1}$, a morphology discussed further in Section~\ref{sec:structure} in the context of the helium lines. In contrast, [Ca\,\textsc{ii}] $0.7291, 0.7324\,\mu$m is comparatively narrow and single-peaked, centered almost exactly at rest velocity. The permitted Ca\,\textsc{ii} near-infrared triplet ($0.8498,0.8542,0.8662\,\mu$m) shows strong separation between the $0.8542$ and $0.8662\,\mu$m lines. The permitted O\textsc{i} $0.7774\,\mu$m feature is strongly affected by telluric absorption blueward of zero velocity and therefore provides no reliable constraint on the intrinsic emission geometry. Taken together, the isolated forbidden lines provide the clearest geometric information: the narrow, single-peaked [Ca\,\textsc{ii}] emission suggests a comparatively compact calcium-emitting region, whereas the double-peaked [O\,\textsc{i}] profile indicates a more asymmetric distribution of the oxygen-rich ejecta.

While the line-profile morphologies probe the velocity-space distribution of the emitting material, the integrated nebular line fluxes provide a separate constraint on the progenitor structure. In particular, the [Ca\,\textsc{ii}] $0.7291,0.7324\,\mu$m/[O\,\textsc{i}] $0.6300,0.6363\,\mu$m integrated flux ratio has been widely used as a diagnostic of progenitor core and ZAMS mass in core-collapse and stripped-envelope supernovae \citep{Fransson1989, Jerkstrand2015, Anderson2015, Jerkstrand_2017, Fang2022, Prentice_2022, Fang2023}. This relation arises because more massive progenitors generally develop larger C/O cores and synthesize larger oxygen masses, whereas the calcium emission is comparatively less sensitive to the progenitor's initial mass \citep{Fransson1989, Jerkstrand2015, Anderson2015, Jerkstrand_2017}. The greater mass of oxygen-rich ejecta in these systems generally produces stronger nebular [O\,\textsc{i}] emission and therefore smaller [Ca\,\textsc{ii}]/[O\,\textsc{i}] ratios. Conversely, larger ratios are generally associated with smaller oxygen-rich cores and lower progenitor masses. The ratio is not a unique mass diagnostic, however, because the emergent line luminosities also depend on the nucleosynthetic structure, excitation conditions, radioactive-energy deposition, mixing, and other properties of the emitting ejecta \citep{Anderson2015, Jerkstrand_2017, Prentice_2022, Fang2023}. To measure this ratio in SN~2025aico, we fit and subtract a local linear continuum using wavelength regions immediately blueward and redward of each emission complex and integrate the residual flux over the full [O\,\textsc{i}] and [Ca\,\textsc{ii}] profiles.

We measure [Ca\,\textsc{ii}]/[O\,\textsc{i}] ratios of $3.303$ at $+133$\,d and $4.869$ at $+166$\,d. The dominance of [Ca\,\textsc{ii}] relative to [O\,\textsc{i}] indicates that SN~2025aico does not exhibit the strong oxygen emission expected from a high-mass progenitor with a massive oxygen-rich core. Instead, the large measured ratios favor a smaller C/O core and place SN~2025aico toward the lower-to-moderate progenitor-mass regime \citep{Fransson1989, Jerkstrand_2017, Fang2023}. Importantly, this interpretation does not imply an absence of oxygen in the ejecta: the prominent nebular [O\,\textsc{i}] emission and its structured profile demonstrate the presence of oxygen-rich material. Rather, it is the strength of [O\,\textsc{i}] relative to [Ca\,\textsc{ii}] that disfavors the substantially larger oxygen-rich core expected from a high-mass progenitor. This interpretation is consistent with the independent light-curve modeling of \citet{zhao2026}, who infer an ejecta mass of $M_{\rm ej}=2.79^{+0.21}_{-0.18}\,M_\odot$ and favor a moderate-mass stripped progenitor for SN~2025aico. Together, the nebular line ratio and light-curve modeling therefore favor a moderate-mass stripped progenitor rather than a high-mass star with a substantially larger oxygen-rich core.

\subsection{Structure of SN~2025aico} \label{sec:structure}

\begin{figure}
    \centering
    \includegraphics[width=1\linewidth]{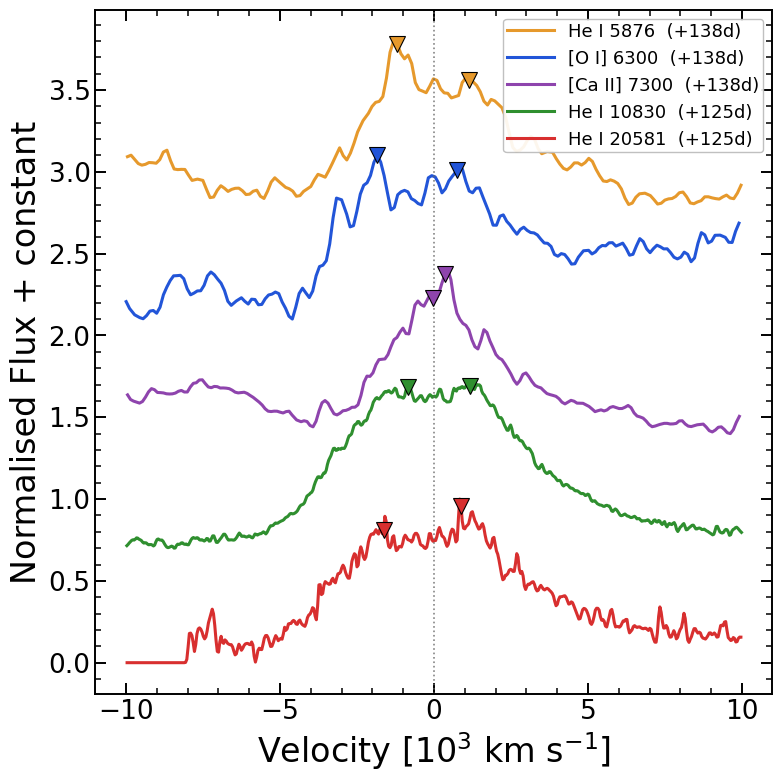}
    \caption{Comparison of the late-time emission-line profiles of [O\,\textsc{i}] $0.6300\,\mu$m at $+138$\,d (blue) and the [Ca~\textsc{ii}] at $0.7300\,\mu$m and He\,\textsc{i} $1.083\,\mu$m (green) and $2.0581\,\mu$m (red) at $+125$\,d post-explosion in SN~2025aico. Arrows represent the peak velocity red and blueward of $v=0$}
    \label{fig:OH}
\end{figure}

Taken together, the hydrogen, helium, oxygen, and calcium velocities and line profiles provide a coherent picture of the ejecta structure in SN~2025aico. The large H$\alpha$ velocities ($\sim10{,}000$--$14{,}000$\,km\,s$^{-1}$), together with the early detached high-velocity absorption component, trace the rapidly expanding outer hydrogen-rich ejecta. In contrast, the systematically lower \ion{He}{1} velocities ($\sim6{,}000$--$9{,}000$\,km\,s$^{-1}$ for the $1.083\,\mu$m transition and $\sim5{,}000$--$10{,}000$\,km\,s$^{-1}$ for the $2.0581\,\mu$m transition) demonstrate that the ejecta are compositionally stratified.

Figure~\ref{fig:OH} compares the He~\textsc{i}, [O~\textsc{i}], and [Ca~\textsc{ii}] emission-line profiles observed later than $+120$\,d. To locate the peaks, each continuum-subtracted profile was smoothed with a Gaussian filter, divided at zero velocity, and searched for the maximum flux within $-8000<v<0$\,km\,s$^{-1}$ and $0<v<+8000$\,km\,s$^{-1}$ for the blue- and redshifted components, respectively. The blue and red peaks occur at $-1190$ and $+1150$\,km\,s$^{-1}$ for He~\textsc{i} $0.5876\,\mu$m, $-840$ and $+1160$\,km\,s$^{-1}$ for He~\textsc{i} $1.083\,\mu$m, $-1640$ and $+870$\,km\,s$^{-1}$ for He~\textsc{i} $2.0581\,\mu$m, and $-1840$ and $+750$\,km\,s$^{-1}$ for [O~\textsc{i}] $0.6300\,\mu$m. The peak velocities are rounded to the nearest $10$\,km\,s$^{-1}$, corresponding to separations of approximately $2340$, $2000$, $2510$, and $2590$\,km\,s$^{-1}$, respectively. The comparable separations and overall velocity extents of the He~\textsc{i} and [O~\textsc{i}] profiles indicate that the helium- and oxygen-rich material likely share a common large-scale kinematic structure. Double-peaked nebular [O~\textsc{i}] profiles have previously been observed in stripped-envelope SNe, including SNe~2003jd and 2008ax, and interpreted as signatures of aspherical inner ejecta, such as toroidal or bipolar distributions \citep{Mazzali_2005,Maeda2008, Milisavljevic2010, Maurer2010, Ergon_2014, modjaz2019}.

\begin{figure*}
    \centering
    \includegraphics[width=\linewidth]{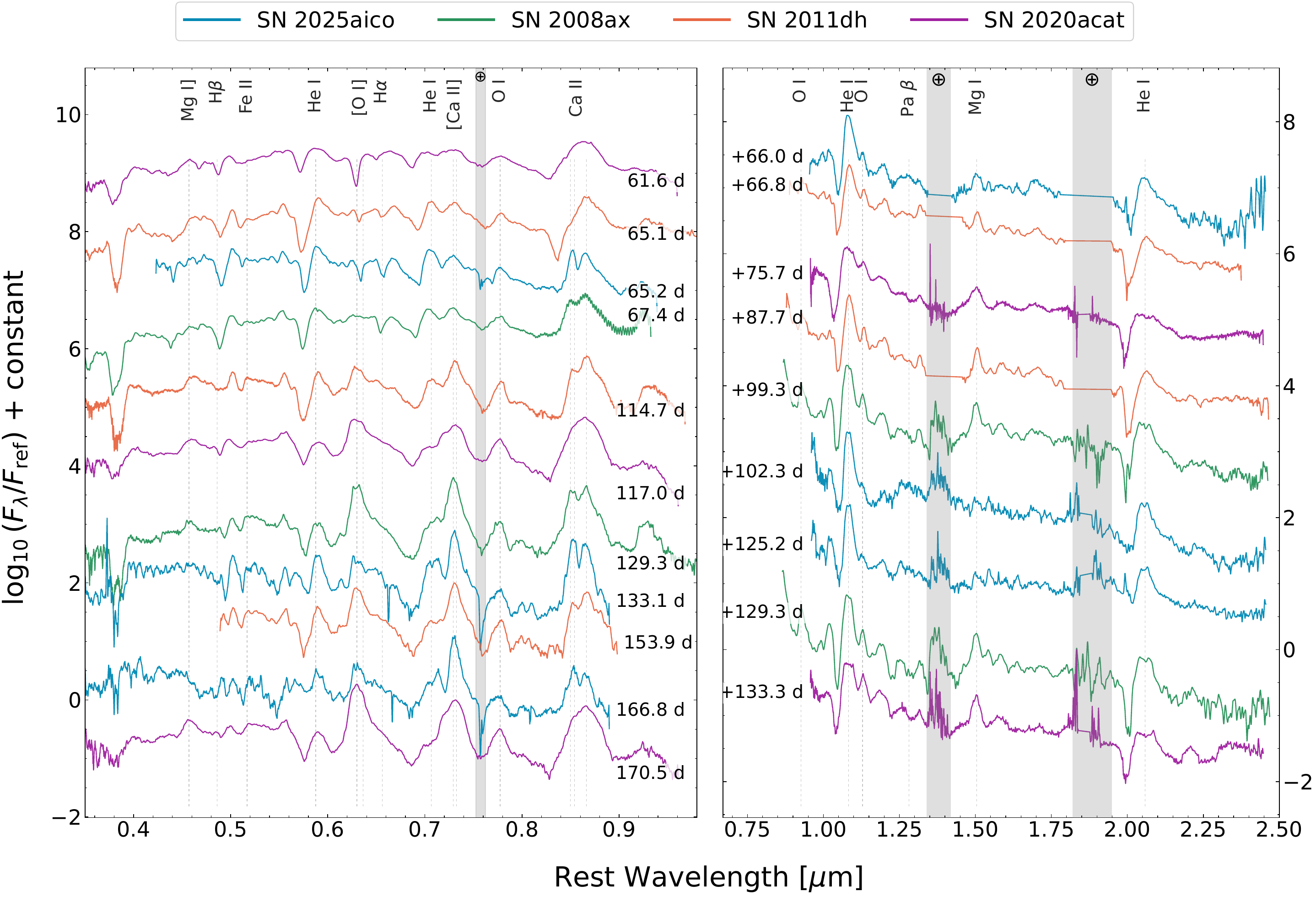}
    \caption{Comparison of optical (left) and NIR (right) spectra of SN~2025aico (blue) with the Type IIb SNe SN~2008ax (green), SN~2011dh (orange), and SN~2020acat (purple), at similar post-explosion phases. Telluric absorption bands are indicated by the $\oplus$ symbol. Key spectral features are marked with dotted lines.}
    \label{fig:comp}
\end{figure*}

Combining these velocity and morphological constraints, SN~2025aico is consistent with a layered ejecta structure comprising a thin, high-velocity hydrogen-rich envelope surrounding deeper, lower-velocity, aspherical helium-rich and oxygen-emitting material. The correspondence between the helium and oxygen line profiles suggests that the large-scale explosion asymmetry extends from the inner oxygen-bearing ejecta into the overlying helium shell and that these regions acquired similar velocity-space geometries during the explosion \citep{Maeda2008,Maurer2010}. This interpretation does not require a large oxygen mass. These observations therefore favor a large-scale asymmetry in the inner ejecta while preserving an overall layered ejecta structure, as indicated by the limited outward mixing of $^{56}$Ni. Together, these findings constrain both the progenitor's internal stratification and the multidimensional nature of the core-collapse explosion \citep{Maeda2008,Fang_2023}.

%%-------------------------------------------------------------
\section{Comparison With Other Type~IIb Supernovae}\label{sec:comparison}

To place SN~2025aico in context, we compare its optical and NIR spectral evolution with three well-observed SNe~IIb: SN~2008ax \citep{Pastorello_2008}, SN~2011dh \citep{Maund_2011}, and SN~2020acat \citep{Medler_2022,Medler_2023}. Figure~\ref{fig:comp} presents spectra of the four objects at comparable post-explosion phases, while the epochs included in the comparison sample are listed in Table~\ref{tab:comparison_epochs}.

\begin{deluxetable}{lcl}
\tablecaption{Spectroscopic epochs included in the SN~IIb comparison
sample.\label{tab:comparison_epochs}}
\tabletypesize{\footnotesize}
\tablehead{
  \colhead{Supernova} &
  \colhead{Phase since explosion (d)} &
  \colhead{References}
}
\startdata
SN~2008ax
  & $+67.4$, $+99.3$, $+129.3$, $+157.2$
  & \citep{Taubenberger_2011} \\
SN~2011dh
  & $+65.1$, $+66.8$, $+87.7$, $+114.7$, $+153.9$, $+198.2$, $+204.8$
  & \citep{Ergon_2014,Ergon2015,Shivvers2019} \\
SN~2020acat
  & $+61.6$, $+75.7$, $+117.0$, $+133.3$, $+170.5$
  & \citep{Medler_2022,Medler_2023} \\
\enddata
\end{deluxetable}

Overall, SN~2025aico follows the characteristic spectral evolution of SNe~IIb, showing residual hydrogen at early epochs and increasingly prominent helium features as the ejecta evolve toward the nebular phase \citep{Gilkis2022,jerkstrand2025}. The principal differences among the objects are the absorption depth, velocity width, and morphology of their hydrogen, helium, calcium, and oxygen features. SN~2025aico retains comparatively strong H$\alpha$ absorption at early epochs, consistent with a residual hydrogen envelope. Its He~\textsc{i} absorption features are deeper than those observed in SN~2008ax and SN~2011dh but weaker than those in SN~2020acat. Despite their relative depth, the He~\textsc{i} profiles of SN~2025aico remain confined to a narrower velocity range than the substantially broader profiles of SN~2020acat, with SN~2008ax and SN~2011dh exhibiting intermediate velocity widths.

The calcium features likewise reveal differences in the characteristic ejecta velocities. The Ca~\textsc{ii} NIR triplet in SN~2020acat remains heavily blended throughout the available observations, consistent with its comparatively large kinetic-energy-to-ejecta-mass ratio, $E_{\rm K}/M_{\rm ej}$. A larger $E_{\rm K}/M_{\rm ej}$ corresponds to higher characteristic expansion velocities, producing broader lines and greater blending among the triplet components \citep{Prentice_2017}. SN~2011dh evolves from a blended profile at early times to a resolved double-peaked morphology during the nebular phase. In contrast, SN~2008ax and SN~2025aico display distinctly separated Ca~\textsc{ii} NIR components at comparatively early epochs, consistent with their lower characteristic velocities and reduced Doppler blending. A similar trend is present in the forbidden [Ca~\textsc{ii}] emission near $0.73\,\mu$m, where SN~2025aico develops a narrow central emission peak that most closely resembles, but is narrower than, the corresponding feature in SN~2008ax and SN~2011dh.

The oxygen features also exhibit systemic differences between the four SNe. Forbidden [O~\textsc{i}] $0.6300\,  \mu$m emission first becomes apparent in SN~2011dh before emerging in the remaining objects, and by approximately $+130$d it is detected in all four SNe. This line is strongest in SN~2020acat and SN~2008ax but comparatively weak in SN~2025aico. Furthermore, SN~2025aico and SN~2008ax both develop pronounced double-peaked [O~\textsc{i}] profiles, whereas SN~2011dh and SN~2020acat have blended, more flat top emission profiles. The permitted O~\textsc{i} $0.7774\mu$m feature is present in all objects and strengthens with time, although it remains weaker in SN~2025aico than in the comparison sample, while the O~\textsc{i} $0.5577\mu$m line remains weak in all four SNe with little variation in relative strength.

An additional diagnostic available at the red end of the NIR spectra is first-overtone emission from carbon monoxide (CO), which produces a characteristic series of rotation-vibrational bands beginning near $2.3\,\mu$m. CO is among the first molecules to form as SN ejecta cool and can provide an efficient radiative cooling channel, helping portions of the ejecta reach temperatures favorable for subsequent dust condensation. CO emission therefore frequently precedes the appearance of newly formed dust and provides an important tracer of the molecular phase of the ejecta \citep{Spyromilio_1988,Ergon2015,Rho2018,Medler2025, Mera_2026}. Within the spectra compared in Figure~\ref{fig:comp}, a rise in flux beyond approximately $2.3\,\mu$m consistent with first-overtone CO emission is most apparent in SN~2020acat, while no comparably strong feature is evident in SN~2025aico at the available epochs. The absence of an obvious first-overtone feature does not rule out CO formation, however, because the strength of the overtone depends on the molecular temperature, excitation, and mass, as well as the wavelength coverage and signal-to-noise ratio of the observations. The lack of a strong CO first-overtone signature in SN~2025aico is particularly noteworthy given the MIR excess detected with \textit{JWST}/MIRI at $+124.4$\,d, which we investigate as possible dust emission in Section~\ref{sec:spectralenergy}.

\begin{deluxetable*}{ccccccccc}
\tabletypesize{\scriptsize}
\tablecaption{Explosion and progenitor properties of the comparison Type IIb supernovae.\label{tab:properties}}
\tablehead{
\colhead{SN} &
\colhead{$M_{\rm ej}$ ($M_\odot$)} &
\colhead{$M_{\rm H}$ ($M_\odot$)} &
\colhead{$E_{\rm K}$ ($10^{51}$ erg)} &
\colhead{$E_{\rm K}/M_{\rm ej}$} &
\colhead{$R_{\rm prog}$ ($R_\odot$)} &
\colhead{$v_{\rm H,peak}$ (km s$^{-1}$)} &
\colhead{Shock breakout} &
\colhead{Phase (d)}
}
\startdata
SN~2008ax\tablenotemark{a} & $2.5 \pm 1.0$ & $0.02$ & $0.8 \pm 0.3$ & 0.32 & $30$--$50$ & 22\,700 & Yes & 129.3 \\
SN~2011dh\tablenotemark{b} & $1.72 \pm 0.19$ & $0.024$ & $0.64 \pm 0.10$ & 0.37 & $200$--$300$ & 17\,500 & No & 198.2 \\
SN~2020acat\tablenotemark{c} & $2.3 \pm 0.3$ & $\sim0.1$ & $1.2 \pm 0.2$ & 0.52 & $<50$ & 22\,800 & No & 133.3 \\
SN~2025aico\tablenotemark{d} & $2.79^{+0.21}_{-0.18}$ & $0.01$ & $0.70^{+0.09}_{-0.08}$ & 0.25 & $10$--$30$ & 18\,500 & No & 125.2 \\
\enddata
\tablenotetext{a}{The hydrogen-envelope mass for SN~2008ax is adopted from \citet{Gilkis2022}}
\tablenotetext{b}{The hydrogen-envelope mass for SN~2011dh is adopted from \citet{Arcavi_2011}.}
\tablenotetext{c}{The hydrogen-envelope mass for SN~2020acat is adopted from \citet{Ergon2024}.}
\tablenotetext{d}{Explosion parameters for SN~2025aico ($M_{\rm ej}$, $M_{\rm H}$, $E_{\rm K}$, and $R_{\rm prog}$) are adopted from the independent light-curve modeling of \citet{zhao2026}.}

\tablecomments{The $E_{\rm K}/M_{\rm ej}$ values are calculated from the listed kinetic energies and ejecta masses and are given in units of $10^{51}$\,erg\,$M_\odot^{-1}$. Phase corresponds to the epoch of the late-time spectra used to measure the double-peaked He\,\textsc{i} profiles listed in Table~\ref{tab:gaussianfits}. Unless otherwise noted, the remaining comparison properties are taken from \citet{Medler_2022}.}

\end{deluxetable*}

%%----------------------
\subsection{Comparing Helium Structures}

\begin{figure*}
    \centering
    \includegraphics[width=0.7\linewidth]{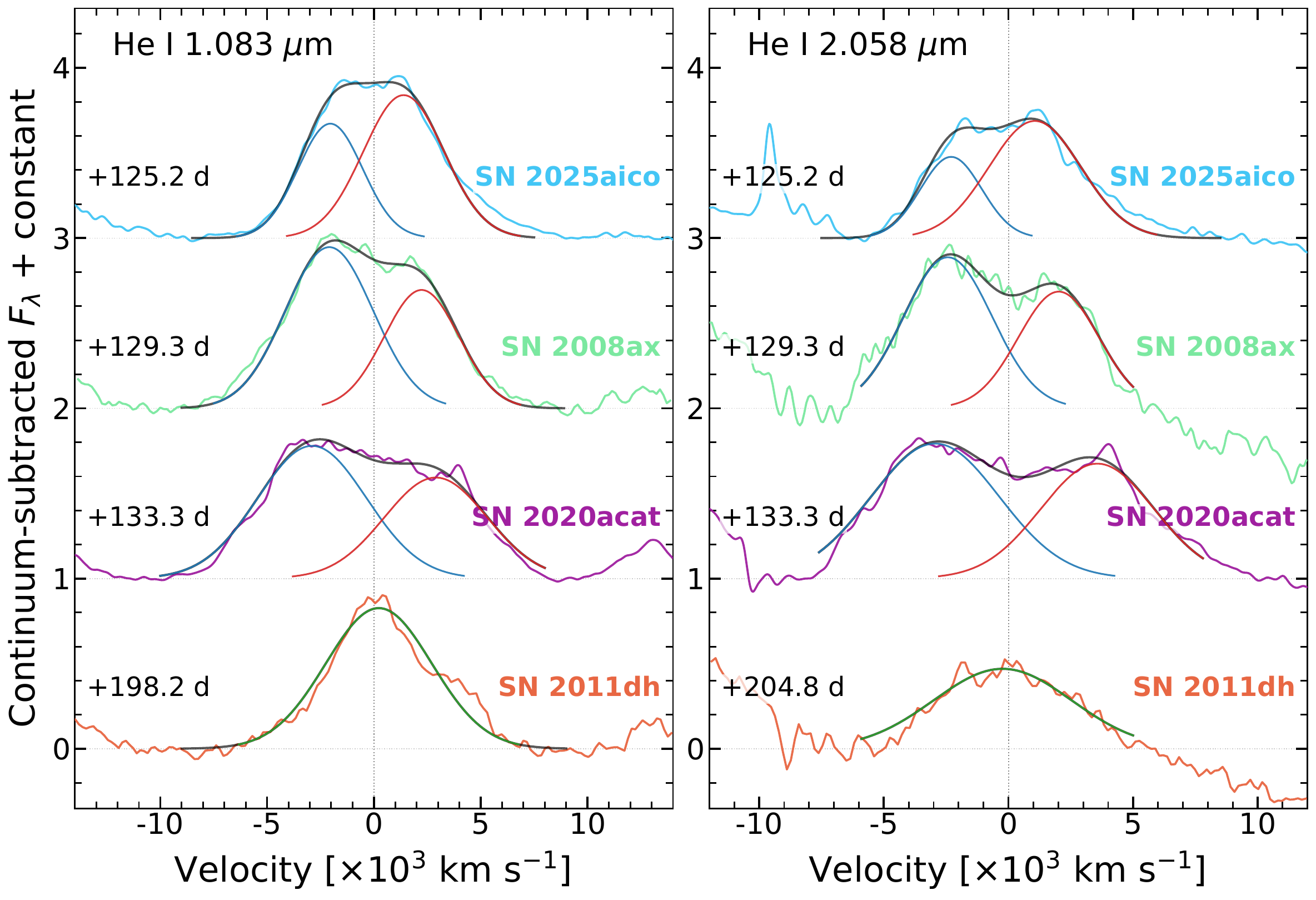}
    \caption{Comparison of the late-time \ion{He}{1} $1.083\,\mu$m (left) and \ion{He}{1} $2.0581\,\mu$m (right) line profiles of SN~2025aico (blue) with those of the Type~IIb SNe SN~2008ax (green) and SN~2011dh (orange), and SN~2020acat (purple). The SN~2011dh spectra were obtained with two different instruments. The emission-line profiles of SN~2025aico, SN~2008ax, and SN~2020acat were fit with double-Gaussian models, whereas the SN~2011dh profiles were fit with a single Gaussian because no statistically significant double-peaked structure is present.}
    \label{fig:comparingHe}
\end{figure*}

\begin{figure*}
    \centering
    \includegraphics[width=0.9\linewidth]{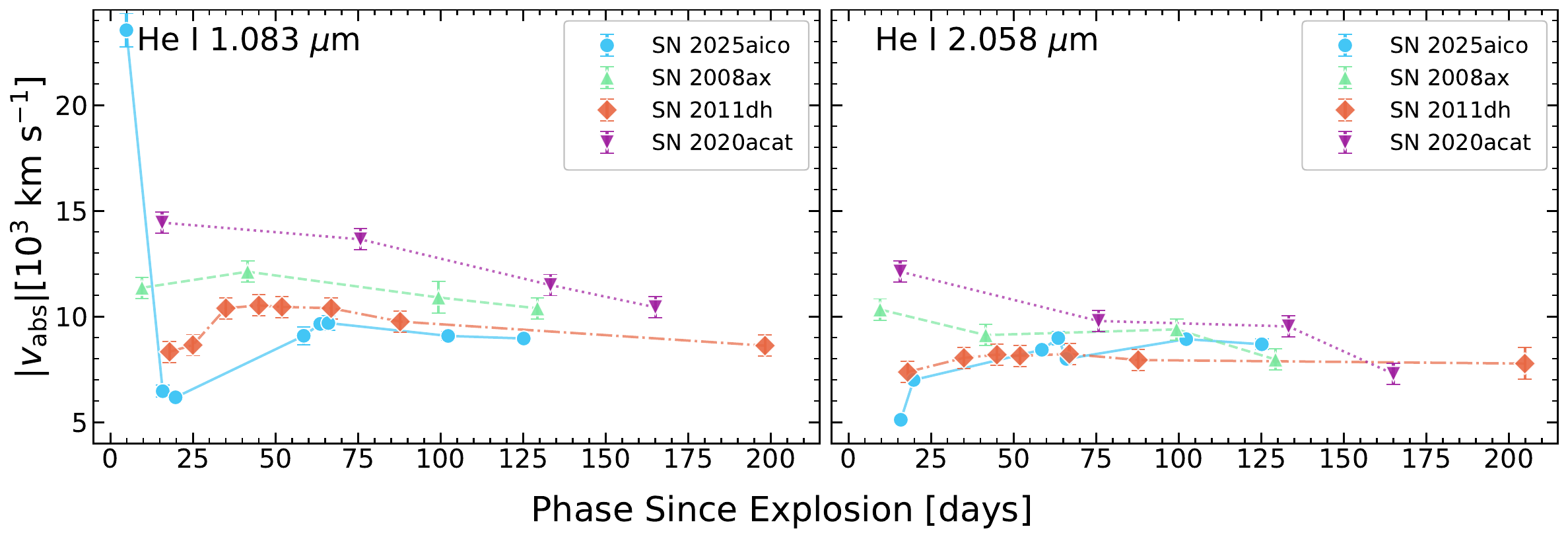}
    \caption{Comparison of \ion{He}{1} $1.083\,\mu$m and \ion{He}{1} $2.0581\,\mu$m absorption velocity evolution in SN~2025aico (cyan circles) against SN~2008ax (green), SN~2011dh (red), and SN~2020acat (yellow).}
    \label{fig:Hecomparisontime}
\end{figure*}

Figure~\ref{fig:comparingHe} shows the continuum-subtracted He~\textsc{i} $1.083\,\mu$m and $2.0581\,\mu$m profiles of the comparison sample at their latest available nebular-phase epochs, spanning $+125.2$ to $+204.8$\,d. Where both transitions are available, they exhibit the same overall morphology within each object. SN~2025aico and SN~2008ax show two partially blended emission components separated by a central depression. The components in SN~2025aico have comparable strengths, producing a nearly flat-topped profile, whereas the SN~2008ax profiles are more strongly weighted toward the blueshifted component. SN~2020acat exhibits broader, more gradually sloping wings and a shallower central depression, with emission extending to substantially higher velocities. In contrast, SN~2011dh displays narrower, rounded He~\textsc{i} profiles without a statistically significant central depression.

To quantify these morphological differences consistently, we fit the He~\textsc{i} $1.083\,\mu$m and $2.0581\,\mu$m profiles of SN~2025aico, SN~2008ax, and SN~2020acat with two Gaussian components. The SN~2011dh profiles are fit with a single Gaussian because the helium emission does not show statistically significant double-peaked structure. The weak red-side excess near the He~\textsc{i} $1.083\,\mu$m feature occurs close to the expected position of H~\textsc{i} Paschen~$\gamma$ at $1.094\,\mu$m and is therefore likely caused by line contamination rather than a second helium component. The absence of a corresponding secondary feature in the more isolated He~\textsc{i} $2.0581\,\mu$m profile supports this interpretation. These Gaussian components provide a phenomenological description of the observed profiles rather than a unique physical reconstruction of the ejecta. Although the double-peaked He~\textsc{i} profiles are consistent with an aspherical helium distribution \citep{Jerkstrand_2017}, asymmetric mixing of $^{56}$Ni provides an alternative explanation. Because He~\textsc{i} emission in stripped-envelope SNe is strongly affected by non-thermal excitation from radioactive decay products, an asymmetric $^{56}$Ni distribution could produce an asymmetric He~\textsc{i} emissivity even if the underlying helium distribution is less strongly aspherical \citep{Lucy_1991,Dessart_2012,Dessart_2015}. The evolution of the velocity separation between the two He~\textsc{i} components therefore provides an additional test of whether the observed profile primarily traces the ejecta geometry or an evolving excitation pattern.

The differences in profile width and velocity extent are also reflected in the photospheric He~\textsc{i} absorption velocities shown in Figure~\ref{fig:Hecomparisontime}. We measured the velocities consistently across the comparison sample by subtracting a local continuum and fitting the P-Cygni absorption minimum with an inverted Gaussian, following the procedure described in Section~\ref{sec:spectroscopy}. SN~2025aico exhibits systematically lower velocities than SN~2020acat and SN~2008ax across most of the observed evolution in both NIR transitions, with SN~2020acat remaining approximately $4000$--$5000$\,km\,s$^{-1}$ faster at comparable phases. This difference is consistent with the broader nebular profiles of SN~2020acat and indicates that its helium-rich ejecta span a larger velocity range. Although SN~2025aico initially reaches an exceptionally high He~\textsc{i} $1.083\,\mu$m absorption velocity of $\sim23{,}000$\,km\,s$^{-1}$, the measured velocity declines rapidly before increasing and approaching those of SN~2008ax and SN~2011dh by approximately $+30$\,d. SN~2011dh is the only other object in the comparison sample that exhibits a similar intermediate velocity increase.

Neither SN~2008ax nor SN~2020acat shows a comparable rise, although sparse early-time coverage may have missed this short-lived phase. Differences in $^{56}$Ni mixing may also affect its prominence. Radiative-transfer models show that stronger outward mixing deposits radioactive energy in the helium-rich ejecta earlier, maintaining the contribution of faster-moving helium and producing a smaller decline from the initial velocity maximum. Consequently, the later increase is weaker or does not appear as a distinct phase. With more limited outward mixing, the measured velocity declines farther before increasing as non-thermal excitation reaches the faster outer helium layers \citep{Dessart_2012,Dessart_2015,Dessart_2016}. The pronounced decline and intermediate increase observed in SN~2025aico are therefore consistent with limited outward $^{56}$Ni mixing. Together with the similar behavior of SN~2011dh, this comparison demonstrates that the three-stage helium-velocity evolution is not unique to SN~2025aico but is a time-sensitive diagnostic that requires densely sampled early spectroscopy to detect.

%%----------------------
\subsection{Peak Separation and Explosion Properties}

\begin{deluxetable}{lccc}
\tablecaption{Gaussian-fit properties of the late-time double-peaked
He\,\textsc{i} emission profiles.\label{tab:gaussianfits}}
\tablehead{
\colhead{Property} &
\colhead{SN~2008ax} &
\colhead{SN~2020acat} &
\colhead{SN~2025aico}
}
\startdata
$v_{\rm sep}^{1.083\,\mu{\rm m}}$ [km\,s$^{-1}$]
  & 4305 & 5808 & 3431 \\
$v_{\rm sep}^{2.0581\,\mu{\rm m}}$ [km\,s$^{-1}$]
  & 4434 & 6512 & 3367 \\
FWHM$^{1.083\,\mu{\rm m}}$ [km\,s$^{-1}$]
  & 7942 & 10482 & 6671 \\
FWHM$^{2.0581\,\mu{\rm m}}$ [km\,s$^{-1}$]
  & 7350 & 10400 & 6497 \\
$v_{\rm peak}^{1.083\,\mu{\rm m}}$ [km\,s$^{-1}$]
  & $-1992$ & $-3333$ & 1137 \\
$v_{\rm peak}^{2.0581\,\mu{\rm m}}$ [km\,s$^{-1}$]
  & $-2377$ & $-3573$ & 1141 \\
\enddata
\tablecomments{
The spectra are from the same phases as those listed in Table~\ref{tab:properties}. $v_{\rm sep}$ is the separation between the fitted blue- and red-peak components of the double-Gaussian model.
FWHM and $v_{\rm peak}$ are measured from a single-component fit to the full continuum-subtracted profile. SN~2011dh is omitted because its profiles show no statistically significant double-peaked structure and were instead fit with a single Gaussian.
}
\end{deluxetable}

\begin{figure*}
    \centering
    \includegraphics[width=\linewidth]{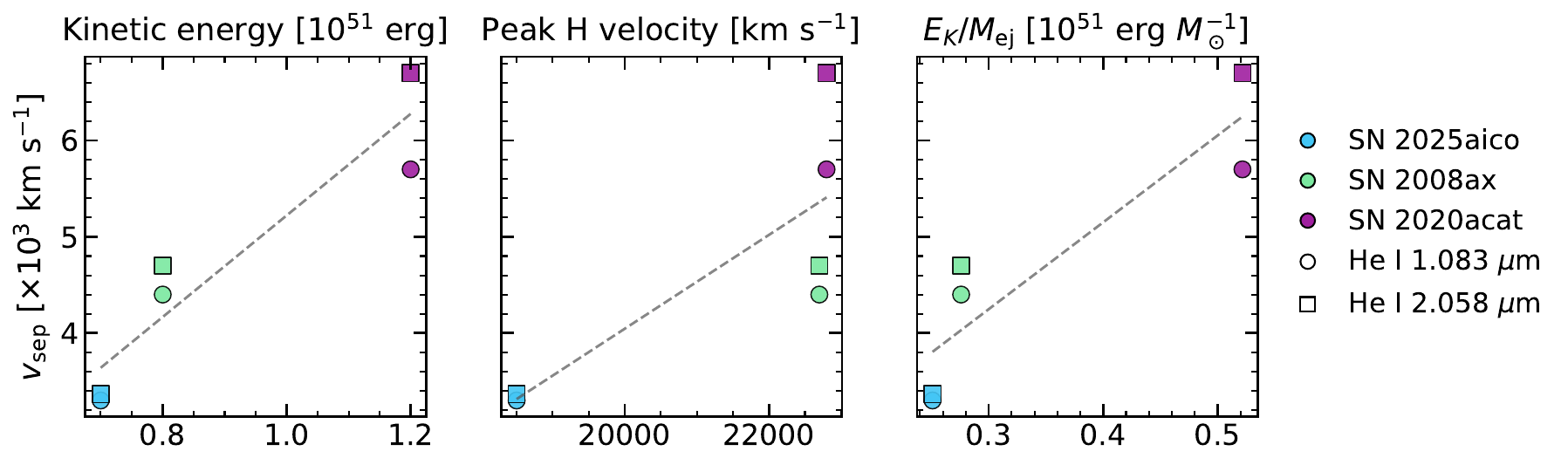}
    \caption{Peak separation of the late-time double-peaked He\,I emission profiles as a function of several explosion properties for SN~2025aico, SN~2008ax, and SN~2020acat. SN~2011dh is not included because it is fit with a single Gaussian. Circles denote measurements from He\,I $1.083\,\mu$m and squares denote measurements from He\,I $2.0581\,\mu$m. Dashed lines indicate least-squares fits and are shown only to guide the eye. A table of these values is available in Table~\ref{tab:gaussianfits}.}
    \label{fig:SEPVSN}
\end{figure*}

The qualitative velocity and profile-shape trends identified above are consistent with the explosion parameters listed in Table~\ref{tab:properties}. Since the characteristic ejecta velocity scales approximately as $v \propto (E_{\rm K}/M_{\rm ej})^{1/2}$ under homologous expansion \citep{deng2001}, the comparatively low He\,\textsc{i} velocities of SN~2025aico point toward a low kinetic-energy-to-ejecta-mass ratio. This is borne out directly: SN~2025aico has the lowest $E_{\rm K}/M_{\rm ej}$ of the sample ($\sim0.25\times10^{51}\,{\rm erg}\,M_\odot^{-1}$), comparable to but somewhat below SN~2008ax ($\sim0.32\times10^{51}\,{\rm erg}\,M_\odot^{-1}$) and SN~2011dh ($\sim0.37\times10^{51}\,{\rm erg}\,M_\odot^{-1}$), while SN~2020acat's ratio is substantially higher ($\sim0.52\times10^{51}\,{\rm erg}\,M_\odot^{-1}$).

This directly tracks the measured \ion{He}{1} peak separations (Table~\ref{tab:gaussianfits}, Figure~\ref{fig:SEPVSN}): SN~2025aico shows the narrowest double-peaked profiles ($v_{\rm sep}\sim3400\,{\rm km\,s^{-1}}$ for both transitions), SN~2008ax intermediate values ($v_{\rm sep}\sim4300$--$4400\,{\rm km\,s^{-1}}$), and SN~2020acat the broadest ($v_{\rm sep}\sim5800$--$6500\,{\rm km\,s^{-1}}$). Rather than reflecting unrelated explosion physics, the low He\,\textsc{i} velocities and narrow peak separations of SN~2025aico therefore point to a low $E_{\rm K}/M_{\rm ej}$ ratio, a somewhat larger ejecta mass, or a more centrally concentrated helium distribution \citep{Dessart_2012, Prentice2016}, placing it, together with SN~2008ax and SN~2011dh, within a physically similar regime of moderate explosion energetics distinct from the more energetic, larger progenitor Type~IIb's like SN~2020acat.

Theoretically, asymmetries in SNe form from a mix of progenitor initial conditions, the explosion mechanism, and instabilities in the envelope. In the central engine context, large samples of stellar core-collapse simulations predict a correlation between the explosion energy and explosion asymmetry \citep{Burrows2024}. Typically (but not always), progenitors with lower-mass, less compact cores explode earlier after gravitational collapse, before turbulence becomes vigorous, producing less energetic and more nearly spherical supernova shocks. By contrast, progenitors with higher-mass, more compact cores generally experience a longer delay before explosion, allowing turbulence to develop and energize the shock, thereby producing more asymmetric explosions with stronger dipolar components \citep{Vartanyan2019}.

Figure~\ref{fig:SEPVSN} shows the peak separation of the double-peaked He\,\textsc{i} profiles as a function of several explosion parameters for our objects. This comparison suggests a tentative positive correlation between the separation of the double-peaked \ion{He}{1} profiles and several explosion parameters: Type~IIb SNe with larger kinetic energies and larger values of $E_{\rm K}/M_{\rm ej}$ exhibit broader peak separations. With our limited data set, we found no correlation between peak luminosity, nickel mass, or peak He velocity. Although the current sample is small and the interpretations should therefore be treated with caution, the trends suggest that the ejecta geometry derived from the nebular helium emission profiles may be connected to the properties of the explosion. Nebular-phase NIR observations of a larger sample of Type~IIb SNe would allow us to further constrain these correlations and determine whether the trends seen in this data set are robust. 3D simulations of core-collapse SNe allow us to quantify observational constraints on the geometry of such explosions \citep{Tanaka_2012}.

%%------------------------------
\section{SN~2025aico Progenitor Constraints} \label{sec:progenitor}
We adopt the explosion parameters derived by \citet{zhao2026} from early optical light-curve modeling and comparisons with spectral-synthesis models. They infer a progenitor radius of $R_{\rm prog}\sim10$--$30\,R_\odot$ and a brief shock-cooling phase lasting less than four days, both signatures of a compact progenitor. The duration of shock-cooling emission depends strongly on the progenitor's pre-explosion radius \citep{Rabinak_2011,Sapir_2017}; therefore, the short cooling timescale is consistent with a compact star rather than an extended envelope. Within the comparison sample, the inferred radius of SN~2025aico is substantially smaller than the $R_{\rm prog}\sim200$--$300\,R_\odot$ estimated for the extended progenitor of SN~2011dh and is more similar to the compact radii inferred for SNe~2008ax and 2020acat \citep{Pastorello_2008,Ergon_2022,Ergon2024}.

\citet{zhao2026} further derive an ejecta mass of $M_{\rm ej}=2.79^{+0.21}_{-0.18}\,M_\odot$, a kinetic energy of $E_{\rm K}=0.70^{+0.09}_{-0.08}\times10^{51}$\,erg, and a nickel mass of $M_{\rm Ni}=0.033^{+0.006}_{-0.004}\,M_\odot$. These values indicate a relatively large ejecta mass paired with a low radioactive yield and a moderate kinetic-energy-to-ejecta-mass ratio. The moderate $E_{\rm K}/M_{\rm ej}$ is also consistent with the comparatively low He~\textsc{i} velocities of SN~2025aico relative to SN~2020acat (Section~\ref{sec:comparison}). Combined with the persistence of hydrogen in our early spectra (Section~\ref{sec:spectroscopy}), which demonstrates that part of the hydrogen envelope survived until explosion, these measurements are consistent with the interpretation of \citet{zhao2026}: a moderate-mass, stripped helium-star progenitor in a compact binary system with only a minimal residual hydrogen envelope. This scenario favors substantial but incomplete binary-envelope stripping rather than either a fully stripped Wolf--Rayet progenitor or a largely intact extended supergiant.

\subsection{Implications for the Stripping Mechanism}

The combination of a compact progenitor radius with a residual hydrogen envelope allows us to distinguish between stripping methods. Single-star line-driven winds follow two main outcomes: at sub-solar to solar metallicity they typically fail to remove the full hydrogen envelope, while only the most massive progenitors ($\gtrsim25$--$30\,M_\odot$) are expected to strip it entirely \citep{Vink2001,Heger2003, Zapartas2021}. Producing an intermediate outcome, a compact star retaining only a thin residual hydrogen layer, is difficult through winds alone. Binary mass transfer, by contrast, can remove any amount of the envelope depending on the mass ratio and separation of the two stars, rather than being limited like single-star line-driven winds are \citep{Sravan2019, long2022}. This flexibility makes binary mass transfer the most natural explanation for a partially stripped progenitor like SN~2025aico, rather than single-star wind-driven loss. 

This interpretation is further supported by population-synthesis modeling calibrated against the current sample of binary companions to stripped-envelope SNe, which finds that 68\% of Type~IIb progenitors explode alongside a surviving main-sequence companion, with the companion population favoring low-to-moderate mass transfer \citep{long2022,Zapartas_2025, Ercolino_2026}. Direct HST companion searches in SN~2008ax and SN~2011dh, two of the comparison objects used in Section~\ref{sec:comparison}, are among the systems used to calibrate this framework, and both are consistent with moderately luminous, still-main-sequence companions rather than isolated Wolf-Rayet progenitors \citep{Zapartas_2025}. Given the compact radius constrained by \citet{zhao2026} and partial-stripping signature established above, this binary-calibrated framework supports a binary-stripping origin for SN~2025aico's small residual hydrogen envelope over a single-star wind-driven channel.
 
%%------------------------------------------------------------------------
\section{Spectral Energy Distribution}\label{sec:spectralenergy}

\subsection{Constructing the Spectral Energy Distribution}
SN~2025aico was serendipitously observed with \textit{JWST}/MIRI at $+124.4$\,d, providing simultaneous imaging in the \textit{F770W}, \textit{F1130W}, and \textit{F1800W} filters. To investigate the origin of the observed MIR excess, we constructed an optical-to-MIR spectral energy distribution (SED) by combining the MIRI photometry with the nearest available optical and NIR spectra, obtained with Keck~I/LRIS at $+138.11$\,d and Keck~II/NIRES at $+125.2$\,d, respectively.

\begin{figure*}
    \centering
    \includegraphics[width=1\linewidth]{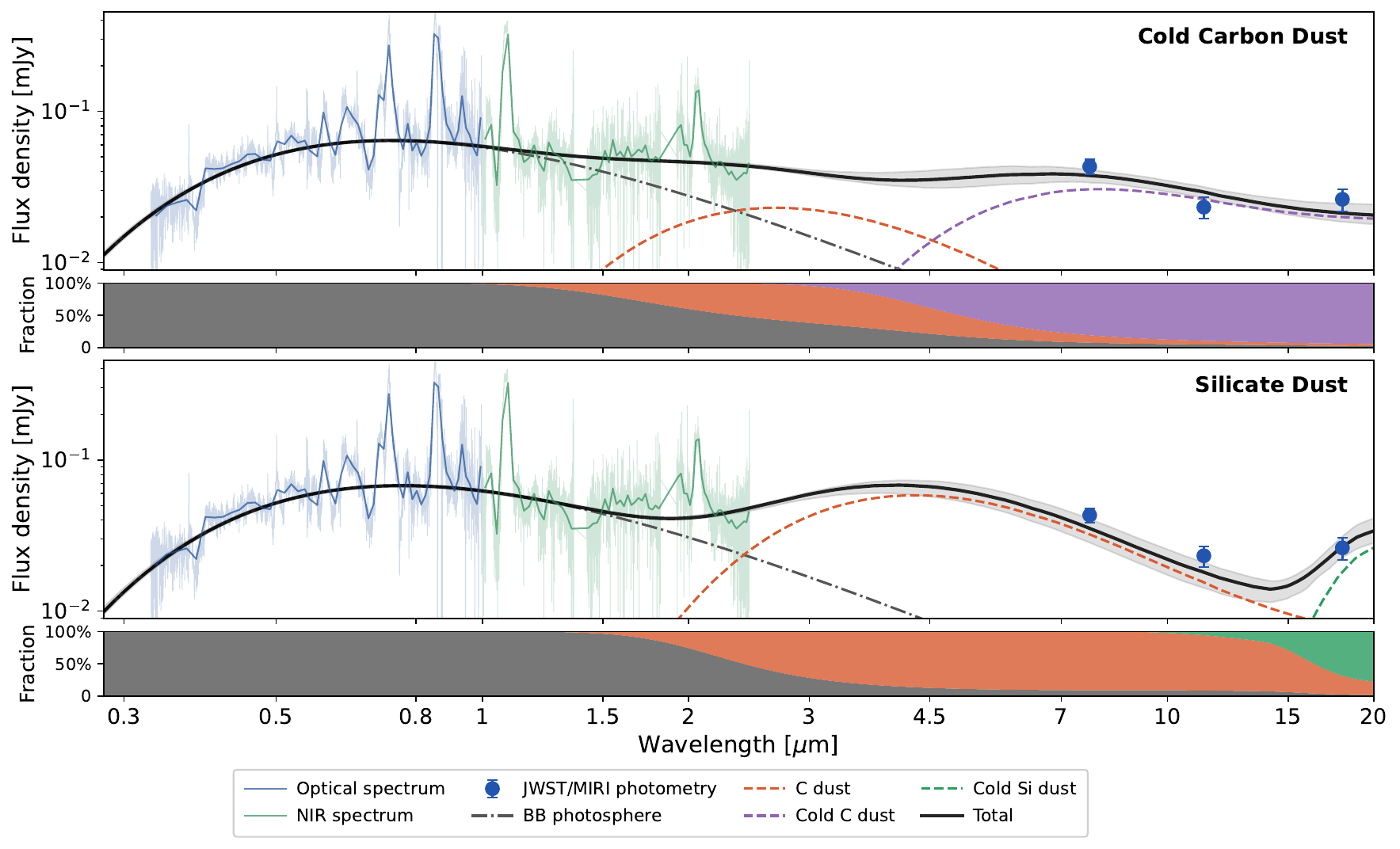}
    \caption{Spectral energy distribution of SN~2025aico at $+124.4$\,d post-explosion, combining the nearest available optical (Keck~I/LRIS, $+138.11$\,d) and near-infrared (Keck~II/NIRES, $+125.2$\,d) spectra with contemporaneous \textit{JWST}/MIRI photometry (Table~\ref{tab:jwst_photometry}). Optical and NIR spectra were flux-calibrated as described in Section~\ref{sec:observations_and_data}.}
    \label{fig:sed}
\end{figure*}

To flux calibrate the LRIS spectrum, we computed a synthetic ATLAS $o$-band flux by integrating the observed-frame spectrum through the ATLAS $o$-band transmission curve \citep{2018PASP..130f4505T}\footnote{The ATLAS transmission curves were downloaded from \url{https://svo2.cab.inta-csic.es/theory/fps/index.php}} and compared it with the inverse-variance-weighted mean of the ATLAS $o$-band forced photometry obtained on the same night\footnote{\url{https://fallingstar-data.com/forcedphot/}}. ATLAS was used because it was the only survey among ATLAS, ZTF, and Pan-STARRS with photometric observations contemporaneous with the LRIS spectrum. The resulting multiplicative scale factor was applied to the LRIS spectrum, preserving its relative spectral shape while placing it on the appropriate absolute flux scale. No contemporaneous NIR photometry was available to independently calibrate the NIRES spectrum. Instead, the NIRES spectrum was scaled to match the median flux of the LRIS spectrum within a $\pm0.02\,\mu$m window centered at $1\,\mu$m. This procedure relies on the relative flux calibration of both spectra and requires only a single multiplicative scaling factor to produce a continuous optical--NIR spectrum. We verified the reliability of the relative flux calibration by inspecting the independently reduced spectrophotometric standard-star observations for each instrument. The LRIS spectrum was adopted for $\lambda_{\rm rest}\leq1\,\mu$m and the rescaled NIRES spectrum for $\lambda_{\rm rest}>1\,\mu$m, yielding a single spectrum spanning approximately $0.32$--$2.5\,\mu$m.

\begin{figure}
    \centering
    \includegraphics[width=\linewidth]{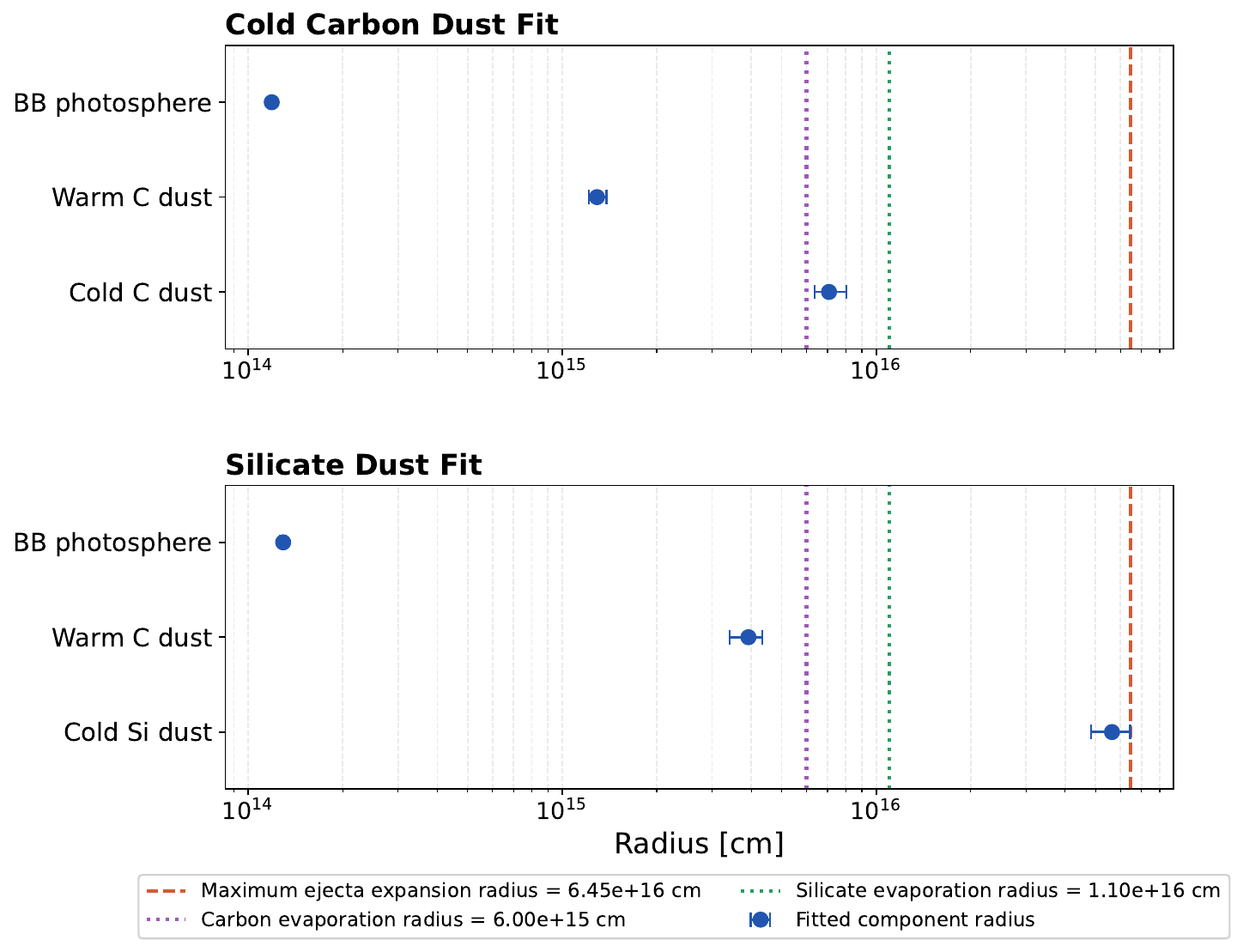}
    \caption{Fitted radii of the photosphere and the warm-carbon, cold-carbon, and cold-silicate dust components compared with the maximum ejecta expansion radius at $+124.4$\,d. The expansion radius, $R_{\rm exp}=6.448896\times10^{16}$\,cm, is calculated using a maximum ejecta velocity of $60{,}000\,\mathrm{km\,s^{-1}}$, measured from the blue wing of the broad He~\textsc{i} profile. The estimated carbon- and silicate-grain evaporation radii are also shown.}
    \label{fig:photospher_radii}
\end{figure}

\begin{table*}[t]
\centering
\caption{Best-fit parameters for the two dust models used to reproduce the SED of SN~2025aico. Unless otherwise noted, uncertainties correspond to the 16th and 84th percentiles of the marginalized posterior distributions. The $50$--$150$\,K range listed for the cold-silicate temperature represents imposed bounds rather than a posterior uncertainty.}
\label{tab:dustfits}

\begin{tabular}{lcccc}
\hline
Component & Temperature (K) & Dust Mass ($M_\odot$) &
Grain Radius ($\mu$m) & Radius (cm) \\
\hline

\multicolumn{5}{c}{\textbf{Cold Carbon Dust Model}} \\
\hline

Blackbody &
$6950 \pm 90$ &
-- &
-- &
$(1.19 \pm 0.03)\times10^{14}$ \\

Warm Carbon Dust &
$1100^{+30}_{-40}$ &
$(9.8^{+2.2}_{-1.4})\times10^{-7}$ &
$0.10$ &
$(1.29^{+0.09}_{-0.08})\times10^{15}$ \\

Cold Carbon Dust &
$400 \pm 30$ &
$(1.6^{+0.6}_{-0.4})\times10^{-4}$ &
$0.10$ &
$(7.1^{+1.0}_{-0.7})\times10^{15}$ \\

\hline
\multicolumn{5}{c}{\textbf{Silicate Dust Model}} \\
\hline

Blackbody &
$6690^{+93}_{-77}$ &
-- &
-- &
$(1.29 \pm 0.03)\times10^{14}$ \\

Warm Carbon Dust &
$709^{+38}_{-25}$ &
$(2.0 \pm 0.6)\times10^{-5}$ &
$0.10$ &
$(3.92^{+0.42}_{-0.51})\times10^{15}$ \\

Cold Silicate Dust &
$100~\mathrm{(fixed;\ 50--150)}$ &
$(1.7^{+0.5}_{-0.5})\times10^{-2}$ &
$0.10$ &
$(5.6 \pm 0.8)\times10^{16}$ \\

\hline
\end{tabular}

\end{table*}

The final SED (Figure~\ref{fig:sed}) combines the flux-calibrated optical--NIR spectrum with the contemporaneous three-band \textit{JWST}/MIRI photometry measured using the $1.0\times$FWHM apertures described in Section~\ref{sec:jwst_photometry}. The resulting rest-frame SED spans approximately $0.32$--$18\,\mu$m and is referenced to the MIRI epoch of $+124.4$\,d post-explosion. The MIRI flux densities decrease from F770W to F1130W and then remain approximately constant, within the uncertainties, between F1130W and F1800W. The MIRI imaging shows SN~2025aico has excess MIR emission above the extrapolated optical--NIR continuum. The observed SED may contain contributions from emission lines in the evolving ejecta, the hot SN continuum, and thermal emission from newly formed or pre-existing dust \citep{Gall_2014,Shahbandeh2023}. We therefore model the full optical-to-MIR SED to investigate the relative contributions of these components.

\subsection{SED Fitting}

We modeled the optical-to-MIR SED shown in Figure~\ref{fig:sed} following the optically thin dust formalism described by \citet{Shahbandeh2023}. The observed flux is represented as the sum of a blackbody continuum tracing the hot SN ejecta and one or more single-temperature dust components. Each dust component is modeled as an optically thin distribution of spherical grains, with the emitted flux calculated from the wavelength-dependent absorption efficiency, $Q_{\rm abs}(\lambda,a)$. We adopt a fixed grain radius of $a=0.10\,\mu$m for all dust components. Graphite optical constants from \citet{Zubko_2004} are used for carbonaceous grains, while astronomical-silicate optical constants are taken from \citet{Draine1984, Draine2007}.

We consider two models that share a photospheric blackbody and a warm carbonaceous component but differ in the composition of the colder material responsible for the F1800W emission. In the Cold Carbon Dust Model, the excess is represented by a second, colder graphite component. In the Silicate Dust Model, it is represented by cold astronomical-silicate grains. The temperatures, dust masses, and emitting radii are fit parameters, except for the cold-silicate temperature. Because only the longest-wavelength MIRI measurement constrains this component, its temperature cannot be independently determined from the SED. We therefore adopt $T_{\rm cold,sil}=100$~K and impose hard lower and upper bounds of $50$ and $150$~K, respectively. The remaining parameters were determined using a Markov Chain Monte Carlo approach. Unless otherwise stated, the quoted uncertainties correspond to the 16th and 84th percentiles of the marginalized posterior distributions, representing a central 68\% credible interval and approximately a $1\sigma$ uncertainty for a Gaussian posterior. The resulting parameters are presented in Table~\ref{tab:dustfits}, and the corresponding model SEDs are shown in Figure~\ref{fig:sed}.

The optical and NIR continuum is well reproduced by a blackbody with a photospheric temperature of $T_{\rm phot}=6950\pm90$~K in the Cold Carbon Dust Model and $T_{\rm phot}=6690^{+93}_{-77}$~K in the Silicate Dust Model. The corresponding photospheric radii are $R_{\rm phot}=(1.19\pm0.03)\times10^{14}$~cm and $R_{\rm phot}=(1.29\pm0.03)\times10^{14}$~cm, respectively (Table~\ref{tab:dustfits}). The MIRI flux densities lie substantially above the long-wavelength extrapolation of the blackbody, revealing a clear MIR excess. In both models, reproducing the flux between approximately $1.5$ and $11.3\,\mu$m requires a warm carbonaceous component with an adopted grain radius of $a_{\rm C}=0.10\,\mu$m. This grain size is consistent with values commonly adopted or inferred for newly formed carbonaceous dust in stripped-envelope SNe, which typically range from approximately $0.01$ to $0.3\,\mu$m \citep{Nozawa_2010,Szalai_2016,Bevan2019,Priestley_2020}. The best-fit warm-carbon temperatures are $T_{\rm C}=1100^{+30}_{-40}$~K in the Cold Carbon Dust Model and $T_{\rm C}=709^{+38}_{-25}$~K in the Silicate Dust Model. The corresponding dust masses are $M_{\rm C}=(9.8^{+2.2}_{-1.4})\times10^{-7},M_\odot$ and $M_{\rm C}=(2.0\pm0.6)\times10^{-5},M_\odot$, while the corresponding emitting radii are $R_{\rm C}=(1.29^{+0.09}_{-0.08})\times10^{15}$~cm and $R_{\rm C}=(3.92^{+0.42}_{-0.51})\times10^{15}$~cm, respectively. We emphasize that the dust radii reported here should be interpreted as minimum characteristic radii within the optically thin framework rather than as uniquely determined outer boundaries of the dust distribution. The emitting dust must therefore lie at or beyond these inferred radii and may extend to substantially larger distances.

An additional colder component is required to reproduce the relatively high F1800W flux density. In the Cold Carbon Dust Model, this component has a temperature of $T_{\rm cold,C}=400\pm30$~K, a mass of $M_{\rm cold,C}=(1.6^{+0.6}_{-0.4})\times10^{-4},M_\odot$, and an emitting radius of $R_{\rm cold,C}=(7.1^{+1.0}_{-0.7})\times10^{15}$~cm for the same adopted grain radius of $0.10\,\mu$m. Alternatively, the Silicate Dust Model is shown for an illustrative cold-silicate temperature of $T_{\rm cold,sil}=100$~K, with hard bounds of $50$--$150$~K rather than a data-driven posterior interval. Conditional on this adopted temperature range and a fixed grain radius of $a_{\rm sil}=0.10\,\mu$m, the model gives a dust mass of $M_{\rm cold,sil}=(1.7^{+0.5}_{-0.5})\times10^{-2},M_\odot$ and an emitting radius of $R_{\rm cold,sil}=(5.6\pm0.8)\times10^{16}$~cm.

Figure~\ref{fig:photospher_radii} compares the fitted radius of each component with the maximum ejecta radius at the SED epoch. We estimate this radius as $R_{\rm exp}=6.45\times10^{16}$~cm using a velocity of $60{,}000\,\mathrm{km\,s^{-1}}$, measured from the blue wing of the broad He~\textsc{i} profile, as shown in Figure~\ref{fig:aicoHe}. Such high velocities are expected in the low-mass outermost ejecta, which are strongly accelerated as the explosion shock propagates through the steeply declining density profile near the stellar surface \citep{Matzner1999}. Numerical models of Type~IIb explosions similarly produce steep high-velocity ejecta tails \citep{Piro2017,Piro2021}. All fitted components lie within this maximum expansion radius, permitting either newly formed dust within the ejecta or pre-existing circumstellar dust heated by the advancing shock. We note that the fitted dust radii depend on the optically thin assumption and should be interpreted as characteristic inner boundaries rather than precise physical radii \citep{Gall_2014,Bevan2019,Priestley_2020}. The radius comparison is therefore consistent with dust located within the ejecta or nearby CSM; however, with only a single epoch of MIR observations, we cannot distinguish between newly formed and pre-existing dust.

To assess whether pre-existing circumstellar grains could have survived the peak SN radiation field, we estimate the dust-evaporation radius as

\begin{equation}
R_{\rm evap}=\left(\frac{L_{\rm peak}}{4\pi\sigma T_{\rm evap}^{4}}\right)^{1/2},
\end{equation}

where $L_{\rm peak}=4.07\times10^{41}\,\mathrm{erg\,s^{-1}}$ is the peak optical pseudo-bolometric luminosity inferred by \citet{zhao2026}, $\sigma$ is the Stefan--Boltzmann constant, and $T_{\rm evap}$ is the adopted grain-evaporation temperature. Adopting $T_{\rm evap,C}=2000$~K for carbonaceous grains and $T_{\rm evap,sil}=1500$~K for silicate grains \citep{Dwek1986}, we obtain $R_{\rm evap,C}\approx6.0\times10^{15}$~cm and $R_{\rm evap,sil}\approx1.1\times10^{16}$~cm. Therefore we find at these radii the hot carbonous component is likely freshly formed. These estimates are approximate because the optical pseudo-bolometric luminosity does not include the complete bolometric output and because grain-equilibrium temperatures depend on wavelength-dependent absorption and emission efficiencies \citep{Dwek1986, Draine2007}.

\subsection{Dust Interpretations}
Type~IIb SNe are an important but still poorly sampled test case for rapid dust production in SN explosions \citep{Gall_2014,Szalai_2016,Tinyanont_2016,Priestley_2020}. In CCSNe, dust is expected to form within the metal-rich inner ejecta, particularly in helium-, carbon-, and oxygen-rich layers, as the ejecta expand and cool \citep{Nozawa_2007,Lee_2009,Gall_2014,Reynolds_2025}. Because Type~IIb progenitors retain only thin hydrogen envelopes, their inner ejecta become observable earlier than those of hydrogen-rich Type~II SNe, potentially revealing molecule formation and dust condensation at earlier phases \citep{Nozawa_2010,Gall_2014}. However, smooth Type~IIb dust models predict that the low envelope mass also produces rapid expansion of the helium core, lower gas densities, and relatively small grains. In the Type~IIb model of \citet{Nozawa_2010}, carbon grains condense at approximately $300$--$350$~d, silicates and oxides at approximately $350$--$500$~d, and typical grain radii are $<0.01\,\mu$m. Dust emission from SN~2025aico at only $\sim125$~d therefore probes an earlier regime than that generally considered by standard Type~IIb dust-formation models.

This phase is also earlier than most quantitative dust constraints for SNe~IIb. In SN~2011dh and SN~2013df, inferred carbon-dust masses range from approximately $3.4\times10^{-6}$ to $1.4\times10^{-3}\,M_\odot$ and $1.3\times10^{-6}$ to $2.8\times10^{-4}\,M_\odot$, respectively, at phases of approximately $250$--$310$~d \citep{Ergon_2014,Szalai_2016,Tinyanont_2016,Priestley_2020}. By comparison, the Cold Carbon Dust Model for SN~2025aico gives $M_{\rm cold,C}=(1.6^{+0.64}_{-0.42})\times10^{-4}\,M_\odot$ at approximately $125$~d, comparable to values inferred for these events despite being measured at roughly half the phase. This mass represents only a small fraction of the available carbon reservoir because Type~IIb models predict that as much as $\sim0.07\,M_\odot$ of carbon dust may ultimately condense \citep{Nozawa_2010}. The significance of the SN~2025aico measurement is therefore primarily temporal: either a small fraction of its carbon-rich ejecta cooled and condensed unusually early, potentially within dense clumps \citep{Gall_2014,Sarangi_2015,Bevan2019}, or previous Type~IIb samples lacked sufficiently early IR coverage to reveal this phase.

The fitted location of the carbonaceous dust does not, by itself, establish that the material condensed within the ejecta. The cold-carbon radius, $R_{\rm cold,C}=(7.076^{+0.957}_{-0.688})\times10^{15}$~cm, lies slightly beyond the estimated carbon-grain evaporation radius of approximately $6.0\times10^{15}$~cm. Pre-existing carbonaceous grains at this location could therefore potentially have survived the peak radiation field. Such grains could subsequently be heated collisionally within or near the ejecta--CSM interaction region or radiatively by the SN emission \citep{Dwek1983,Dwek1986,Fox2011}. Emission from unrelated host-galaxy dust within the MIRI aperture may also contribute, particularly to the F1800W measurement. The current spatial, wavelength, and temporal coverage cannot uniquely distinguish newly condensed ejecta dust from collisionally or radiatively heated circumstellar material, including a possible infrared light echo, or unrelated host-galaxy emission.

The Silicate Dust Model would imply a substantially different physical regime. Its cold-silicate component has an inferred mass of $M_{\rm cold,sil}=(1.7^{+0.52}_{-0.46})\times10^{-2}\,M_\odot$, corresponding to approximately 20\% of the $\sim0.08\,M_\odot$ silicate-dust yield predicted by the Type~IIb model of \citet{Nozawa_2010}. Producing this much silicate dust by approximately $125$~d would be difficult to reconcile with the nominal silicate-condensation window of approximately $350$--$500$~d. Its radius, $R_{\rm cold,sil}=(5.638^{+0.800}_{-0.809})\times10^{16}$~cm, also lies well outside the estimated silicate-grain evaporation radius of approximately $1.1\times10^{16}$~cm, so pre-existing silicate grains could have survived at this location \citep{Dwek1986}. If the cold-silicate component instead represents newly condensed ejecta dust, its early formation would require substantial departures from smooth, one-dimensional ejecta conditions, such as dense chemically enriched clumps, rapid cooling within a shocked dense shell, or a temperature structure that permits localized regions to reach condensation conditions much earlier than the average ejecta \citep{Gall_2014,Sarangi_2015,Bevan2019}. Given its large inferred mass and poorly constrained temperature, however, the cold-silicate component may more plausibly trace pre-existing circumstellar material or unrelated host-galaxy dust. Pre-existing circumstellar silicate grains could be heated either collisionally by the ejecta--CSM interaction. But due to the low radius of the emission we rule out that it radiatively heated through an infrared light echo \citep{Dwek1983,Fox2011}. The available observations do not provide sufficient wavelength coverage to locate the peak of the cold component or sufficient temporal coverage to measure its evolution. We therefore cannot distinguish among newly formed dust, collisionally heated circumstellar dust, and unrelated host-galaxy emission as the origin of the fitted silicate component.

Host-galaxy dust does not favor only one of the fitted compositions. Carbon-rich asymptotic giant branch stars primarily produce carbonaceous grains, whereas oxygen-rich asymptotic giant branch stars predominantly produce silicate grains \citep{Hofner_2018}. Either cold component could therefore represent emission from material along the line of sight, consistent with the nearby sources visible in Figure~\ref{fig:miri_cutout}. Dust composition alone cannot establish whether the cold component is physically associated with SN~2025aico.

The warm carbonaceous component is more plausibly associated with the SN because it dominates much of the NIR and MIR excess and has a substantially smaller inferred radius. Its radius is $R_{\rm C}=(1.287^{+0.090}_{-0.075})\times10^{15}$~cm in the Cold Carbon Dust Model and $R_{\rm C}=(3.920^{+0.419}_{-0.509})\times10^{15}$~cm in the Silicate Dust Model, placing it inside the estimated carbon-grain evaporation radius in both cases. Unshielded pre-existing carbonaceous grains would be unlikely to survive the peak radiation field at these locations, so the warm component may represent dust that condensed after the ejecta expanded and cooled \citep{Dwek1986,Nozawa_2007,Gall_2014}. Nevertheless, its physical origin is not uniquely established by the available observations. The inferred temperatures, masses, and radii should consequently be regarded as model-dependent, particularly for the cold component, whose temperature and luminosity are constrained primarily by the F1800W measurement.

An additional uncertainty is the possible contribution of free--free emission from ionized ejecta or circumstellar gas. Free--free and free--bound emission can contribute to the NIR and MIR continua of CCSNe and has been included alongside thermal dust emission in previous SN SED analyses \citep{Wooden1993,Kotak2009}. If present in SN~2025aico, this emission could account for part of the flux currently attributed to the warm dust component, reducing the amount of emission that must be reproduced by thermal dust and altering the inferred dust temperature and mass. Because the available SED does not independently constrain a free--free component, the fitted dust properties are conditional on the assumption that the IR excess is dominated by thermal dust emission.

If dust has formed within the ejecta, its long-term survival will depend strongly on its subsequent interaction with the reverse shock \citep{Nozawa_2010,Silvia_2010,Bocchio2016}. A delayed or asymmetric reverse shock may allow grains to remain in dense, cooling gas for longer, increasing the time available for grain growth before sputtering becomes efficient. Larger grains are generally more resistant to sputtering and are therefore more likely to survive shock processing and contribute to the final dust yield \citep{Nozawa_2010,Silvia_2010,Micelotta2016}. This behavior is consistent with dust-formation modeling of the well-studied Type~IIb remnant Cassiopeia~A. Its low hydrogen-envelope mass exposed grains formed in the metal-rich core to reverse-shock processing earlier than in more extended Type~IIP events, with the resulting grain sizes and destruction efficiencies depending strongly on the circumstellar density structure \citep{Nozawa_2010}.

In the broader CCSNe context, SN~2025aico occupies the gap between early warm-dust detections, which commonly imply approximately $10^{-6}$--$10^{-4}\,M_\odot$ of carbonaceous dust \citep{Gall_2014,Szalai_2016,Tinyanont_2016,Priestley_2020}, and the larger dust masses measured in decades-old SNe and remnants. These later measurements include approximately $3.5$--$4.7\times10^{-3}\,M_\odot$ for SN~1993J \citep{Zs_ros_2021}, approximately $3.8\times10^{-3}\,M_\odot$ for the Type~IIb SN~2021afdx \citep{Hosseinzadeh_2023}, and comparably large dust reservoirs in SN~1987A and Cas~A \citep{Wesson2015,De_Looze_2016,Szalai_2021,Szalai_2025}. CCSNe may provide an important source of dust on timescales shorter than those required for lower-mass stellar populations to evolve, making their dust yields particularly relevant to the enrichment of the early Universe \citep{Todini_2001,Nozawa_2003,Gall_2011}. Type~IIb-like explosions therefore provide an important test of how efficiently stripped massive stars can expose carbon-rich material, form dust seeds, and retain those grains through subsequent shock processing. SN~2025aico provides a rare early-phase constraint on the initial development of dust emission and its potential connection to the larger dust reservoirs observed in older CCSNe and remnants.

%%--------------------------------------------------------------
\section{Conclusion}\label{sec:Conclusion}

We have presented extensive optical, NIR, and MIR observations of SN~2025aico, a nearby Type~IIb SN in LEDA~35384. Our spectroscopic sequence extends from $+1.64$\,d, during the shock-cooling phase, to $+166.82$\,d in the nebular phase and combines observations from Keck, IRTF, Gemini, MMT, SOAR, and Magellan with serendipitous \textit{JWST}/MIRI imaging at $+124.4$\,d. The dense early-time cadence, broad wavelength coverage, and extended nebular sequence make SN~2025aico one of the most comprehensively observed nearby SNe~IIb at optical and NIR wavelengths.

The H$\alpha$ and NIR He~\textsc{i} absorption velocities reveal stratified ejecta, with a high-velocity hydrogen-rich envelope at approximately $10{,}000$--$14{,}000\,\mathrm{km\,s^{-1}}$ surrounding helium-rich material at approximately $5{,}000$--$9{,}000\,\mathrm{km\,s^{-1}}$. The He~\textsc{i} velocity evolution occurs in three stages: a rapid early decline, an intermediate increase, and a late plateau. We interpret the intermediate increase as the onset of non-thermal excitation in faster helium-rich ejecta as radioactive energy deposition extends outward. Because strong outward $^{56}$Ni mixing is expected to weaken or erase this increase, its detection favors limited mixing of radioactive material into the helium-rich ejecta \citep{Dessart_2012,Dessart_2015,Dessart_2016,Ergon_2022}. Among the comparison sample, only SN~2011dh shows a similar intermediate increase, highlighting the importance of densely sampled spectroscopy during this phase.

At late times, the He~\textsc{i} and [O~\textsc{i}] lines develop similar double-peaked profiles, indicating that the helium- and oxygen-rich ejecta share a related large-scale asymmetric structure. The narrower, predominantly single-peaked [Ca~\textsc{ii}] emission has a different velocity distribution, suggesting that the asymmetry is not identical across all compositional layers. These observations favor an intrinsically aspherical inner ejecta geometry over a spherically symmetric structure \citep{Maeda2008,Maurer2010,Jerkstrand_2017}. Three-dimensional explosion models predict that large-scale asymmetries produced during the explosion can persist into the homologous ejecta and that the resulting distribution depends on both the explosion dynamics and the progenitor structure \citep{Wongwathanarat_2015}. More recent calculations of stripped-envelope SNe show that these differences can also appear directly in nebular line profiles. In particular, \citet{vanbaal2024} find that [Ca~\textsc{ii}] can show substantially different velocity shifts and viewing-angle dependence from [O~\textsc{i}], with calcium tracing the explosion asymmetry more strongly than oxygen. The different He~\textsc{i}, [O~\textsc{i}], and [Ca~\textsc{ii}] profiles in SN~2025aico therefore provide an observational comparison with these predictions and suggest that different compositional layers do not share the same ejecta geometry.

We compare SN~2025aico with SNe~2008ax, 2011dh, and 2020acat, which comprise the other SNe~IIb with publicly available nebular-phase NIR spectroscopy. SN~2025aico has the lowest kinetic-energy-to-ejecta-mass ratio and the narrowest He~\textsc{i} peak separations in this sample, while the more energetic SN~2020acat has the broadest profiles and largest separations. This suggests a possible connection between the velocity scale of the helium-rich structure and explosion energetics, although the four-object sample is too small to establish a correlation. Together with the compact progenitor radius of $R_{\rm prog}\sim10$--$30\,R_\odot$, brief shock-cooling phase, and residual hydrogen envelope \citep{zhao2026}, these properties favor a compact, binary-stripped progenitor. These properties are consistent with a moderate-mass helium star that lost most, but not all, of its hydrogen envelope through interaction with a binary companion. 

The \textit{JWST}/MIRI observations reveal a MIR excess whose flux density decreases from F770W to F1130W and then remains approximately constant between F1130W and F1800W. Both SED models require a warm carbonaceous component, with fitted temperatures of approximately $1100$~K and $710$~K in the Cold Carbon and Silicate Dust Models, respectively. An additional cold component is required to reproduce the F1800W flux. The Cold Carbon Dust Model gives $T_{\rm cold,C}\approx399$~K and $M_{\rm cold,C}\approx1.6\times10^{-4}\,M_\odot$, whereas the Silicate Dust Model gives $M_{\rm cold,sil}\approx1.7\times10^{-2}\,M_\odot$ for an adopted temperature of $100$~K with imposed bounds of $50$--$150$~K. The warm component is plausibly associated with the SN, but the origin of the cold component remains uncertain. The available wavelength and temporal coverage cannot distinguish among newly formed dust, collisionally heated circumstellar dust, an infrared light echo, and unrelated host-galaxy emission \citep{Dwek1983,Fox2011}.

SN~2025aico therefore provides a rare combined view of radioactive mixing, ejecta geometry, and early dust emission in a Type~IIb explosion. Its three-stage helium velocity evolution favors limited outward $^{56}$Ni mixing, while the matching He~\textsc{i} and [O~\textsc{i}] profiles reveal coherent asymmetry across the ejecta. Its early MIR excess offers an important constraint on dust in stripped-envelope SNe. Where it is likely that at these phases there is some carbonous dist formation, although additional MIR wavelength coverage and multi-epoch observations are needed to determine the origin of the colder emission. Looking forward, a larger sample with densely sampled early-time NIR spectroscopy and nebular-phase NIR and MIR observations would allow us to determine how common these properties are among SNe~IIb, and its implications for early universe dust formation and mass loss mechanisms in massive stars. 

%%------------------------------------------------------------------------
\section{Acknowledgments}
This material is based upon work supported by the National Science Foundation (NSF) under Grant Number 2548017. Disclaimer: Any opinions, findings, and conclusions or recommendations expressed in this material are those of the author(s) and do not necessarily reflect the views of the National Science Foundation. We acknowledge support from the NSF Research Experiences for Undergraduates (REU) Program at the University of Hawai‘i Institute for Astronomy. G.D acknowledges program support from Roy Gal and Finn Giddings, as well as fellow cohort members.

K.M., C.A., and C.P., acknowledge support from NASA grants JWST-GO-04522, JWST-GO-04217, JWST-GO-04436, JWST-GO-03726, JWST-GO-05057, JWST-GO-05290, JWST-GO-06023, JWST-GO-06677, JWST-GO-06213, JWST-GO-06583. Support for programs \#3726, \#4217, \#4436, \#4522, \#5057, \#6023, \#6213, \#6583, and \#6677 were provided by NASA through a grant from the Space Telescope Science Institute, which is operated by the Association of Universities for Research in Astronomy, Inc., under NASA contract NAS 5-03127.

D.O.J. acknowledges support from NSF grants AST-2407632, AST-2429450, and AST-2510993, NASA grants 80NSSC24M0023 and 80NSSC24K0353, and HST/JWST grants HST-GO-17128.028 and JWST-GO-05324.031, awarded by the Space Telescope Science Institute (STScI), which is operated by the Association of Universities for Research in Astronomy, Inc., for NASA, under contract NAS5-26555. This work is also funded in part by the Gordon and Betty Moore Foundation through Grant GBMF13900 to D.O.J.

We are grateful to the telescope operators, instrument scientists, and support staff at the NASA Infrared Telescope Facility (IRTF), SOAR Observatory, Gemini Observatory, MMT Observatory, Las Campanas Observatory, and the W.~M.~Keck Observatory for their assistance in obtaining the data presented in this paper. This work is based in part on observations obtained with IRTF/SpeX, SOAR/GHTS, SOAR/TripleSpec, Gemini North/GMOS-N, MMT/Binospec, Magellan I/IMACS, Keck I/LRIS, Keck II/NIRES, and Keck II/KCWI. 

We recognize and acknowledge the Indigenous peoples and local communities with longstanding cultural and historical connections to the lands on which these ground-based observatories are located. The authors wish to recognize and acknowledge the very significant cultural role and reverence that the summit of Maunakea has within the Native Hawaiian community, and we are most fortunate to have the opportunity to conduct observations from this mountain. We acknowledge the O'odham and Yaqui peoples of southern Arizona, where the MMT Observatory is located. We also recognize the Indigenous peoples and local communities with historical and ongoing connections to the regions of Chile in which the Cerro Pachón and Las Campanas Observatory are located.

This paper makes use of observations obtained with the NASA/ESA/CSA James Webb Space Telescope. These observations are associated with program 7041 (PI: D.~Berg). The authors made use of the Mikulski Archive for Space Telescopes (MAST) at the Space Telescope Science Institute (STScI). STScI is operated by the Association of Universities for Research in Astronomy, Inc., under NASA contract NAS~5-03127. Machine learning tools were used to suggest language improvements within the manuscript. AI tools were used to assist with refinement of the presentation of code used in the research process and reported in the manuscript. The GAI tool used was ChatGPT.

\facility{JWST (MIRI), Keck (LRIS, NIRES, KCWI), IRTF (SpeX), SOAR (Goodman, TripleSpec), Gemini (GMOS-N), MMT (Binospec), Magellan (IMACS)}.

\software{
Astropy \citep{astropy2013,astropy2018,astropy2022},
PypeIt \citep{Prochaska2020},
Photutils \citep{Bradley2025},
NumPy \citep{numpy2020},
SciPy \citep{scipy2020},
Matplotlib \citep{Hunter2007},
space\_phot \citep{spacephot},
pandas \citep{mckinney-proc-scipy-2010, reback2020pandas},
ipywidgets \citep{ipywidgets}
}
%%-------------------------------------------------------------------------
\bibliography{references}

@article{Andrews2025,
  author  = {Andrews, J. and Hsu, B. and Shrestha, M. and Sand, D. and Bostroem, A.},
  title   = {{SN 2025aico}},
  journal = {Transient Name Server Classification Report},
  year    = {2025},
  volume  = {2025-5179},
  pages   = {1}
}

@misc{barmentloo2024,
      title={Nebular Nitrogen Line Emission in Stripped-Envelope Supernovae -- a New Progenitor Mass Diagnostic}, 
      author={Stan Barmentloo and Anders Jerkstrand and Koichi Iwamoto and Izumi Hachisu and Ken'ichi Nomoto and Jesper Sollerman and Stan Woosley},
      year={2024},
      eprint={2403.08911},
      archivePrefix={arXiv},
      primaryClass={astro-ph.HE},
      url={https://arxiv.org/abs/2403.08911}, 
}

@MISC{Berg_2025jwst,
       author = {{Berg}, Danielle and {James}, Bethan Lesley and {Amorin}, Ricardo and {Arellano Cordova}, Karla Ziboney and {Bolatto}, Alberto and {Brinchmann}, Jarle and {Carr}, Cody Andrew and {Chisholm}, John and {Dinerstein}, Harriet L. and {Erb}, Dawn K. and {Feltre}, Anna and {Fisher}, Deanne B. and {Hayes}, Matthew James and {Heckman}, Timothy M. and {Henry}, Alaina L. and {Hernandez}, Svea S. and {Hu}, Weida and {Hunt}, Leslie and {Kumari}, Nimisha and {Lai}, Thomas and {Leitherer}, Claus and {Martin}, Crystal Linn and {Martinez}, Zorayda and {Maseda}, Michael and {McQuinn}, Kristen B.~W. and {Mingozzi}, Matilde and {Parker}, Kaelee S. and {Pogge}, Richard W. and {Ravindranath}, Swara and {Rigby}, Jane R. and {Rogers}, Noah Sidney James and {Roy}, Namrata and {Sandstrom}, Karin Marie and {Scarlata}, Claudia and {Senchyna}, Peter and {Skillman}, Evan D. and {Smith}, JD and {Stark}, Daniel P. and {Strom}, Allison L. and {Wofford}, Aida and {Xu}, Xinfeng},
        title = "{The CLASSYIR Treasury: Unveiling the Cosmic Engines Powering Galaxies with JWST/MIRI}",
 howpublished = {JWST Proposal. Cycle 4, ID. \#7041},
         year = 2025,
        month = mar,
        pages = {7041},
       adsurl = {https://ui.adsabs.harvard.edu/abs/2025jwst.prop.7041B}
}

@article{Bianco_2014,
   title={MULTI-COLOR OPTICAL AND NEAR-INFRARED LIGHT CURVES OF 64 STRIPPED-ENVELOPE CORE-COLLAPSE SUPERNOVAE},
   volume={213},
   ISSN={1538-4365},
   url={http://dx.doi.org/10.1088/0067-0049/213/2/19},
   DOI={10.1088/0067-0049/213/2/19},
   number={2},
   journal={The Astrophysical Journal Supplement Series},
   publisher={American Astronomical Society},
   author={Bianco, F. B. and Modjaz, M. and Hicken, M. and Friedman, A. and Kirshner, R. P. and Bloom, J. S. and Challis, P. and Marion, G. H. and Wood-Vasey, W. M. and Rest, A.},
   year={2014},
   month="July", pages={19} }

@ARTICLE{Bradley2025,
       author = {{Bradley}, L. and others},
        title = "{Photutils: Photometry tools for Python}",
      journal = {Zenodo},
         year = 2025
}

@ARTICLE{Burns2021,
       author = {{Burns}, C. and {Hsiao}, E. and {Suntzeff}, N. and {Baron}, E. and {Shappee}, B. and {Aldoroty}, L. and {Anderson}, J. and {Ashall}, C. and {Bersten}, M. and {Brown}, P. and {Burrow}, A. and {Clochiatti}, Alejandro and {Davis}, S. and {DerKacy}, J. and {Do}, A. and {Folatelli}, G. and {Forster Buron}, F. and {Galbany}, L. and {Hoeflich}, P. and {Holmbo}, S. and {Karamehmetoglu}, E. and {Krisciunas}, K. and {Kumar}, S. and {Lu}, J. and {Mazzali}, P. and {Morrell}, N. and {Pessi}, P. and {Phillips}, M. and {Pignata}, G. and {Piro}, A.~L. and {Polin}, A. and {Shahbandeh}, M. and {Stangl}, S. and {Stritzinger}, M. and {Teffs}, J. and {Tonry}, J. and {Tucker}, M. and {Uddin}, S. and {Yang}, J.},
        title = "{Introducing POISE: Precision Observations of Infant Supernova Explosions}",
      journal = {The Astronomer's Telegram},
         year = 2021,
        month = mar,
       volume = {14441},
        pages = {1},
       adsurl = {https://ui.adsabs.harvard.edu/abs/2021ATel14441....1B}
}

@ARTICLE{Cao2013,
       author = {{Cao}, Yi and {Kasliwal}, Mansi M. and {Arcavi}, Iair and {Horesh}, Assaf and {Hancock}, Paul and {Valenti}, Stefano and {Cenko}, S. Bradley and {Kulkarni}, S.~R. and {Gal-Yam}, Avishay and {Gorbikov}, Evgeny and {Ofek}, Eran O. and {Sand}, David and {Yaron}, Ofer and {Graham}, Melissa and {Silverman}, Jeffrey M. and {Wheeler}, J. Craig and {Marion}, G.~H. and {Walker}, Emma S. and {Mazzali}, Paolo and {Howell}, D. Andrew and {Li}, K.~L. and {Kong}, A.~K.~H. and {Bloom}, Joshua S. and {Nugent}, Peter E. and {Surace}, Jason and {Masci}, Frank and {Carpenter}, John and {Degenaar}, Nathalie and {Gelino}, Christopher R.},
        title = "{Discovery, Progenitor and Early Evolution of a Stripped Envelope Supernova iPTF13bvn}",
      journal = {\apjl},
         year = 2013,
        month = sep,
       volume = {775},
       number = {1},
          eid = {L7},
        pages = {L7},
          doi = {10.1088/2041-8205/775/1/L7},
archivePrefix = {arXiv},
       eprint = {1307.1470},
 primaryClass = {astro-ph.SR},
       adsurl = {https://ui.adsabs.harvard.edu/abs/2013ApJ...775L...7C}
}

@ARTICLE{Chevalier2010,
       author = {{Chevalier}, Roger A. and {Soderberg}, Alicia M.},
        title = "{Type IIb Supernovae with Compact and Extended Progenitors}",
      journal = {\apjl},
         year = 2010,
        month = mar,
       volume = {711},
       number = {1},
        pages = {L40-L43},
          doi = {10.1088/2041-8205/711/1/L40},
archivePrefix = {arXiv},
       eprint = {0911.3408},
 primaryClass = {astro-ph.HE},
       adsurl = {https://ui.adsabs.harvard.edu/abs/2010ApJ...711L..40C}
}

@ARTICLE{Ciabattari2013,
       author = {{Ciabattari}, F. and {Mazzoni}, E. and {Donati}, S. and {Petroni}, G. and {Foglia}, S. and {Galli}, G. and {Cenko}, S.~B. and {Clubb}, K.~I. and {Zheng}, W. and {Kelly}, P.~L. and {Filippenko}, A.~V. and {Van Dyk}, S.~D.},
        title = "{Supernova 2013df in NGC 4414 = Psn J12262933+3113383}",
      journal = {Central Bureau Electronic Telegrams},
         year = 2013,
        month = "June",
       volume = {3557},
        pages = {1},
       adsurl = {https://ui.adsabs.harvard.edu/abs/2013CBET.3557....1C}
}

@ARTICLE{astropy2013,
       author = {{Astropy Collaboration} and {Robitaille}, T.~P. and
                 {Tollerud}, E.~J. and others},
        title = "{Astropy: A community Python package for astronomy}",
      journal = {A\&A},
         year = 2013,
       volume = {558},
          eid = {A33},
        pages = {A33},
          doi = {10.1051/0004-6361/201322068}
}

@ARTICLE{astropy2018,
       author = {{Astropy Collaboration} and {Price-Whelan}, A.~M. and
                 {Sip{\H o}cz}, B.~M. and others},
        title = "{The Astropy Project: Building an inclusive, open-science project and status of the v2.0 core package}",
      journal = {AJ},
         year = 2018,
       volume = {156},
       number = {3},
          eid = {123},
        pages = {123},
          doi = {10.3847/1538-3881/aabc4f}
}

@ARTICLE{astropy2022,
       author = {{Astropy Collaboration} and {Price-Whelan}, A.~M. and
                 {Lim}, P.~L. and others},
        title = "{The Astropy Project: Sustaining and Growing a Community-oriented Open-source Project and the Latest Major Release (v5.0) of the Core Package}",
      journal = {ApJ},
         year = 2022,
       volume = {935},
       number = {2},
          eid = {167},
        pages = {167},
          doi = {10.3847/1538-4357/ac7c74}
}

@misc{ipywidgets,
  author       = {{Jupyter Widgets Community}},
  title        = {ipywidgets: Interactive Widgets for the Jupyter Notebook},
  year         = {2015},
  publisher    = {GitHub},
  url          = {https://github.com/jupyter-widgets/ipywidgets}
}

@article{Davis2019,
   title={Carnegie Supernova Project-II: Near-infrared Spectroscopic Diversity of Type II Supernovae},
   volume={887},
   ISSN={1538-4357},
   url={http://dx.doi.org/10.3847/1538-4357/ab4c40},
   DOI={10.3847/1538-4357/ab4c40},
   number={1},
   journal={The Astrophysical Journal},
   publisher={American Astronomical Society},
   author={Davis, S. and Hsiao, E. Y. and Ashall, C. and Hoeflich, P. and Phillips, M. M. and Marion, G. H. and Kirshner, R. P. and Morrell, N. and Sand, D. J. and Burns, C. and Contreras, C. and Stritzinger, M. and Anderson, J. P. and Baron, E. and Diamond, T. and Gutiérrez, C. P. and Hamuy, M. and Holmbo, S. and Kasliwal, M. M. and Krisciunas, K. and Kumar, S. and Lu, J. and Pessi, P. J. and Piro, A. L. and Prieto, J. L. and Shahbandeh, M. and Suntzeff, N. B.},
   year={2019},
   month=Dec, pages={4} }

@article{Davis_2021,
   title={SN 2013ai: A Link between Hydrogen-rich and Hydrogen-poor Core-collapse Supernovae},
   volume={909},
   ISSN={1538-4357},
   url={http://dx.doi.org/10.3847/1538-4357/abdd36},
   DOI={10.3847/1538-4357/abdd36},
   number={2},
   journal={The Astrophysical Journal},
   publisher={American Astronomical Society},
   author={Davis, S. and Pessi, P. J. and Fraser, M. and Ertini, K. and Martinez, L. and Hoeflich, P. and Hsiao, E. Y. and Folatelli, G. and Ashall, C. and Phillips, M. M. and Anderson, J. P. and Bersten, M. and Englert, B. and Fisher, A. and Benetti, S. and Bunzel, A. and Burns, C. and Chen, T. W. and Contreras, C. and Elias-Rosa, N. and Falco, E. and Galbany, L. and Kirshner, R. P. and Kumar, S. and Lu, J. and Lyman, J. D. and Marion, G. H. and Mattila, S. and Maund, J. and Morrell, N. and Serón, J. and Stritzinger, M. and Shahbandeh, M. and Sullivan, M. and Suntzeff, N. B. and Young, D. R.},
   year={2021},
   month=Mar, pages={145} }

@ARTICLE{Davis2026,
       author = {{Davis}, Kyle W. and {Taggart}, Kirsty and {Tinyanont}, Samaporn and {Foley}, Ryan J. and {Rho}, Jeonghee and {Auchettl}, Katie and {Farias}, Diego and {Fox}, Ori D. and {Johansson}, Joel and {Kilpatrick}, Charles D. and {Patra}, Kishore C. and {Pellegrino}, Craig and {Ramirez-Ruiz}, Enrico and {Coulter}, David A. and {Dong}, Yize and {Gagliano}, Alexander T. and {Geballe}, T.~R. and {Jacobson-Gal{\'a}n}, Wynn V. and {Karcheski}, Jenna and {Kaur}, Ravjit and {Lau}, Ryan M. and {Moore}, Thomas and {Park}, Seong Hyun and {Rest}, Armin and {Szalai}, Tam{\'a}s and {Wang}, Qinan},
        title = "{JWST Reveals Large Reservoirs of Dust and Ongoing Circumstellar Interaction in SN Ibn/Icn 2023xgo over a Year Post-Explosion}",
      journal = {arXiv e-prints},
         year = 2026,
        month = may,
          eid = {arXiv:2606.00208},
        pages = {arXiv:2606.00208},
          doi = {10.48550/arXiv.2606.00208},
archivePrefix = {arXiv},
       eprint = {2606.00208},
 primaryClass = {astro-ph.HE},
       adsurl = {https://ui.adsabs.harvard.edu/abs/2026arXiv260600208D}
}

@misc{deng2001,
      title={Spectrum Analysis of Type IIb Supernova 1996cb}, 
      author={Jinsong Deng and Yulei Qiu and Jinyao Hu},
      year={2001},
      eprint={astro-ph/0106404},
      archivePrefix={arXiv},
      primaryClass={astro-ph},
      url={https://arxiv.org/abs/astro-ph/0106404}, 
}

@article{Dessart_2012,
   title={On the nature of supernovae Ib and Ic: \textit{Radiative transfer of SN Ib/Ic ejecta}},
   volume={424},
   ISSN={0035-8711},
   url={http://dx.doi.org/10.1111/j.1365-2966.2012.21374.x},
   DOI={10.1111/j.1365-2966.2012.21374.x},
   number={3},
   journal={Monthly Notices of the Royal Astronomical Society},
   publisher={Oxford University Press (OUP)},
   author={Dessart, Luc and Hillier, D. John and Li, Chengdong and Woosley, Stan},
   year={2012},
   month="July", pages={2139–2159} }

@article{Dessart_2015,
   title={Radiative-transfer models for supernovae IIb/Ib/Ic from binary-star progenitors},
   volume={453},
   ISSN={1365-2966},
   url={http://dx.doi.org/10.1093/mnras/stv1747},
   DOI={10.1093/mnras/stv1747},
   number={2},
   journal={Monthly Notices of the Royal Astronomical Society},
   publisher={Oxford University Press (OUP)},
   author={Dessart, Luc and Hillier, D. John and Woosley, Stan and Livne, Eli and Waldman, Roni and Yoon, Sung-Chul and Langer, Norbert},
   year={2015},
   month=Aug, pages={2189–2213} }

@article{Dessart_2022,
   title={Using LSST late-time photometry to constrain Type Ibc supernovae and their progenitors},
   volume={666},
   ISSN={1432-0746},
   url={http://dx.doi.org/10.1051/0004-6361/202244413},
   DOI={10.1051/0004-6361/202244413},
   journal={Astronomy \& Astrophysics},
   publisher={EDP Sciences},
   author={Dessart, Luc and Prieto, Jose L. and Hillier, D. John and Kuncarayakti, Hanindyo and Hueichapan, Emilio D.},
   year={2022},
   month=Oct, pages={L14} }

@ARTICLE{Draine1984,
       author = {{Draine}, B.~T. and {Lee}, H.~M.},
        title = "{Optical Properties of Interstellar Graphite and Silicate Grains}",
      journal = {\apj},
         year = 1984,
        month = oct,
       volume = {285},
        pages = {89},
          doi = {10.1086/162480},
       adsurl = {https://ui.adsabs.harvard.edu/abs/1984ApJ...285...89D}
}

@ARTICLE{Draine2007,
       author = {{Draine}, B.~T. and {Li}, Aigen},
        title = "{Infrared Emission from Interstellar Dust. IV. The Silicate-Graphite-PAH Model in the Post-Spitzer Era}",
      journal = {\apj},
         year = 2007,
        month = mar,
       volume = {657},
       number = {2},
        pages = {810-837},
          doi = {10.1086/511055},
archivePrefix = {arXiv},
       eprint = {astro-ph/0608003},
 primaryClass = {astro-ph},
       adsurl = {https://ui.adsabs.harvard.edu/abs/2007ApJ...657..810D}
}

@ARTICLE{Dressler2011,
       author = {{Dressler}, Alan and {Bigelow}, Bruce and {Hare}, Tyson and {Sutin}, Brian and {Thompson}, Ian and {Burley}, Greg and {Epps}, Harland and {Oemler}, Jr., Augustus and {Bagish}, Alan and {Birk}, Christoph and {Clardy}, Ken and {Gunnels}, Steve and {Kelson}, Daniel and {Shectman}, Stephen and {Osip}, David},
        title = "{IMACS: The Inamori-Magellan Areal Camera and Spectrograph on Magellan-Baade}",
      journal = {\pasp},
         year = 2011,
        month = mar,
       volume = {123},
       number = {901},
        pages = {288},
          doi = {10.1086/658908}
}

@article{Dwek1983,
  author  = {Dwek, Eli},
  title   = {The Infrared Echo of a Type II Supernova with a Circumstellar Dust Shell},
  journal = {The Astrophysical Journal},
  volume  = {274},
  pages   = {175--183},
  year    = {1983}
}

@article{Ercolino_2026,
   title={The demographics of core-collapse supernovae: The role of binary evolution and interaction with the circumstellar medium},
   volume={706},
   ISSN={1432-0746},
   url={http://dx.doi.org/10.1051/0004-6361/202557572},
   DOI={10.1051/0004-6361/202557572},
   journal={Astronomy \& Astrophysics},
   publisher={EDP Sciences},
   author={Ercolino, Andrea and Jin, Harim and Langer, Norbert and Gal-Yam, Avishay and Schootemeijer, Abel and Mannes, Caroline},
   year={2026},
   month=Feb, pages={A169} }

@article{Ergon_2014,
   title={Optical and near-infrared observations of SN 2011dh – The first 100 days},
   volume={562},
   ISSN={1432-0746},
   url={http://dx.doi.org/10.1051/0004-6361/201321850},
   DOI={10.1051/0004-6361/201321850},
   journal={Astronomy \& Astrophysics},
   publisher={EDP Sciences},
   author={Ergon, M. and Sollerman, J. and Fraser, M. and Pastorello, A. and Taubenberger, S. and Elias-Rosa, N. and Bersten, M. and Jerkstrand, A. and Benetti, S. and Botticella, M. T. and Fransson, C. and Harutyunyan, A. and Kotak, R. and Smartt, S. and Valenti, S. and Bufano, F. and Cappellaro, E. and Fiaschi, M. and Howell, A. and Kankare, E. and Magill, L. and Mattila, S. and Maund, J. and Naves, R. and Ochner, P. and Ruiz, J. and Smith, K. and Tomasella, L. and Turatto, M.},
   year={2014},
   month=Feb, pages={A17} }

@ARTICLE{Ergon2015,
       author = {{Ergon}, M. and {Jerkstrand}, A. and {Sollerman}, J. and {Elias-Rosa}, N. and {Fransson}, C. and {Fraser}, M. and {Pastorello}, A. and {Kotak}, R. and {Taubenberger}, S. and {Tomasella}, L. and {Valenti}, S. and {Benetti}, S. and {Helou}, G. and {Kasliwal}, M.~M. and {Maund}, J. and {Smartt}, S.~J. and {Spyromilio}, J.},
        title = "{The Type IIb SN 2011dh: Two years of observations and modelling of the lightcurves}",
      journal = {\aap},
         year = 2015,
        month = aug,
       volume = {580},
          eid = {A142},
        pages = {A142},
          doi = {10.1051/0004-6361/201424592},
archivePrefix = {arXiv},
       eprint = {1408.0731},
 primaryClass = {astro-ph.SR},
       adsurl = {https://ui.adsabs.harvard.edu/abs/2015A&A...580A.142E}
}

@article{Ergon_2022,
   title={Spectral modelling of Type IIb supernovae: Comparison with SN 2011dh and the effect of macroscopic mixing},
   volume={666},
   ISSN={1432-0746},
   url={http://dx.doi.org/10.1051/0004-6361/202243448},
   DOI={10.1051/0004-6361/202243448},
   journal={Astronomy \& Astrophysics},
   publisher={EDP Sciences},
   author={Ergon, Mattias and Fransson, Claes},
   year={2022},
   month=Oct, pages={A104} }

@ARTICLE{Ergon2024,
       author = {{Ergon}, Mattias and {Lundqvist}, Peter and {Fransson}, Claes and {Kuncarayakti}, Hanindyo and {Das}, Kaustav K. and {De}, Kishalay and {Ferrari}, Lucia and {Fremling}, Christoffer and {Medler}, Kyle and {Maeda}, Keiichi and {Pastorello}, Andrea and {Sollerman}, Jesper and {Stritzinger}, Maximilian D.},
        title = "{Light curve and spectral modelling of the type IIb SN 2020acat. Evidence for a strong Ni bubble effect on the diffusion time}",
      journal = {\aap},
         year = 2024,
        month = mar,
       volume = {683},
          eid = {A241},
        pages = {A241},
          doi = {10.1051/0004-6361/202346718},
archivePrefix = {arXiv},
       eprint = {2308.07158},
 primaryClass = {astro-ph.HE},
       adsurl = {https://ui.adsabs.harvard.edu/abs/2024A&A...683A.241E}
}

@article{Fabricant_2019,
   title={Binospec: A Wide-field Imaging Spectrograph for the MMT},
   volume={131},
   ISSN={1538-3873},
   url={http://dx.doi.org/10.1088/1538-3873/ab1d78},
   DOI={10.1088/1538-3873/ab1d78},
   number={1001},
   journal={Publications of the Astronomical Society of the Pacific},
   publisher={IOP Publishing},
   author={Fabricant, Daniel and Fata, Robert and Epps, Harland and Gauron, Thomas and Mueller, Mark and Zajac, Joseph and Amato, Stephen and Barberis, Jack and Bergner, Henry and Brennan, Patricia and Brown, Warren and Chilingarian, Igor and Geary, John and Kradinov, Vladimir and McLeod, Brian and Smith, Matthew and Woods, Deborah},
   year={2019},
   month="June", pages={075004} }

@ARTICLE{Fang2019,
       author = {{Fang}, Qiliang and {Maeda}, Keiichi and {Kuncarayakti}, Hanindyo and {Sun}, Fengwu and {Gal-Yam}, Avishay},
        title = "{A hybrid envelope-stripping mechanism for massive stars from supernova nebular spectroscopy}",
      journal = {Nature Astronomy},
         year = 2019,
        month = mar,
       volume = {3},
        pages = {434-439},
          doi = {10.1038/s41550-019-0710-6},
archivePrefix = {arXiv},
       eprint = {1808.04834},
 primaryClass = {astro-ph.HE},
       adsurl = {https://ui.adsabs.harvard.edu/abs/2019NatAs...3..434F}
}

@ARTICLE{Fang2023,
       author = {{Fang}, Qiliang and {Maeda}, Keiichi},
        title = "{Inferring the Progenitor Mass-Kinetic Energy Relation of Stripped-envelope Core-collapse Supernovae from Nebular Spectroscopy}",
      journal = {\apj},
         year = 2023,
        month = "June",
       volume = {949},
       number = {2},
          eid = {93},
        pages = {93},
          doi = {10.3847/1538-4357/acc5e7},
archivePrefix = {arXiv},
       eprint = {2303.12432},
 primaryClass = {astro-ph.HE},
       adsurl = {https://ui.adsabs.harvard.edu/abs/2023ApJ...949...93F}
}

@article{Fang_2023,
   title={An aspherical distribution for the explosive burning ash of core-collapse supernovae},
   volume={8},
   ISSN={2397-3366},
   url={http://dx.doi.org/10.1038/s41550-023-02120-8},
   DOI={10.1038/s41550-023-02120-8},
   number={1},
   journal={Nature Astronomy},
   publisher={Springer Science and Business Media LLC},
   author={Fang, Qiliang and Maeda, Keiichi and Kuncarayakti, Hanindyo and Nagao, Takashi},
   year={2023},
   month=Oct, pages={111–118} }

@ARTICLE{Ferrari2024,
       author = {{Ferrari}, Luc{\'\i}a and {Folatelli}, Gast{\'o}n and {Kuncarayakti}, Hanindyo and {Stritzinger}, Maximilian and {Maeda}, Keiichi and {Bersten}, Melina and {Rom{\'a}n Aguilar}, Lili M. and {S{\'a}ez}, M. Manuela and {Dessart}, Luc and {Lundqvist}, Peter and {Mazzali}, Paolo and {Nagao}, Takashi and {Ashall}, Chris and {Bose}, Subhash and {Brennan}, Se{\'a}n J. and {Cai}, Yongzhi and {Handberg}, Rasmus and {Holmbo}, Simon and {Karamehmetoglu}, Emir and {Pastorello}, Andrea and {Reguitti}, Andrea and {Anderson}, Joseph and {Chen}, Ting-Wan and {Galbany}, Llu{\'\i}s and {Gromadzki}, Mariusz and {Guti{\'e}rrez}, Claudia P. and {Inserra}, Cosimo and {Kankare}, Erkki and {M{\"u}ller Bravo}, Tom{\'a}s E. and {Mattila}, Seppo and {Nicholl}, Matt and {Pignata}, Giuliano and {Sollerman}, Jesper and {Srivastav}, Shubham and {Young}, David R.},
        title = "{The metamorphosis of the Type Ib SN 2019yvr: late-time interaction}",
      journal = {\mnras},
         year = 2024,
        month = mar,
       volume = {529},
       number = {1},
        pages = {L33-L40},
          doi = {10.1093/mnrasl/slad195},
archivePrefix = {arXiv},
       eprint = {2401.15052},
 primaryClass = {astro-ph.HE},
       adsurl = {https://ui.adsabs.harvard.edu/abs/2024MNRAS.529L..33F}
}

@article{Finn_2016,
   title={COMPARISON OF DIVERSITY OF TYPE IIB SUPERNOVAE WITH ASYMMETRY IN CASSIOPEIA A USING LIGHT ECHOES},
   volume={830},
   ISSN={1538-4357},
   url={http://dx.doi.org/10.3847/0004-637X/830/2/73},
   DOI={10.3847/0004-637x/830/2/73},
   number={2},
   journal={The Astrophysical Journal},
   publisher={American Astronomical Society},
   author={Finn, Kieran and Bianco, Federica B. and Modjaz, Maryam and Liu, Yu-Qian and Rest, Armin},
   year={2016},
   month=Oct, pages={73} }

@ARTICLE{Folatelli2013,
       author = {{Folatelli}, Gast{\'o}n and {Morrell}, Nidia and {Phillips}, Mark M. and {Hsiao}, Eric and {Campillay}, Abdo and {Contreras}, Carlos and {Castell{\'o}n}, Sergio and {Hamuy}, Mario and {Krzeminski}, Wojtek and {Roth}, Miguel and {Stritzinger}, Maximilian and {Burns}, Christopher R. and {Freedman}, Wendy L. and {Madore}, Barry F. and {Murphy}, David and {Persson}, S.~E. and {Prieto}, Jos{\'e} L. and {Suntzeff}, Nicholas B. and {Krisciunas}, Kevin and {Anderson}, Joseph P. and {F{\"o}rster}, Francisco and {Maza}, Jos{\'e} and {Pignata}, Giuliano and {Rojas}, P. Andrea and {Boldt}, Luis and {Salgado}, Francisco and {Wyatt}, Pamela and {Olivares E.}, Felipe and {Gal-Yam}, Avishay and {Sako}, Masao},
        title = "{Spectroscopy of Type Ia Supernovae by the Carnegie Supernova Project}",
      journal = {\apj},
         year = 2013,
        month = aug,
       volume = {773},
       number = {1},
          eid = {53},
        pages = {53},
          doi = {10.1088/0004-637X/773/1/53},
archivePrefix = {arXiv},
       eprint = {1305.6997},
 primaryClass = {astro-ph.CO}
}

@article{Fox2011,
   title={A \textit{Spitzer} Survey for Dust in Type IIn Supernovae},
   volume={741},
   ISSN={1538-4357},
   url={http://dx.doi.org/10.1088/0004-637X/741/1/7},
   DOI={10.1088/0004-637x/741/1/7},
   number={1},
   journal={The Astrophysical Journal},
   publisher={American Astronomical Society},
   author={Fox, Ori D. and Chevalier, Roger A. and Skrutskie, Michael F. and Soderberg, Alicia M. and Filippenko, Alexei V. and Ganeshalingam, Mohan and Silverman, Jeffrey M. and Smith, Nathan and Steele, Thea N.},
   year={2011},
   month=oct, pages={7} }

@ARTICLE{Fransson1989,
       author = {{Fransson}, Claes and {Chevalier}, Roger A.},
        title = "{Late Emission from Supernovae: A Window on Stellar Nucleosynthesis}",
      journal = {\apj},
         year = 1989,
        month = aug,
       volume = {343},
        pages = {323},
          doi = {10.1086/167707},
       adsurl = {https://ui.adsabs.harvard.edu/abs/1989ApJ...343..323F}
}

@article{Gall_2014,
   title={Rapid formation of large dust grains in the luminous supernova 2010jl},
   volume={511},
   ISSN={1476-4687},
   url={http://dx.doi.org/10.1038/nature13558},
   DOI={10.1038/nature13558},
   number={7509},
   journal={Nature},
   publisher={Springer Science and Business Media LLC},
   author={Gall, Christa and Hjorth, Jens and Watson, Darach and Dwek, Eli and Maund, Justyn R. and Fox, Ori and Leloudas, Giorgos and Malesani, Daniele and Day-Jones, Avril C.},
   year={2014},
   month="July", pages={326–329} }

@ARTICLE{Gangopadhyay2023,
       author = {{Gangopadhyay}, Anjasha and {Maeda}, Keiichi and {Singh}, Avinash and {Nayana}, A.~J. and {Nakaoka}, Tatsuya and {Kawabata}, Koji S. and {Taguchi}, Kenta and {Singh}, Mridweeka and {Chandra}, Poonam and {Ryder}, Stuart D. and {Dastidar}, Raya and {Yamanaka}, Masayuki and {Kawabata}, Miho and {Alsaberi}, Rami Z.~E. and {Dukiya}, Naveen and {Teja}, Rishabh Singh and {Ailawadhi}, Bhavya and {Dutta}, Anirban and {Sahu}, D.~K. and {Moriya}, Takashi J. and {Misra}, Kuntal and {Tanaka}, Masaomi and {Chevalier}, Roger and {Tominaga}, Nozomu and {Uno}, Kohki and {Imazawa}, Ryo and {Hamada}, Taisei and {Hori}, Tomoya and {Isogai}, Keisuke},
        title = "{Bridging between Type IIb and Ib Supernovae: SN IIb 2022crv with a Very Thin Hydrogen Envelope}",
      journal = {\apj},
         year = 2023,
        month = nov,
       volume = {957},
       number = {2},
          eid = {100},
        pages = {100},
          doi = {10.3847/1538-4357/acfa94},
archivePrefix = {arXiv},
       eprint = {2309.07463},
 primaryClass = {astro-ph.HE},
       adsurl = {https://ui.adsabs.harvard.edu/abs/2023ApJ...957..100G}
}

@ARTICLE{Gilkis2022,
       author = {{Gilkis}, Avishai and {Arcavi}, Iair},
        title = "{How much hydrogen is in Type Ib and IIb supernova progenitors?}",
      journal = {\mnras},
         year = 2022,
        month = mar,
       volume = {511},
       number = {1},
        pages = {691-712},
          doi = {10.1093/mnras/stac088},
archivePrefix = {arXiv},
       eprint = {2111.04432},
 primaryClass = {astro-ph.SR},
       adsurl = {https://ui.adsabs.harvard.edu/abs/2022MNRAS.511..691G}
}

@misc{giudici2025,
      title={Hydrodynamic instabilities in long-term three-dimensional simulations of neutrino-driven supernovae of 13 red supergiant progenitors}, 
      author={Beatrice Giudici and Michael Gabler and Hans-Thomas Janka},
      year={2025},
      eprint={2511.11796},
      archivePrefix={arXiv},
      primaryClass={astro-ph.HE},
      url={https://arxiv.org/abs/2511.11796}, 
}

@ARTICLE{2016A&C....16...41G,
       author = {{Greenfield}, P. and {Miller}, T.},
        title = "{The Calibration Reference Data System}",
      journal = {Astronomy and Computing},
         year = 2016,
        month = "July",
       volume = {16},
        pages = {41-53},
          doi = {10.1016/j.ascom.2016.04.001},
       adsurl = {https://ui.adsabs.harvard.edu/abs/2016A&C....16...41G}
}

@article{Groh2013,
	author = {{Groh}, Jose H. and {Georgy}, Cyril and {Ekstr{\"o}m}, Sylvia},
        title = "{Progenitors of supernova Ibc: a single Wolf-Rayet star as the possible progenitor of the SN Ib iPTF13bvn}",
      journal = {\aap},
         year = 2013,
        month = oct,
       volume = {558},
          eid = {L1},
        pages = {L1},
          doi = {10.1051/0004-6361/201322369},
archivePrefix = {arXiv},
       eprint = {1307.8434},
 primaryClass = {astro-ph.SR},
       adsurl = {https://ui.adsabs.harvard.edu/abs/2013A&A...558L...1G}
}

@ARTICLE{Hachinger2012,
       author = {{Hachinger}, S. and {Mazzali}, P.~A. and {Taubenberger}, S. and {Hillebrandt}, W. and {Nomoto}, K. and {Sauer}, D.~N.},
        title = "{How much H and He is 'hidden' in SNe Ib/c? - I. Low-mass objects}",
      journal = {\mnras},
         year = 2012,
        month = may,
       volume = {422},
       number = {1},
        pages = {70-88},
          doi = {10.1111/j.1365-2966.2012.20464.x},
archivePrefix = {arXiv},
       eprint = {1201.1506},
 primaryClass = {astro-ph.SR},
       adsurl = {https://ui.adsabs.harvard.edu/abs/2012MNRAS.422...70H}
}

@ARTICLE{Hamuy2006,
       author = {{Hamuy}, Mario and {Folatelli}, Gast{\'o}n and {Morrell}, Nidia I. and {Phillips}, Mark M. and {Suntzeff}, Nicholas B. and {Persson}, S.~E. and {Roth}, Miguel and {Gonzalez}, Sergio and {Krzeminski}, Wojtek and {Contreras}, Carlos and {Freedman}, Wendy L. and {Murphy}, D.~C. and {Madore}, Barry F. and {Wyatt}, P. and {Maza}, Jos{\'e} and {Filippenko}, Alexei V. and {Li}, Weidong and {Pinto}, P.~A.},
        title = "{The Carnegie Supernova Project: The Low-Redshift Survey}",
      journal = {\pasp},
         year = 2006,
        month = jan,
       volume = {118},
       number = {839},
        pages = {2-20},
          doi = {10.1086/500228},
archivePrefix = {arXiv},
       eprint = {astro-ph/0512039},
 primaryClass = {astro-ph}
}

@ARTICLE{numpy2020,
       author = {{Harris}, C.~R. and others},
        title = "{Array programming with NumPy}",
      journal = {Nature},
         year = 2020,
       volume = {585},
        pages = {357-362},
          doi = {10.1038/s41586-020-2649-2}
}

@ARTICLE{Heger2003,
       author = {{Heger}, A. and {Fryer}, C.~L. and {Woosley}, S.~E. and {Langer}, N. and {Hartmann}, D.~H.},
        title = "{How Massive Single Stars End Their Life}",
      journal = {\apj},
         year = 2003,
        month = "July",
       volume = {591},
       number = {1},
        pages = {288-300},
          doi = {10.1086/375341},
archivePrefix = {arXiv},
       eprint = {astro-ph/0212469},
 primaryClass = {astro-ph},
       adsurl = {https://ui.adsabs.harvard.edu/abs/2003ApJ...591..288H}
}

@INPROCEEDINGS{Herter2020,
       author = {{Herter}, Terry and {Henderson}, Charles and {Bonati}, Marco and {Wilson}, John and {Allers}, Katelyn and {David}, Nicole and {Elias}, Jonathan and {James}, David and {Piraces}, Jose and {Points}, Sean and {Probst}, Ron and {Schlawin}, Everett and {Schurter}, Patricio and {Tighe}, Roberto and {Warner}, Michael},
        title = "{TSPEC4: near-IR spectroscopy for the SOAR telescope}",
    booktitle = {Ground-based and Airborne Instrumentation for Astronomy VIII},
         year = 2020,
       editor = {{Evans}, Christopher J. and {Bryant}, Julia J. and {Motohara}, Kentaro},
       series = {Society of Photo-Optical Instrumentation Engineers (SPIE) Conference Series},
       volume = {11447},
        month = dec,
          eid = {114476L},
        pages = {114476L},
          doi = {10.1117/12.2563035},
       adsurl = {https://ui.adsabs.harvard.edu/abs/2020SPIE11447E..6LH}
}

@article{Hoflich1993,
  author  = {H{\"o}flich, P. and Langer, N. and Duschinger, M.},
  title   = "{Supernova 1993J: One Year Later}",
  journal = aap,
  year    = {1993},
  volume  = {275},
  pages   = {L29}
}

@ARTICLE{Hook2004,
       author = {{Hook}, I.~M. and {J{\o}rgensen}, Inger and {Allington-Smith}, J.~R. and {Davies}, R.~L. and {Metcalfe}, N. and {Murowinski}, R.~G. and {Crampton}, D.},
        title = "{The Gemini-North Multi-Object Spectrograph: Performance in Imaging, Long-Slit, and Multi-Object Spectroscopic Modes}",
      journal = {\pasp},
         year = 2004,
        month = may,
       volume = {116},
       number = {819},
        pages = {425-440},
          doi = {10.1086/383624},
       adsurl = {https://ui.adsabs.harvard.edu/abs/2004PASP..116..425H}
}

@article{Hosseinzadeh_2023,
   title={JWST Imaging of the Cartwheel Galaxy Reveals Dust Associated with SN 2021afdx},
   volume={942},
   ISSN={2041-8213},
   url={http://dx.doi.org/10.3847/2041-8213/aca64e},
   DOI={10.3847/2041-8213/aca64e},
   number={1},
   journal={The Astrophysical Journal Letters},
   publisher={American Astronomical Society},
   author={Hosseinzadeh, Griffin and Sand, David J. and Jencson, Jacob E. and Andrews, Jennifer E. and Shivaei, Irene and Bostroem, K. Azalee and Valenti, Stefano and Szalai, Tamás and Burke, Jamison and Howell, D. Andrew and McCully, Curtis and Newsome, Megan and Gonzalez, Estefania Padilla and Pellegrino, Craig and Terreran, Giacomo},
   year={2023},
   month=Jan, pages={L18} }

@ARTICLE{Hosseinzadeh2025,
       author = {{Hosseinzadeh}, G. and {Andrews}, J. and {Sand}, D.~J. and {Andreoni}, I. and {Jha}, S.~W. and {Modjaz}, M. and {Andrews}, M. and {Arcavi}, I. and {Baer-Way}, R. and {Carney}, J. and {Cartier}, R. and {Dong}, Y. and {Franz}, N. and {Freeburn}, J. and {Gomez}, S. and {Hiramatsu}, D. and {Howell}, D.~A. and {Hsu}, B. and {Kilpatrick}, C.~D. and {Kumar}, S. and {Kwok}, L.~A. and {Li}, W.~X. and {Pearson}, J. and {Ransome}, C.~L. and {Ravi}, A.~P. and {Rho}, J. and {Shrestha}, M. and {Subrayan}, B. and {Tucker}, B.~E. and {Valenti}, S. and {Vasylyev}, S.~S. and {Wang}, X.},
        title = "{PASSTA: The Public AEON Spectroscopic Survey for Transient Astronomy}",
      journal = {Transient Name Server AstroNote},
         year = 2025,
        month = sep,
       volume = {280},
        pages = {1},
       adsurl = {https://ui.adsabs.harvard.edu/abs/2025TNSAN.280....1H}
}

@ARTICLE{Hunter2007,
       author = {{Hunter}, J.~D.},
        title = "{Matplotlib: A 2D Graphics Environment}",
      journal = {Computing in Science and Engineering},
         year = 2007,
       volume = {9},
       number = {3},
        pages = {90-95},
          doi = {10.1109/MCSE.2007.55}
}

@misc{hwangbo2026,
      title={Near-Infrared and Optical Observations of SN 2024rbc: The First Early Detection of CO and Dust in a Type Ib Supernova}, 
      author={Ryan Hwangbo and Jeonghee Rho and Aravind P. Ravi and Seong Hyun Park and Harim Jin and Sung-Chul Yoon and T. R. Geballe and Ryan Foley and Kirsty Taggart and Kyle W. Davis and Kishore C. Patra and S. Tinyanont and Jesper Sollerman and Steve Schulze and Natalie LeBaron and Chang Liu and Charles D. Kilpatrick},
      year={2026},
      eprint={2603.23877},
      archivePrefix={arXiv},
      primaryClass={astro-ph.HE},
      url={https://arxiv.org/abs/2603.23877}, 
}

@article{Janka_2012,
   title={Explosion Mechanisms of Core-Collapse Supernovae},
   volume={62},
   ISSN={1545-4134},
   url={http://dx.doi.org/10.1146/annurev-nucl-102711-094901},
   DOI={10.1146/annurev-nucl-102711-094901},
   number={1},
   journal={Annual Review of Nuclear and Particle Science},
   publisher={Annual Reviews},
   author={Janka, Hans-Thomas},
   year={2012},
   month=Nov, pages={407–451} }

@article{Janka_2017,
   title={Neutron Star Kicks by the Gravitational Tug-boat Mechanism in Asymmetric Supernova Explosions: Progenitor and Explosion Dependence},
   volume={837},
   ISSN={1538-4357},
   url={http://dx.doi.org/10.3847/1538-4357/aa618e},
   DOI={10.3847/1538-4357/aa618e},
   number={1},
   journal={The Astrophysical Journal},
   publisher={American Astronomical Society},
   author={Janka, Hans-Thomas},
   year={2017},
   month=Mar, pages={84} }

@inbook{Jerkstrand_2017,
   title={Spectra of Supernovae in the Nebular Phase},
   ISBN={9783319218465},
   url={http://dx.doi.org/10.1007/978-3-319-21846-5_29},
   DOI={10.1007/978-3-319-21846-5_29},
   booktitle={Handbook of Supernovae},
   publisher={Springer International Publishing},
   author={Jerkstrand, Anders},
   year={2017},
   pages={795–842} }

@misc{jerkstrand2025,
      title={Core-collapse supernovae}, 
      author={Jerkstrand, Anders and Milisavljevic, Dan and Müller, Bernhard},
      year={2025},
      eprint={2503.01321},
      archivePrefix={arXiv},
      primaryClass={astro-ph.HE},
      url={https://arxiv.org/abs/2503.01321}, 
}

@misc{khakpash2024,
      title={Multi-filter UV to NIR Data-driven Light Curve Templates for Stripped Envelope Supernovae}, 
      author={Somayeh Khakpash and Federica B. Bianco and Maryam Modjaz and Willow F. Fortino and Alexander Gagliano and Conor Larison and Tyler A. Pritchard},
      year={2024},
      eprint={2405.01672},
      archivePrefix={arXiv},
      primaryClass={astro-ph.HE},
      url={https://arxiv.org/abs/2405.01672}, 
}

@article{Kifonidis_2003,
   title={Non-spherical core collapse supernovae: I. Neutrino-driven convection, Rayleigh-Taylor instabilities, and the formation and propagation of metal clumps},
   volume={408},
   ISSN={1432-0746},
   url={http://dx.doi.org/10.1051/0004-6361:20030863},
   DOI={10.1051/0004-6361:20030863},
   number={2},
   journal={Astronomy \& Astrophysics},
   publisher={EDP Sciences},
   author={Kifonidis, K. and Plewa, T. and Janka, H.-Th. and Müller, E.},
   year={2003},
   month="Sept", pages={621–649} }

@article{Kilpatrick_2016,
   title={On the progenitor of the Type IIb supernova 2016gkg},
   volume={465},
   ISSN={1365-2966},
   url={http://dx.doi.org/10.1093/mnras/stw3082},
   DOI={10.1093/mnras/stw3082},
   number={4},
   journal={Monthly Notices of the Royal Astronomical Society},
   publisher={Oxford University Press (OUP)},
   author={Kilpatrick, Charles D. and Foley, Ryan J. and Abramson, Louis E. and Pan, Yen-Chen and Lu, Cicero-Xinyu and Williams, Peter and Treu, Tommaso and Siebert, Matthew R. and Fassnacht, Christopher D. and Max, Claire E.},
   year={2016},
   month=Nov, pages={4650–4657} }

@ARTICLE{Kumar2013,
       author = {{Kumar}, Brajesh and {Pandey}, S.~B. and {Sahu}, D.~K. and {Vinko}, J. and {Moskvitin}, A.~S. and {Anupama}, G.~C. and {Bhatt}, V.~K. and {Ordasi}, A. and {Nagy}, A. and {Sokolov}, V.~V. and {Sokolova}, T.~N. and {Komarova}, V.~N. and {Kumar}, Brijesh and {Bose}, Subhash and {Roy}, Rupak and {Sagar}, Ram},
        title = "{Light curve and spectral evolution of the Type IIb supernova 2011fu}",
      journal = {\mnras},
         year = 2013,
        month = may,
       volume = {431},
       number = {1},
        pages = {308-321},
          doi = {10.1093/mnras/stt162},
archivePrefix = {arXiv},
       eprint = {1301.6538},
 primaryClass = {astro-ph.HE},
       adsurl = {https://ui.adsabs.harvard.edu/abs/2013MNRAS.431..308K}
}

@article{Lee_2009,
   title={\textit{AKARI} INFRARED OBSERVATIONS OF THE SUPERNOVA REMNANT G292.0+1.8: UNVEILING CIRCUMSTELLAR MEDIUM AND SUPERNOVA EJECTA},
   volume={706},
   ISSN={1538-4357},
   url={http://dx.doi.org/10.1088/0004-637X/706/1/441},
   DOI={10.1088/0004-637x/706/1/441},
   number={1},
   journal={The Astrophysical Journal},
   publisher={American Astronomical Society},
   author={Lee, Ho-Gyu and Koo, Bon-Chul and Moon, Dae-Sik and Sakon, Itsuki and Onaka, Takashi and Jeong, Woong-Seob and Kaneda, Hidehiro and Nozawa, Takaya and Kozasa, Takashi},
   year={2009},
   month=Oct, pages={441–453} }

@ARTICLE{Li1995,
       author = {{Li}, Hongwei and {McCray}, Richard},
        title = "{The He i Emission Lines of SN 1987A}",
      journal = {\apj},
         year = 1995,
        month = mar,
       volume = {441},
        pages = {821},
          doi = {10.1086/175405},
       adsurl = {https://ui.adsabs.harvard.edu/abs/1995ApJ...441..821L}
}

@article{long2022,
  title={The Formation of the Stripped-envelope Type IIb Supernova Progenitors: Rotation, Metallicity, and Overshooting},
  volume={262},
  DOI={10.3847/1538-4365/ac7ffe},
  number={1},
  journal={The Astrophysical Journal Supplement Series},
  publisher={The American Astronomical Society},
  author={Long, Gang and Song, Hanfeng and Meynet, Georges and Maeder, Andre and Zhang, Ruiyu and Qin, Ying and Ekström, Sylvia and Georgy, Cyril and Zhao, Liuyan},
  year={2022},
  month={Sep},
  pages={26}
}

@ARTICLE{Lucy_1999,
       author = {{Lucy}, L.~B.},
        title = "{Improved Monte Carlo techniques for the spectral synthesis of supernovae}",
      journal = {\aap},
         year = 1999,
        month = may,
       volume = {345},
        pages = {211-220},
       adsurl = {https://ui.adsabs.harvard.edu/abs/1999A&A...345..211L}
}

@ARTICLE{Maeda2008,
       author = {{Maeda}, Keiichi and {Kawabata}, Koji and {Mazzali}, Paolo A. and {Tanaka}, Masaomi and {Valenti}, Stefano and {Nomoto}, Ken'ichi and {Hattori}, Takashi and {Deng}, Jinsong and {Pian}, Elena and {Taubenberger}, Stefan and {Iye}, Masanori and {Matheson}, Thomas and {Filippenko}, Alexei V. and {Aoki}, Kentaro and {Kosugi}, George and {Ohyama}, Youichi and {Sasaki}, Toshiyuki and {Takata}, Tadafumi},
        title = "{Asphericity in Supernova Explosions from Late-Time Spectroscopy}",
      journal = {Science},
         year = 2008,
        month = feb,
       volume = {319},
       number = {5867},
        pages = {1220},
          doi = {10.1126/science.1149437},
archivePrefix = {arXiv},
       eprint = {0801.1100},
 primaryClass = {astro-ph},
       adsurl = {https://ui.adsabs.harvard.edu/abs/2008Sci...319.1220M}
}

@article{Maund_2007,
   title={Spectropolarimetry of the Type IIb Supernova 2001ig},
   volume={671},
   ISSN={1538-4357},
   url={http://dx.doi.org/10.1086/523261},
   DOI={10.1086/523261},
   number={2},
   journal={The Astrophysical Journal},
   publisher={American Astronomical Society},
   author={Maund, Justyn R. and Wheeler, J. Craig and Patat, Ferdinando and Wang, Lifan and Baade, Dietrich and Hoflich, Peter A.},
   year={2007},
   month=Dec, pages={1944–1958} }

@article{Maund_2011,
   title={THE YELLOW SUPERGIANT PROGENITOR OF THE TYPE II SUPERNOVA 2011dh IN M51},
   volume={739},
   ISSN={2041-8213},
   url={http://dx.doi.org/10.1088/2041-8205/739/2/L37},
   DOI={10.1088/2041-8205/739/2/l37},
   number={2},
   journal={The Astrophysical Journal},
   publisher={American Astronomical Society},
   author={Maund, J. R. and Fraser, M. and Ergon, M. and Pastorello, A. and Smartt, S. J. and Sollerman, J. and Benetti, S. and Botticella, M.-T. and Bufano, F. and Danziger, I. J. and Kotak, R. and Magill, L. and Stephens, A. W. and Valenti, S.},
   year={2011},
   month="Sept", pages={L37} }

@ARTICLE{Maurer2010,
       author = {{Maurer}, I. and {Mazzali}, P.~A. and {Taubenberger}, S. and {Hachinger}, S.},
        title = "{Hydrogen and helium in the late phase of supernovae of Type IIb}",
      journal = {\mnras},
         year = 2010,
        month = dec,
       volume = {409},
       number = {4},
        pages = {1441-1454},
          doi = {10.1111/j.1365-2966.2010.17186.x},
archivePrefix = {arXiv},
       eprint = {1007.1881},
 primaryClass = {astro-ph.HE},
       adsurl = {https://ui.adsabs.harvard.edu/abs/2010MNRAS.409.1441M}
}

@InProceedings{mckinney-proc-scipy-2010,
  author    = {Wes McKinney},
  title     = {{Data Structures for Statistical Computing in Python}},
  booktitle = {Proceedings of the 9th Python in Science Conference},
  pages     = {56--61},
  year      = {2010},
  editor    = {St\'efan van der Walt and Jarrod Millman},
  doi       = {10.25080/Majora-92bf1922-00a}
}

@ARTICLE{Medler_2022,
       author = {{Medler}, K. and {Mazzali}, P.~A. and {Teffs}, J. and {Ashall}, C. and {Anderson}, J.~P. and {Arcavi}, I. and {Benetti}, S. and {Bostroem}, K.~A. and {Burke}, J. and {Cai}, Y.-Z. and {Charalampopoulos}, P. and {Elias-Rosa}, N. and {Ergon}, M. and {Galbany}, L. and {Gromadzki}, M. and {Hiramatsu}, D. and {Howell}, D.~A. and {Inserra}, C. and {Lundqvist}, P. and {McCully}, C. and {M{\"u}ller-Bravo}, T. and {Newsome}, M. and {Nicholl}, M. and {Padilla Gonzalez}, E. and {Paraskeva}, E. and {Pastorello}, A. and {Pellegrino}, C. and {Pessi}, P.~J. and {Reguitti}, A. and {Reynolds}, T.~M. and {Roy}, R. and {Terreran}, G. and {Tomasella}, L. and {Young}, D.~R.},
        title = "{SN 2020acat: an energetic fast rising Type IIb supernova}",
      journal = {\mnras},
         year = 2022,
        month = "July",
       volume = {513},
       number = {4},
        pages = {5540-5558},
          doi = {10.1093/mnras/stac1192},
archivePrefix = {arXiv},
       eprint = {2201.06991},
 primaryClass = {astro-ph.HE},
       adsurl = {https://ui.adsabs.harvard.edu/abs/2022MNRAS.513.5540M}
}

@ARTICLE{Medler_2023,
       author = {{Medler}, K. and {Mazzali}, P.~A. and {Ashall}, C. and {Teffs}, J. and {Shahbandeh}, M. and {Shappee}, B.},
        title = "{Flat-topped NIR profiles originating from an unmixed helium shell in the Type IIb SN 2020acat}",
      journal = {\mnras},
         year = 2023,
        month = jan,
       volume = {518},
       number = {1},
        pages = {L40-L44},
          doi = {10.1093/mnrasl/slac127},
       adsurl = {https://ui.adsabs.harvard.edu/abs/2023MNRAS.518L..40M}
}

@ARTICLE{2025ApJ...993..191M,
       author = {{Medler}, K. and {Ashall}, C. and {Hoeflich}, P. and {Baron}, E. and {DerKacy}, J.~M. and {Shahbandeh}, M. and {Mera}, T. and {Pfeffer}, C.~M. and {Hoogendam}, W.~B. and {Jones}, D.~O. and {Shiber}, S. and {Fereidouni}, E. and {Fox}, O.~D. and {Jencson}, J. and {Galbany}, L. and {Hinkle}, J.~T. and {Tucker}, M.~A. and {Shappee}, B.~J. and {Huber}, M.~E. and {Auchettl}, K. and {Angus}, C.~R. and {Desai}, D.~D. and {Do}, A. and {Payne}, A.~V. and {Shi}, J. and {Kong}, M.~Y. and {Romagnoli}, S. and {Syncatto}, A. and {Burns}, C.~R. and {Clayton}, G. and {Dulude}, M. and {Engesser}, M. and {Filippenko}, A.~V. and {Gomez}, S. and {Hsiao}, E.~Y. and {de Jaeger}, T. and {Johansson}, J. and {Krisciunas}, K. and {Kumar}, S. and {Lu}, J. and {Matsuura}, M. and {Mazzali}, P.~A. and {Milisavljevic}, D. and {Morrell}, N. and {O'Steen}, R. and {Park}, S. and {Phillips}, M.~M. and {Ravi}, A.~P. and {Rest}, A. and {Rho}, J. and {Suntzeff}, N.~B. and {Sarangi}, A. and {Smith}, N. and {Stritzinger}, M.~D. and {Strolger}, L. and {Szalai}, T. and {Temim}, T. and {Tinyanont}, S. and {Van Dyk}, S.~D. and {Wang}, L. and {Wang}, Q. and {Wesson}, R. and {Yang}, Y. and {Zs{\'\i}ros}, S.},
        title = "{JWST Observations of SN 2023ixf. II. The Panchromatic Evolution between 250 and 720 Days after the Explosion}",
      journal = {\apj},
         year = 2025,
        month = nov,
       volume = {993},
       number = {2},
          eid = {191},
        pages = {191},
          doi = {10.3847/1538-4357/ae0736},
archivePrefix = {arXiv},
       eprint = {2507.19727},
 primaryClass = {astro-ph.SR},
       adsurl = {https://ui.adsabs.harvard.edu/abs/2025ApJ...993..191M}
}

@ARTICLE{Medler2025,
       author = {{Medler}, K. and {Ashall}, C. and {Shahbandeh}, M. and {DerKacy}, J.~M. and {Hoogendam}, W.~B. and {Jones}, D.~O. and {Shappee}, B.~J. and {Hinkle}, J.~T. and {Pfeffer}, C.~M. and {Baron}, E. and {Hoeflich}, P. and {Hsiao}, E.},
        title = "{The Hawaii Infrared Supernova Study (HISS): Spectroscopic Data Release 1}",
      journal = {\apjs},
         year = 2025,
        month = dec,
       volume = {281},
       number = {2},
          eid = {28},
        pages = {28},
          doi = {10.3847/1538-4365/ae092c},
archivePrefix = {arXiv},
       eprint = {2505.18507},
 primaryClass = {astro-ph.HE},
       adsurl = {https://ui.adsabs.harvard.edu/abs/2025ApJS..281...28M}
}

@misc{modjaz2019,
      title={New Regimes in the Observation of Core-Collapse Supernovae}, 
      author={Maryam Modjaz and Claudia P. Gutierrez and Iair Arcavi},
      year={2019},
      eprint={1908.02476},
      archivePrefix={arXiv},
      primaryClass={astro-ph.HE},
      url={https://arxiv.org/abs/1908.02476}, 
}

@article{Mazzali_2005,
   title={An Asymmetric Energetic Type Ic Supernova Viewed Off-Axis, and a Link to Gamma Ray Bursts},
   volume={308},
   ISSN={1095-9203},
   url={http://dx.doi.org/10.1126/science.1111384},
   DOI={10.1126/science.1111384},
   number={5726},
   journal={Science},
   publisher={American Association for the Advancement of Science (AAAS)},
   author={Mazzali, Paolo A. and Kawabata, Koji S. and Maeda, Keiichi and Nomoto, Ken’ichi and Filippenko, Alexei V. and Ramirez-Ruiz, Enrico and Benetti, Stefano and Pian, Elena and Deng, Jinsong and Tominaga, Nozomu and Ohyama, Youichi and Iye, Masanori and Foley, Ryan J. and Matheson, Thomas and Wang, Lifan and Gal-Yam, Avishay},
   year={2005},
   month=May, pages={1284–1287} }

@article{Neopane2024Long,
	author = {Neopane, Sudarshan and Sandoval, Michael and Hix, William and Messer, O. E. Bronson and Lentz, Eric and Harris, James},
	journal = {Bulletin of the AAS},
	number = {2},
	year = {2024},
	month = {feb 7},
	note = {https://baas.aas.org/pub/2024n2i260p09},
	publisher = {},
	title = {Long-time evolution of {Core}-collapse {Supernovae}},
	volume = {56},
}

@article{Nomoto2013,
title = "Nucleosynthesis in stars and the chemical enrichment of galaxies",
author = "Ken'Ichi Nomoto and Chiaki Kobayashi and Nozomu Tominaga",
year = "2013",
month = aug,
day = "1",
doi = "10.1146/annurev-astro-082812-140956",
language = "English",
volume = "51",
pages = "457--509",
journal = "Annual Review of Astronomy and Astrophysics",
issn = "0066-4146",
publisher = "Annual Reviews Inc.",
}

@article{Nozawa_2007,
   title={Evolution of Dust in Primordial Supernova Remnants: Can Dust Grains Formed in the Ejecta Survive and Be Injected into the Early Interstellar Medium?},
   volume={666},
   ISSN={1538-4357},
   url={http://dx.doi.org/10.1086/520621},
   DOI={10.1086/520621},
   number={2},
   journal={The Astrophysical Journal},
   publisher={American Astronomical Society},
   author={Nozawa, Takaya and Kozasa, Takashi and Habe, Asao and Dwek, Eli and Umeda, Hideyuki and Tominaga, Nozomu and Maeda, Keiichi and Nomoto, Ken’ichi},
   year={2007},
   month="Sept", pages={955–966} }

@article{Nozawa_2010,
   title={FORMATION AND EVOLUTION OF DUST IN TYPE IIb SUPERNOVAE WITH APPLICATION TO THE CASSIOPEIA A SUPERNOVA REMNANT},
   volume={713},
   ISSN={1538-4357},
   url={http://dx.doi.org/10.1088/0004-637X/713/1/356},
   DOI={10.1088/0004-637x/713/1/356},
   number={1},
   journal={The Astrophysical Journal},
   publisher={American Astronomical Society},
   author={Nozawa, Takaya and Kozasa, Takashi and Tominaga, Nozomu and Maeda, Keiichi and Umeda, Hideyuki and Nomoto, Ken’ichi and Krause, Oliver},
   year={2010},
   month=Mar, pages={356–373} }

@ARTICLE{Park2025,
       author = {{Park}, Seong Hyun and {Rho}, Jeonghee and {Yoon}, Sung-Chul and {Pearson}, Jeniveve and {Shrestha}, Manisha and {Tinyanont}, Samaporn and {Geballe}, T.~R. and {Foley}, Ryan J. and {Ravi}, Aravind P. and {Andrews}, Jennifer and {Sand}, David J. and {Azalee Bostroem}, K. and {Ashall}, Chris and {Hoeflich}, Peter and {Valenti}, Stefano and {Dong}, Yize and {Retamal}, Nicolas Meza and {Hoang}, Emily and {Mehta}, Darshana and {Andrew Howell}, D. and {Farah}, Joseph R. and {Terreran}, Giacomo and {Padilla Gonzalez}, Estefania and {Andrews}, Moira and {Newsome}, Megan and {Shahbandeh}, Melissa and {Smith}, Nathan and {Hwan Kang}, Jae and {Suntzeff}, Nick and {Baron}, Eddie and {Medler}, Kyle and {Mera Evans}, Tyco and {DerKacy}, James M. and {Larison}, Conor and {Galbany}, Llu{\'\i}s and {Jacobson-Gal{\'a}n}, Wynn},
        title = "{Near-infrared spectroscopy and detection of carbon monoxide in the Type II supernova SN 2023ixf}",
      journal = {\aap},
         year = 2025,
        month = nov,
       volume = {703},
          eid = {A227},
        pages = {A227},
          doi = {10.1051/0004-6361/202555244},
archivePrefix = {arXiv},
       eprint = {2507.11877},
 primaryClass = {astro-ph.HE},
       adsurl = {https://ui.adsabs.harvard.edu/abs/2025A&A...703A.227P}
}

@article{Pastorello_2008,
   title={The Type IIb SN 2008ax: spectral and light curve evolution},
   volume={389},
   ISSN={1365-2966},
   url={http://dx.doi.org/10.1111/j.1365-2966.2008.13618.x},
   DOI={10.1111/j.1365-2966.2008.13618.x},
   number={2},
   journal={Monthly Notices of the Royal Astronomical Society},
   publisher={Oxford University Press (OUP)},
   author={Pastorello, A. and Kasliwal, M. M. and Crockett, R. M. and Valenti, S. and Arbour, R. and Itagaki, K. and Kaspi, S. and Gal-Yam, A. and Smartt, S. J. and Griffith, R. and Maguire, K. and Ofek, E. O. and Seymour, N. and Stern, D. and Wiethoff, W.},
   year={2008},
   month="Sept", pages={955–966} }

@misc{spacephot,
  author       = {Justin Pierel},
  title        = {{space\_phot}: Aperture and PSF Photometry for HST and JWST},
  year         = {2026},
  howpublished = {\url{https://github.com/jpierel14/space_phot}},
  note         = {GitHub repository, accessed 2026 July 10}
}

@ARTICLE{Prentice2016,
       author = {{Prentice}, S.~J. and {Mazzali}, P.~A. and {Pian}, E. and {Gal-Yam}, A. and {Kulkarni}, S.~R. and {Rubin}, A. and {Corsi}, A. and {Fremling}, C. and {Sollerman}, J. and {Yaron}, O. and {Arcavi}, I. and {Zheng}, W. and {Kasliwal}, M.~M. and {Filippenko}, A.~V. and {Cenko}, S.~B. and {Cao}, Y. and {Nugent}, P.~E.},
        title = "{The bolometric light curves and physical parameters of stripped-envelope supernovae}",
      journal = {\mnras},
         year = 2016,
        month = may,
       volume = {458},
       number = {3},
        pages = {2973-3002},
          doi = {10.1093/mnras/stw299},
archivePrefix = {arXiv},
       eprint = {1602.01736},
 primaryClass = {astro-ph.HE},
       adsurl = {https://ui.adsabs.harvard.edu/abs/2016MNRAS.458.2973P}
}

@article{Prentice_2017,
   title={A physically motivated classification of stripped-envelope supernovae},
   volume={469},
   ISSN={1365-2966},
   url={http://dx.doi.org/10.1093/mnras/stx980},
   DOI={10.1093/mnras/stx980},
   number={3},
   journal={Monthly Notices of the Royal Astronomical Society},
   publisher={Oxford University Press (OUP)},
   author={Prentice, S. J. and Mazzali, P. A.},
   year={2017},
   month=Apr, pages={2672–2694} }

@article{Priestley_2020,
   title={Constraining early-time dust formation in core-collapse supernovae},
   volume={497},
   ISSN={1365-2966},
   url={http://dx.doi.org/10.1093/mnras/staa2121},
   DOI={10.1093/mnras/staa2121},
   number={2},
   journal={Monthly Notices of the Royal Astronomical Society},
   publisher={Oxford University Press (OUP)},
   author={Priestley, F D and Bevan, A and Barlow, M J and De Looze, I},
   year={2020},
   month="July", pages={2227–2238} }

@article{Prochaska2020,
  doi       = {10.21105/joss.02308},
  url       = {https://doi.org/10.21105/joss.02308},
  year      = {2020},
  publisher = {The Open Journal},
  volume    = {5},
  number    = {56},
  pages     = {2308},
  author    = {Prochaska, J. Xavier and Hennawi, Joseph F. and Westfall, Kyle B. and Cooke, Ryan J. and Wang, Feige and Hsyu, Tiffany and Davies, Frederick B. and Farina, Emanuele Paolo and Pelliccia, Debora},
  title     = {PypeIt: The Python Spectroscopic Data Reduction Pipeline},
  journal   = {Journal of Open Source Software}
}

@article{Reynolds_2025,
   title={The bright long-lived Type II SN 2021irp powered by aspherical circumstellar material interaction: II. Estimating the CSM mass and geometry with polarimetry and light curve modeling},
   volume={702},
   ISSN={1432-0746},
   url={http://dx.doi.org/10.1051/0004-6361/202553793},
   DOI={10.1051/0004-6361/202553793},
   journal={Astronomy \& Astrophysics},
   publisher={EDP Sciences},
   author={Reynolds, T. M. and Nagao, T. and Maeda, K. and Elias-Rosa, N. and Fraser, M. and Gutiérrez, C. and Kangas, T. and Kuncarayakti, H. and Mattila, S. and Pessi, P. J.},
   year={2025},
   month=Oct, pages={A213} }

@ARTICLE{Rho2018,
       author = {{Rho}, J. and {Geballe}, T.~R. and {Banerjee}, D.~P.~K. and {Dessart}, L. and {Evans}, A. and {Joshi}, V.},
        title = "{Near-infrared Spectroscopy of Supernova 2017eaw in 2017: Carbon Monoxide and Dust Formation in a Type II-P Supernova}",
      journal = {\apjl},
         year = 2018,
        month = sep,
       volume = {864},
       number = {1},
          eid = {L20},
        pages = {L20},
          doi = {10.3847/2041-8213/aad77f},
archivePrefix = {arXiv},
       eprint = {1808.00683},
 primaryClass = {astro-ph.GA},
       adsurl = {https://ui.adsabs.harvard.edu/abs/2018ApJ...864L..20R}
}

@ARTICLE{Sana2012,
       author = {{Sana}, H. and {de Mink}, S.~E. and {de Koter}, A. and {Langer}, N. and {Evans}, C.~J. and {Gieles}, M. and {Gosset}, E. and {Izzard}, R.~G. and {Le Bouquin}, J.-B. and {Schneider}, F.~R.~N.},
        title = "{Binary Interaction Dominates the Evolution of Massive Stars}",
      journal = {Science},
         year = 2012,
        month = "July",
       volume = {337},
       number = {6093},
        pages = {444},
          doi = {10.1126/science.1223344},
archivePrefix = {arXiv},
       eprint = {1207.6397},
 primaryClass = {astro-ph.SR},
       adsurl = {https://ui.adsabs.harvard.edu/abs/2012Sci...337..444S}
}

@misc{TNS2025aico,
  title        = {{TNS Object 2025aico}},
  author       = {{Transient Name Server}},
  year         = {2025},
  howpublished = {\url{https://www.wis-tns.org/object/2025aico}},
  note         = {Accessed: \today}
}

@ARTICLE{Shahbandeh2023,
       author = {{Shahbandeh}, Melissa and {Sarangi}, Arkaprabha and {Temim}, Tea and {Szalai}, Tam{\'a}s and {Fox}, Ori D. and {Tinyanont}, Samaporn and {Dwek}, Eli and {Dessart}, Luc and {Filippenko}, Alexei V. and {Brink}, Thomas G. and {Foley}, Ryan J. and {Jencson}, Jacob and {Pierel}, Justin and {Zs{\'\i}ros}, Szanna and {Rest}, Armin and {Zheng}, WeiKang and {Andrews}, Jennifer and {Clayton}, Geoffrey C. and {De}, Kishalay and {Engesser}, Michael and {Gezari}, Suvi and {Gomez}, Sebastian and {Gonzaga}, Shireen and {Johansson}, Joel and {Kasliwal}, Mansi and {Lau}, Ryan and {De Looze}, Ilse and {Marston}, Anthony and {Milisavljevic}, Dan and {O'Steen}, Richard and {Siebert}, Matthew and {Skrutskie}, Michael and {Smith}, Nathan and {Strolger}, Lou and {Van Dyk}, Schuyler D. and {Wang}, Qinan and {Williams}, Brian and {Williams}, Robert and {Xiao}, Lin and {Yang}, Yi},
        title = "{JWST observations of dust reservoirs in type IIP supernovae 2004et and 2017eaw}",
      journal = {\mnras},
         year = 2023,
        month = aug,
       volume = {523},
       number = {4},
        pages = {6048-6060},
          doi = {10.1093/mnras/stad1681},
archivePrefix = {arXiv},
       eprint = {2301.10778},
 primaryClass = {astro-ph.HE},
       adsurl = {https://ui.adsabs.harvard.edu/abs/2023MNRAS.523.6048S}
}

@misc{sharma2024,
      title={Dramatic rebrightening of the type-changing stripped-envelope supernova SN 2023aew}, 
      author={Yashvi Sharma and Jesper Sollerman and Shrinivas R. Kulkarni and Takashi J. Moriya and Steve Schulze and Stan Barmentloo and Michael Fausnaugh and Avishay Gal-Yam and Anders Jerkstrand and Tomás Ahumada and Eric C. Bellm and Kaustav K. Das and Andrew Drake and Christoffer Fremling and Saarah Hall and K. R. Hinds and Theophile Jegou du Laz and Viraj Karambelkar and Mansi M. Kasliwal and Frank J. Masci and Adam A. Miller and Guy Nir and Daniel A. Perley and Josiah N. Purdum and Yu-Jing Qin and Nabeel Rehemtulla and R. Michael Rich and Reed L. Riddle and Antonio C. Rodriguez and Sam Rose and Jean Somalwar and Jacob L. Wise and Avery Wold and Lin Yan and Yuhan Yao},
      year={2024},
      eprint={2402.02780},
      archivePrefix={arXiv},
      primaryClass={astro-ph.HE},
      url={https://arxiv.org/abs/2402.02780}, 
}

@ARTICLE{Shivvers2019,
       author = {{Shivvers}, Isaac and {Filippenko}, Alexei V. and {Silverman}, Jeffrey M. and {Zheng}, WeiKang and {Foley}, Ryan J. and {Chornock}, Ryan and {Barth}, Aaron J. and {Cenko}, S. Bradley and {Clubb}, Kelsey I. and {Fox}, Ori D. and {Ganeshalingam}, Mohan and {Graham}, Melissa L. and {Kelly}, Patrick L. and {Kleiser}, Io K.~W. and {Leonard}, Douglas C. and {Li}, Weidong and {Matheson}, Thomas and {Mauerhan}, Jon C. and {Modjaz}, Maryam and {Serduke}, Franklin J.~D. and {Shields}, Joseph C. and {Steele}, Thea N. and {Swift}, Brandon J. and {Wong}, Diane S. and {Yuk}, Heechan},
        title = "{The Berkeley sample of stripped-envelope supernovae}",
      journal = {\mnras},
         year = 2019,
        month = jan,
       volume = {482},
       number = {2},
        pages = {1545-1556},
          doi = {10.1093/mnras/sty2719},
archivePrefix = {arXiv},
       eprint = {1810.03650},
 primaryClass = {astro-ph.SR},
       adsurl = {https://ui.adsabs.harvard.edu/abs/2019MNRAS.482.1545S}
}

@ARTICLE{Smartt2009,
       author = {{Smartt}, Stephen J.},
        title = "{Progenitors of Core-Collapse Supernovae}",
      journal = {\araa},
         year = 2009,
        month = sep,
       volume = {47},
       number = {1},
        pages = {63-106},
          doi = {10.1146/annurev-astro-082708-101737},
archivePrefix = {arXiv},
       eprint = {0908.0700},
 primaryClass = {astro-ph.SR},
       adsurl = {https://ui.adsabs.harvard.edu/abs/2009ARA&A..47...63S}
}

@article{Sravan2019,
   title={Progenitors of Type IIb Supernovae. I. Evolutionary Pathways and Rates},
   volume={885},
   ISSN={1538-4357},
   url={http://dx.doi.org/10.3847/1538-4357/ab4ad7},
   DOI={10.3847/1538-4357/ab4ad7},
   number={2},
   journal={The Astrophysical Journal},
   publisher={American Astronomical Society},
   author={Sravan, Niharika and Marchant, Pablo and Kalogera, Vassiliki},
   year={2019},
   month=nov, pages={130} }

@misc{stevance2019,
      title={Spectropolarimetry of stripped envelope core collapse supernovae and their progenitors}, 
      author={H. F. Stevance},
      year={2019},
      eprint={1906.07184},
      archivePrefix={arXiv},
      primaryClass={astro-ph.SR},
      url={https://arxiv.org/abs/1906.07184}, 
}

@article{Szalai_2016,
   title={The continuing story of SN IIb 2013df: new optical and IR observations and analysis},
   volume={460},
   ISSN={1365-2966},
   url={http://dx.doi.org/10.1093/mnras/stw1031},
   DOI={10.1093/mnras/stw1031},
   number={2},
   journal={Monthly Notices of the Royal Astronomical Society},
   publisher={Oxford University Press (OUP)},
   author={Szalai, Tamás and Vinkó, József and Nagy, Andrea P. and Silverman, Jeffrey M. and Wheeler, J. Craig and Dhungana, Govinda and Marion, G. Howie and Kehoe, Robert and Fox, Ori D. and Sárneczky, Krisztián and Marschalkó, Gábor and Bíró, Barna I. and Borkovits, Tamás and Hegedüs, Tibor and Szakáts, Róbert and Ferrante, Farley V. and Bányai, Evelin and Hodosán, Gabriella and Kelemen, János and Pál, András},
   year={2016},
   month=May, pages={1500–1518} }

@article{Szalai_2021,
   title={Spitzer’s Last Look at Extragalactic Explosions: Long-term Evolution of Interacting Supernovae},
   volume={919},
   ISSN={1538-4357},
   url={http://dx.doi.org/10.3847/1538-4357/ac0e2b},
   DOI={10.3847/1538-4357/ac0e2b},
   number={1},
   journal={The Astrophysical Journal},
   publisher={American Astronomical Society},
   author={Szalai, Tamás and Fox, Ori D. and Arendt, Richard G. and Dwek, Eli and Andrews, Jennifer E. and Clayton, Geoffrey C. and Filippenko, Alexei V. and Johansson, Joel and Kelly, Patrick L. and Krafton, Kelsie and Marston, A. P. and Mauerhan, Jon C. and Van Dyk, Schuyler D.},
   year={2021},
   month="Sept", pages={17} }

@article{Szalai_2025,
   title={JWST/MIRI detects the dusty SN1993J about 30 years after explosion},
   volume={697},
   ISSN={1432-0746},
   url={http://dx.doi.org/10.1051/0004-6361/202451470},
   DOI={10.1051/0004-6361/202451470},
   journal={Astronomy \& Astrophysics},
   publisher={EDP Sciences},
   author={Szalai, Tamás and Zsíros, Szanna and Jencson, Jacob and Fox, Ori D. and Shahbandeh, Melissa and Sarangi, Arkaprabha and Temim, Tea and De Looze, Ilse and Smith, Nathan and Filippenko, Alexei V. and Van Dyk, Schuyler D. and Andrews, Jennifer and Ashall, Chris and Clayton, Geoffrey C. and Dessart, Luc and Dulude, Michael and Dwek, Eli and Gomez, Sebastian and Johansson, Joel and Milisavljevic, Dan and Pierel, Justin and Rest, Armin and Tinyanont, Samaporn and Brink, Thomas G. and De, Kishalay and Engesser, Michael and Foley, Ryan J. and Gezari, Suvi and Kasliwal, Mansi and Lau, Ryan and Marston, Anthony and O’Steen, Richard and Siebert, Matthew and Skrutskie, Michael and Strolger, Lou and Wang, Qinan and Williams, Brian J. and Williams, Robert and Xiao, Lin and Zheng, WeiKang},
   year={2025},
   month=May, pages={A132} }

@article{Tanaka_2012,
   title={THREE-DIMENSIONAL EXPLOSION GEOMETRY OF STRIPPED-ENVELOPE CORE-COLLAPSE SUPERNOVAE. I. SPECTROPOLARIMETRIC OBSERVATIONS},
   volume={754},
   ISSN={1538-4357},
   url={http://dx.doi.org/10.1088/0004-637X/754/1/63},
   DOI={10.1088/0004-637x/754/1/63},
   number={1},
   journal={The Astrophysical Journal},
   publisher={American Astronomical Society},
   author={Tanaka, Masaomi and Kawabata, Koji S. and Hattori, Takashi and Mazzali, Paolo A. and Aoki, Kentaro and Iye, Masanori and Maeda, Keiichi and Nomoto, Ken’ichi and Pian, Elena and Sasaki, Toshiyuki and Yamanaka, Masayuki},
   year={2012},
   month="July", pages={63} }

@article{Taubenberger_2011,
   title={The He-rich stripped-envelope core-collapse supernova 2008ax★: SN 2008ax},
   volume={413},
   ISSN={0035-8711},
   url={http://dx.doi.org/10.1111/j.1365-2966.2011.18287.x},
   DOI={10.1111/j.1365-2966.2011.18287.x},
   number={3},
   journal={Monthly Notices of the Royal Astronomical Society},
   publisher={Oxford University Press (OUP)},
   author={Taubenberger, S. and Navasardyan, H. and Maurer, J. I. and Zampieri, L. and Chugai, N. N. and Benetti, S. and Agnoletto, I. and Bufano, F. and Elias-Rosa, N. and Turatto, M. and Patat, F. and Cappellaro, E. and Mazzali, P. A. and Iijima, T. and Valenti, S. and Harutyunyan, A. and Claudi, R. and Dolci, M.},
   year={2011},
   month=Mar, pages={2140–2156} }

@software{reback2020pandas,
  author    = {The pandas development team},
  title     = {pandas-dev/pandas: Pandas},
  month     = feb,
  year      = 2020,
  publisher = {Zenodo},
  version   = {latest},
  doi       = {10.5281/zenodo.3509134},
  url       = {https://doi.org/10.5281/zenodo.3509134}
}

@article{Tinyanont_2016,
   title={A SYSTEMATIC STUDY OF MID-INFRARED EMISSION FROM CORE-COLLAPSE SUPERNOVAE WITH SPIRITS},
   volume={833},
   ISSN={1538-4357},
   url={http://dx.doi.org/10.3847/1538-4357/833/2/231},
   DOI={10.3847/1538-4357/833/2/231},
   number={2},
   journal={The Astrophysical Journal},
   publisher={American Astronomical Society},
   author={Tinyanont, Samaporn and Kasliwal, Mansi M. and Fox, Ori D. and Lau, Ryan and Smith, Nathan and Williams, Robert and Jencson, Jacob and Perley, Daniel and Dykhoff, Devin and Gehrz, Robert and Johansson, Joel and Van Dyk, Schuyler D. and Masci, Frank and Cody, Ann Marie and Prince, Thomas},
   year={2016},
   month=Dec, pages={231} }

@ARTICLE{2018PASP..130f4505T,
       author = {{Tonry}, J.~L. and {Denneau}, L. and {Heinze}, A.~N. and {Stalder}, B. and {Smith}, K.~W. and {Smartt}, S.~J. and {Stubbs}, C.~W. and {Weiland}, H.~J. and {Rest}, A.},
        title = "{ATLAS: A High-cadence All-sky Survey System}",
      journal = {\pasp},
         year = 2018,
        month = jun,
       volume = {130},
       number = {988},
        pages = {064505},
          doi = {10.1088/1538-3873/aabadf},
archivePrefix = {arXiv},
       eprint = {1802.00879},
 primaryClass = {astro-ph.IM},
       adsurl = {https://ui.adsabs.harvard.edu/abs/2018PASP..130f4505T}
}

@article{Tsujimoto_2022,
   title={From Galactic chemical evolution to cosmic supernova rates synchronized with core-collapse supernovae limited to the narrow progenitor mass range},
   volume={518},
   ISSN={1365-2966},
   url={http://dx.doi.org/10.1093/mnras/stac3351},
   DOI={10.1093/mnras/stac3351},
   number={3},
   journal={Monthly Notices of the Royal Astronomical Society},
   publisher={Oxford University Press (OUP)},
   author={Tsujimoto, T},
   year={2022},
   month=Nov, pages={3475–3481} }

@article{gilkis2025,
 title={The landscape of binary core-collapse supernova progenitors and the late emergence of Wolf–Rayet winds}, volume={540}, url={https://doi.org/10.1093/mnras/staf884}, DOI={10.1093/mnras/staf884}, number={4}, journal={Monthly Notices of the Royal Astronomical Society}, publisher={Oxford University Press (OUP)}, author={Gilkis, Avishai and Laplace, Eva and Arcavi, Iair and Shenar, Tomer and Schneider, Fabian R N}, year={2025}, month={May}, pages={3094–3120} }

@article{filippenko_1997, 
 title={OPTICAL SPECTRA OF SUPERNOVAE}, 
 volume={35}, 
 DOI={10.1146/annurev.astro.35.1.309}, 
 number={1}, 
 journal={Annual Review of Astronomy and Astrophysics}, 
 author={Filippenko, Alexei V.}, 
 year={1997}, 
 month={Sep},
 pages={309–355} }

@inbook{Turatto_2003,
   title={Classification of Supernovae},
   ISBN={9783540458630},
   ISSN={0075-8450},
   url={http://dx.doi.org/10.1007/3-540-45863-8_3},
   DOI={10.1007/3-540-45863-8_3},
   booktitle={Supernovae and Gamma-Ray Bursters},
   publisher={Springer Berlin Heidelberg},
   author={Turatto, Massimo},
   year={2003},
   pages={21–36} }

@article{Vink2001,
  author = {Vink, J. S. and de Koter, A. and Lamers, H. J. G. L. M.},
  title = {Mass-loss predictions for O and B stars as a function of metallicity},
  journal = {Astronomy \& Astrophysics},
  volume = {369},
  pages = {574--588},
  year = {2001},
  doi = {10.1051/0004-6361:20010127}
}

@ARTICLE{scipy2020,
       author = {{Virtanen}, P. and others},
        title = "{SciPy 1.0: Fundamental Algorithms for Scientific Computing in Python}",
      journal = {Nature Methods},
         year = 2020,
       volume = {17},
        pages = {261-272},
          doi = {10.1038/s41592-019-0686-2}
}

@article{Wang_2008,
   title={Spectropolarimetry of Supernovae},
   volume={46},
   ISSN={1545-4282},
   url={http://dx.doi.org/10.1146/annurev.astro.46.060407.145139},
   DOI={10.1146/annurev.astro.46.060407.145139},
   number={1},
   journal={Annual Review of Astronomy and Astrophysics},
   publisher={Annual Reviews},
   author={Wang, Lifan and Wheeler, J. Craig},
   year={2008},
   month="Sept", 
   pages={433–474} }

@INPROCEEDINGS{Weiner2026,
       author = {{Weiner}, Benjamin and {Williams}, George and {Pickering}, Timothy},
        title = "{The MMT Observatory: Instrumentation, Queue Observing, and Flexible Scheduling}",
    booktitle = {American Astronomical Society Meeting Abstracts},
         year = 2026,
       series = {American Astronomical Society Meeting Abstracts},
       volume = {247},
        month = feb,
          eid = {311.04},
        pages = {311.04},
       adsurl = {https://ui.adsabs.harvard.edu/abs/2026AAS...24731104W}
}

@ARTICLE{Wesson2015,
       author = {{Wesson}, R. and {Barlow}, M.~J. and {Matsuura}, M. and {Ercolano}, B.},
        title = "{The timing and location of dust formation in the remnant of SN 1987A}",
      journal = {\mnras},
         year = 2015,
        month = jan,
       volume = {446},
       number = {2},
        pages = {2089-2101},
          doi = {10.1093/mnras/stu2250},
archivePrefix = {arXiv},
       eprint = {1410.7386},
 primaryClass = {astro-ph.SR},
       adsurl = {https://ui.adsabs.harvard.edu/abs/2015MNRAS.446.2089W}
}

@misc{WISeREP2025aico,
  title        = {{WISeREP Object 29614: SN 2025aico}},
  author       = {{WISeREP}},
  year         = {2025},
  howpublished = {\url{https://www.wiserep.org/object/29614}},
  note         = {Accessed 2026 June 16}
}

@ARTICLE{Woolsey1994,
       author = {{Woosley}, S.~E. and {Eastman}, Ronald G. and {Weaver}, Thomas A. and {Pinto}, Philip A.},
        title = "{SN 1993J: A Type IIb Supernova}",
      journal = {\apj},
         year = 1994,
        month = "July",
       volume = {429},
        pages = {300},
          doi = {10.1086/174319},
       adsurl = {https://ui.adsabs.harvard.edu/abs/1994ApJ...429..300W}
}

@ARTICLE{Woosley1995,
       author = {{Woosley}, S.~E. and {Weaver}, Thomas A.},
        title = "{The Evolution and Explosion of Massive Stars. II. Explosive Hydrodynamics and Nucleosynthesis}",
      journal = {\apjs},
         year = 1995,
        month = nov,
       volume = {101},
        pages = {181},
          doi = {10.1086/192237},
       adsurl = {https://ui.adsabs.harvard.edu/abs/1995ApJS..101..181W}
}

@ARTICLE{Woosley2002,
       author = {{Woosley}, S.~E. and {Heger}, A. and {Weaver}, T.~A.},
        title = "{The evolution and explosion of massive stars}",
      journal = {Reviews of Modern Physics},
         year = 2002,
        month = nov,
       volume = {74},
       number = {4},
        pages = {1015-1071},
          doi = {10.1103/RevModPhys.74.1015},
       adsurl = {https://ui.adsabs.harvard.edu/abs/2002RvMP...74.1015W}
}

@misc{yamanaka2026,
      title={SN 2023dbc in M108: Optical and Near-Infrared Observations of a Highly-Obscured, Moderately Energetic Stripped-Envelope Supernova}, 
      author={Masayuki Yamanaka and Takahiro Nagayama and Akari Kumano and Devendra Kumar Sahu and Avinash Singh and Hrishav Das and G. C. Anupama},
      year={2026},
      eprint={2605.16916},
      archivePrefix={arXiv},
      primaryClass={astro-ph.HE},
      url={https://arxiv.org/abs/2605.16916}, 
}

@article{Zapartas2021,
   title={Revisiting the explodability of single massive star progenitors of stripped-envelope supernovae},
   volume={656},
   ISSN={1432-0746},
   url={http://dx.doi.org/10.1051/0004-6361/202141506},
   DOI={10.1051/0004-6361/202141506},
   journal={Astronomy \& Astrophysics},
   publisher={EDP Sciences},
   author={Zapartas, E. and Renzo, M. and Fragos, T. and Dotter, A. and Andrews, J. J. and Bavera, S. S. and Coughlin, S. and Misra, D. and Kovlakas, K. and Román-Garza, J. and Serra, J. G. and Qin, Y. and Rocha, K. A. and Tran, N. H. and Xing, Z. P.},
   year={2021},
   month=dec, pages={L19} }

@article{Zapartas_2025,
   title={The demographics of binary companions to stripped-envelope supernovae: confronting population synthesis models with observations},
   volume={546},
   ISSN={1365-2966},
   url={http://dx.doi.org/10.1093/mnras/staf2208},
   DOI={10.1093/mnras/staf2208},
   number={2},
   journal={Monthly Notices of the Royal Astronomical Society},
   publisher={Oxford University Press (OUP)},
   author={Zapartas, E and Fox, O D and Su, J and Souropanis, D and Drout, M R and Rocha, K A and vanDyk, S D and Williams, B F and Briel, M and Renzo, M and Andrews, J J and Fragos, T and Gossage, S and Kruckow, M U and Liotine, C and Ryder, S D and Srivastava, P M and Teng, E},
   year={2025},
   month=Dec }

@misc{zhao2026,
      title={SN 2025aico: Early observations of a faint Type IIb supernova with a low-mass envelope}, 
      author={J. -W. Zhao and A. Pastorello and B. Kumar and Y. -Z. Cai and A. Dutta and D. K. Sahu and A. Reguitti and R. S. Teja and H. Das and T. J. Moriya and N. Pyykkinen and K. Valeckas and G. Valerin and X. -Z. Zou and C. Ashall and S. Bijavara Seshashayana and G. -W. Du and G. C. Anupama and A. L. Bouquin and S. Campana and K. Chatterjee and X. -L. Chen and X. -L. Du and N. Elias-Rosa and Y. Fang and M. Fraser and W. Hoogendam and E. Hsiao and E. Kankare and E. P. Lagioia and W. -Y. Li and X. -K. Liu and P. Lundqvist and K. Matilainen and J. Martikainen and K. Medler and N. Morrell and Y. Pan and C. Pfeffer and G. Rameshan and T. M. Reynolds and M. D. Stritzinger and V. Vuolteenaho and Z. -Y. Wang and H. -F. Xiao and J. -H. Zhang and X. -W. Liu and Y. -P. Yang},
      year={2026},
      eprint={2607.10671},
      archivePrefix={arXiv},
      primaryClass={astro-ph.SR},
      url={https://arxiv.org/abs/2607.10671}, 
}

@article{Zhao_2026,
   title={SN 2022ngb: A faint, slowly evolving Type IIb supernova with a low-mass envelope},
   volume={706},
   ISSN={1432-0746},
   url={http://dx.doi.org/10.1051/0004-6361/202557619},
   DOI={10.1051/0004-6361/202557619},
   journal={Astronomy \& Astrophysics},
   publisher={EDP Sciences},
   author={Zhao, J.-W. and Benetti, S. and Cai, Y.-Z. and Pastorello, A. and Elias-Rosa, N. and Reguitti, A. and Valerin, G. and Wang, Z.-Y. and Cappellaro, E. and Feng, G.-F. and Fiore, A. and Fitzpatrick, B. and Fraser, M. and Isern, J. and Kankare, E. and Kravtsov, T. and Kumar, B. and Lundqvist, P. and Matilainen, K. and Mattila, S. and Mazzali, P. A. and Moran, S. and Ochner, P. and Peng, Z.-H. and Reynolds, T. M. and Salmaso, I. and Srivastav, S. and Stritzinger, M. D. and Taubenberger, S. and Tomasella, L. and Vinkó, J. and Wheeler, J. C. and Williams, S. and Pei, S.-P. and Yang, Y.-J. and Liu, X.-K. and Liu, X.-W. and Yang, Y.-P.},
   year={2026},
   month=Feb, pages={A271} }

@article{Zubko_2004,
   title={Interstellar Dust Models Consistent with Extinction, Emission, and Abundance Constraints},
   volume={152},
   ISSN={1538-4365},
   url={http://dx.doi.org/10.1086/382351},
   DOI={10.1086/382351},
   number={2},
   journal={The Astrophysical Journal Supplement Series},
   publisher={American Astronomical Society},
   author={Zubko, Viktor and Dwek, Eli and Arendt, Richard G.},
   year={2004},
   month="June", pages={211–249} }

@ARTICLE{Milisavljevic2010,
       author = {{Milisavljevic}, Dan and {Fesen}, Robert A. and {Gerardy}, Christopher L. and {Kirshner}, Robert P. and {Challis}, Peter},
        title = "{Doublets and Double Peaks: Late-Time [O I] {\ensuremath{\lambda}}{\ensuremath{\lambda}}6300, 6364 Line Profiles of Stripped-Envelope, Core-Collapse Supernovae}",
      journal = {\apj},
         year = 2010,
        month = feb,
       volume = {709},
       number = {2},
        pages = {1343-1355},
          doi = {10.1088/0004-637X/709/2/1343},
archivePrefix = {arXiv},
       eprint = {0904.4256},
 primaryClass = {astro-ph.CO},
       adsurl = {https://ui.adsabs.harvard.edu/abs/2010ApJ...709.1343M}
}

@article{Rabinak_2011,
   title={THE EARLY UV/OPTICAL EMISSION FROM CORE-COLLAPSE SUPERNOVAE},
   volume={728},
   ISSN={1538-4357},
   url={http://dx.doi.org/10.1088/0004-637X/728/1/63},
   DOI={10.1088/0004-637x/728/1/63},
   number={1},
   journal={The Astrophysical Journal},
   publisher={American Astronomical Society},
   author={Rabinak, Itay and Waxman, Eli},
   year={2011},
   month=Jan, pages={63} }

@article{Sapir_2017,
   title={UV/Optical Emission from the Expanding Envelopes of Type II Supernovae},
   volume={838},
   ISSN={1538-4357},
   url={http://dx.doi.org/10.3847/1538-4357/aa64df},
   DOI={10.3847/1538-4357/aa64df},
   number={2},
   journal={The Astrophysical Journal},
   publisher={American Astronomical Society},
   author={Sapir, Nir and Waxman, Eli},
   year={2017},
   month=Apr, pages={130} }

@ARTICLE{Bevan2019,
       author = {{Bevan}, A. and {Wesson}, R. and {Barlow}, M.~J. and {De Looze}, I. and {Andrews}, J.~E. and {Clayton}, G.~C. and {Krafton}, K. and {Matsuura}, M. and {Milisavljevic}, D.},
        title = "{A decade of ejecta dust formation in the Type IIn SN 2005ip}",
      journal = {\mnras},
         year = 2019,
        month = jun,
       volume = {485},
       number = {4},
        pages = {5192-5206},
          doi = {10.1093/mnras/stz679},
archivePrefix = {arXiv},
       eprint = {1809.09055},
 primaryClass = {astro-ph.SR},
       adsurl = {https://ui.adsabs.harvard.edu/abs/2019MNRAS.485.5192B}
}

@INPROCEEDINGS{Rayner1998,
       author = {{Rayner}, John T. and {Toomey}, Douglas W. and {Onaka}, Peter M. and {Denault}, Anthony J. and {Stahlberger}, Werner E. and {Watanabe}, Darryl Y. and {Wang}, Shu-I.},
        title = "{SpeX: a medium-resolution IR spectrograph for IRTF}",
    booktitle = {Infrared Astronomical Instrumentation},
         year = 1998,
       editor = {{Fowler}, Albert M.},
       series = {Society of Photo-Optical Instrumentation Engineers (SPIE) Conference Series},
       volume = {3354},
        month = aug,
        pages = {468-479},
          doi = {10.1117/12.317273},
       adsurl = {https://ui.adsabs.harvard.edu/abs/1998SPIE.3354..468R}
}

@article{Zs_ros_2021,
   title={Rescued from oblivion: detailed analysis of archival <i>Spitzer</i> data of SN 1993J},
   volume={509},
   ISSN={1365-2966},
   url={http://dx.doi.org/10.1093/mnras/stab3075},
   DOI={10.1093/mnras/stab3075},
   number={3},
   journal={Monthly Notices of the Royal Astronomical Society},
   publisher={Oxford University Press (OUP)},
   author={Zsíros, Szanna and Nagy, Andrea P and Szalai, Tamás},
   year={2021},
   month=Nov, pages={3235–3246} }

@ARTICLE{Burrows2024,
       author = {{Burrows}, Adam and {Wang}, Tianshu and {Vartanyan}, David},
        title = "{Physical Correlations and Predictions Emerging from Modern Core-collapse Supernova Theory}",
      journal = {\apjl},
         year = 2024,
        month = mar,
       volume = {964},
       number = {1},
          eid = {L16},
        pages = {L16},
          doi = {10.3847/2041-8213/ad319e},
archivePrefix = {arXiv},
       eprint = {2401.06840},
 primaryClass = {astro-ph.HE},
       adsurl = {https://ui.adsabs.harvard.edu/abs/2024ApJ...964L..16B}
}

@ARTICLE{Vartanyan2019,
       author = {{Vartanyan}, David and {Burrows}, Adam and {Radice}, David},
        title = "{Temporal and angular variations of 3D core-collapse supernova emissions and their physical correlations}",
      journal = {\mnras},
         year = 2019,
        month = oct,
       volume = {489},
       number = {2},
        pages = {2227-2246},
          doi = {10.1093/mnras/stz2307},
archivePrefix = {arXiv},
       eprint = {1906.08787},
 primaryClass = {astro-ph.HE},
       adsurl = {https://ui.adsabs.harvard.edu/abs/2019MNRAS.489.2227V}
}

@ARTICLE{Lucy_1991,
       author = {{Lucy}, L.~B.},
        title = "{Nonthermal Excitation of Helium in Type Ib Supernovae}",
      journal = {\apj},
         year = 1991,
        month = dec,
       volume = {383},
        pages = {308},
          doi = {10.1086/170787},
       adsurl = {https://ui.adsabs.harvard.edu/abs/1991ApJ...383..308L}
}

@article{Dessart_2016,
   title={Inferring supernova IIb/Ib/Ic ejecta properties from light curves and spectra: correlations from radiative-transfer models},
   volume={458},
   ISSN={1365-2966},
   url={http://dx.doi.org/10.1093/mnras/stw418},
   DOI={10.1093/mnras/stw418},
   number={2},
   journal={Monthly Notices of the Royal Astronomical Society},
   publisher={Oxford University Press (OUP)},
   author={Dessart, Luc and Hillier, D. John and Woosley, Stan and Livne, Eli and Waldman, Roni and Yoon, Sung-Chul and Langer, Norbert},
   year={2016},
   month=Feb, pages={1618–1635} }

@ARTICLE{Lyman_2016,
       author = {{Lyman}, J.~D. and {Bersier}, D. and {James}, P.~A. and {Mazzali}, P.~A. and {Eldridge}, J.~J. and {Fraser}, M. and {Pian}, E.},
        title = "{Bolometric light curves and explosion parameters of 38 stripped-envelope core-collapse supernovae}",
      journal = {\mnras},
         year = 2016,
        month = mar,
       volume = {457},
       number = {1},
        pages = {328-350},
          doi = {10.1093/mnras/stv2983},
archivePrefix = {arXiv},
       eprint = {1406.3667},
 primaryClass = {astro-ph.SR},
       adsurl = {https://ui.adsabs.harvard.edu/abs/2016MNRAS.457..328L}
}

@ARTICLE{Dwek1986,
       author = {{Dwek}, E.},
        title = "{Temperature Fluctuations and Infrared Emission from Dust Particles in a Hot Gas}",
      journal = {\apj},
         year = 1986,
        month = mar,
       volume = {302},
        pages = {363},
          doi = {10.1086/163995},
       adsurl = {https://ui.adsabs.harvard.edu/abs/1986ApJ...302..363D}
}

@ARTICLE{Sarangi_2015,
       author = {{Sarangi}, Arkaprabha and {Cherchneff}, Isabelle},
        title = "{Condensation of dust in the ejecta of Type II-P supernovae}",
      journal = {\aap},
         year = 2015,
        month = mar,
       volume = {575},
          eid = {A95},
        pages = {A95},
          doi = {10.1051/0004-6361/201424969},
archivePrefix = {arXiv},
       eprint = {1412.5522},
 primaryClass = {astro-ph.SR},
       adsurl = {https://ui.adsabs.harvard.edu/abs/2015A&A...575A..95S}
}

@ARTICLE{Hofner_2018,
       author = {{H{\"o}fner}, Susanne and {Olofsson}, Hans},
        title = "{Mass loss of stars on the asymptotic giant branch. Mechanisms, models and measurements}",
      journal = {\aapr},
         year = 2018,
        month = jan,
       volume = {26},
       number = {1},
          eid = {1},
        pages = {1},
          doi = {10.1007/s00159-017-0106-5},
       adsurl = {https://ui.adsabs.harvard.edu/abs/2018A&ARv..26....1H}
}

@ARTICLE{Bocchio2016,
       author = {{Bocchio}, M. and {Marassi}, S. and {Schneider}, R. and {Bianchi}, S. and {Limongi}, M. and {Chieffi}, A.},
        title = "{Dust grains from the heart of supernovae}",
      journal = {\aap},
         year = 2016,
        month = mar,
       volume = {587},
          eid = {A157},
        pages = {A157},
          doi = {10.1051/0004-6361/201527432},
archivePrefix = {arXiv},
       eprint = {1601.06770},
 primaryClass = {astro-ph.HE},
       adsurl = {https://ui.adsabs.harvard.edu/abs/2016A&A...587A.157B}
}

@ARTICLE{Micelotta2016,
       author = {{Micelotta}, Elisabetta R. and {Dwek}, Eli and {Slavin}, Jonathan D.},
        title = "{Dust destruction by the reverse shock in the Cassiopeia A supernova remnant}",
      journal = {\aap},
         year = 2016,
        month = may,
       volume = {590},
          eid = {A65},
        pages = {A65},
          doi = {10.1051/0004-6361/201527350},
archivePrefix = {arXiv},
       eprint = {1602.02754},
 primaryClass = {astro-ph.GA},
       adsurl = {https://ui.adsabs.harvard.edu/abs/2016A&A...590A..65M}
}

@article{Silvia_2010,
   title={NUMERICAL SIMULATIONS OF SUPERNOVA DUST DESTRUCTION. I. CLOUD-CRUSHING AND POST-PROCESSED GRAIN SPUTTERING},
   volume={715},
   ISSN={1538-4357},
   url={http://dx.doi.org/10.1088/0004-637X/715/2/1575},
   DOI={10.1088/0004-637x/715/2/1575},
   number={2},
   journal={The Astrophysical Journal},
   publisher={American Astronomical Society},
   author={Silvia, Devin W. and Smith, Britton D. and Michael Shull, J.},
   year={2010},
   month=May, pages={1575–1590} }

@article{Todini_2001,
   title={Dust formation in primordial Type II supernovae},
   volume={325},
   ISSN={1365-2966},
   url={http://dx.doi.org/10.1046/j.1365-8711.2001.04486.x},
   DOI={10.1046/j.1365-8711.2001.04486.x},
   number={2},
   journal={Monthly Notices of the Royal Astronomical Society},
   publisher={Oxford University Press (OUP)},
   author={Todini, P. and Ferrara, A.},
   year={2001},
   month=Aug, pages={726–736} }

@article{Nozawa_2003,
   title={Dust in the Early Universe: Dust Formation in the Ejecta of Population III Supernovae},
   volume={598},
   ISSN={1538-4357},
   url={http://dx.doi.org/10.1086/379011},
   DOI={10.1086/379011},
   number={2},
   journal={The Astrophysical Journal},
   publisher={American Astronomical Society},
   author={Nozawa, Takaya and Kozasa, Takashi and Umeda, Hideyuki and Maeda, Keiichi and Nomoto, Ken’ichi},
   year={2003},
   month=Dec, pages={785–803} }

@article{Gall_2011,
   title={Production of dust by massive stars at high redshift},
   volume={19},
   ISSN={1432-0754},
   url={http://dx.doi.org/10.1007/s00159-011-0043-7},
   DOI={10.1007/s00159-011-0043-7},
   number={1},
   journal={The Astronomy and Astrophysics Review},
   publisher={Springer Science and Business Media LLC},
   author={Gall, C. and Hjorth, J. and Andersen, A. C.},
   year={2011},
   month=Sept }

@ARTICLE{Jerkstrand2015,
       author = {{Jerkstrand}, A. and {Ergon}, M. and {Smartt}, S.~J. and {Fransson}, C. and {Sollerman}, J. and {Taubenberger}, S. and {Bersten}, M. and {Spyromilio}, J.},
        title = "{Late-time spectral line formation in Type IIb supernovae, with application to SN 1993J, SN 2008ax, and SN 2011dh}",
      journal = {\aap},
         year = 2015,
        month = jan,
       volume = {573},
          eid = {A12},
        pages = {A12},
          doi = {10.1051/0004-6361/201423983},
archivePrefix = {arXiv},
       eprint = {1408.0732},
 primaryClass = {astro-ph.HE},
       adsurl = {https://ui.adsabs.harvard.edu/abs/2015A&A...573A..12J}
}

@ARTICLE{Anderson2015,
       author = {{Anderson}, Joseph P. and {James}, Phil A. and {Habergham}, Stacey M. and {Galbany}, Llu{\'\i}s and {Kuncarayakti}, Hanindyo},
        title = "{Statistical Studies of Supernova Environments}",
      journal = {\pasa},
         year = 2015,
        month = may,
       volume = {32},
          eid = {e019},
        pages = {e019},
          doi = {10.1017/pasa.2015.19},
archivePrefix = {arXiv},
       eprint = {1504.04043},
 primaryClass = {astro-ph.HE},
       adsurl = {https://ui.adsabs.harvard.edu/abs/2015PASA...32...19A}
}

@article{Prentice_2022,
   title={Oxygen and calcium nebular emission line relationships in core-collapse supernovae and Ca-rich transients},
   volume={514},
   ISSN={1365-2966},
   url={http://dx.doi.org/10.1093/mnras/stac1657},
   DOI={10.1093/mnras/stac1657},
   number={4},
   journal={Monthly Notices of the Royal Astronomical Society},
   publisher={Oxford University Press (OUP)},
   author={Prentice, S J and Maguire, K and Siebenaler, L and Jerkstrand, A},
   year={2022},
   month=June, pages={5686–5705} }

@ARTICLE{Fang2022,
       author = {{Fang}, Qiliang and {Maeda}, Keiichi and {Kuncarayakti}, Hanindyo and {Tanaka}, Masaomi and {Kawabata}, Koji S. and {Hattori}, Takashi and {Aoki}, Kentaro and {Moriya}, Takashi J. and {Yamanaka}, Masayuki},
        title = "{Statistical Properties of the Nebular Spectra of 103 Stripped-envelope Core-collapse Supernovae}",
      journal = {\apj},
         year = 2022,
        month = apr,
       volume = {928},
       number = {2},
          eid = {151},
        pages = {151},
          doi = {10.3847/1538-4357/ac4f60},
archivePrefix = {arXiv},
       eprint = {2201.11467},
 primaryClass = {astro-ph.HE},
       adsurl = {https://ui.adsabs.harvard.edu/abs/2022ApJ...928..151F}
}

@article{Arcavi_2011,
   title={SN 2011dh: DISCOVERY OF A TYPE IIb SUPERNOVA FROM A COMPACT PROGENITOR IN THE NEARBY GALAXY M51},
   volume={742},
   ISSN={2041-8213},
   url={http://dx.doi.org/10.1088/2041-8205/742/2/L18},
   DOI={10.1088/2041-8205/742/2/l18},
   number={2},
   journal={The Astrophysical Journal},
   publisher={American Astronomical Society},
   author={Arcavi, Iair and Gal-Yam, Avishay and Yaron, Ofer and Sternberg, Assaf and Rabinak, Itay and Waxman, Eli and Kasliwal, Mansi M. and Quimby, Robert M. and Ofek, Eran O. and Horesh, Assaf and Kulkarni, Shrinivas R. and Filippenko, Alexei V. and Silverman, Jeffrey M. and Cenko, S. Bradley and Li, Weidong and Bloom, Joshua S. and Sullivan, Mark and Nugent, Peter E. and Poznanski, Dovi and Gorbikov, Evgeny and Fulton, Benjamin J. and Howell, D. Andrew and Bersier, David and Riou, Amedee and Lamotte-Bailey, Stephane and Griga, Thomas and Cohen, Judith G. and Hachinger, Stephan and Polishook, David and Xu, Dong and Ben-Ami, Sagi and Manulis, Ilan and Walker, Emma S. and Maguire, Kate and Pan, Yen-Chen and Matheson, Thomas and Mazzali, Paolo A. and Pian, Elena and Fox, Derek B. and Gehrels, Neil and Law, Nicholas and James, Philip and Marchant, Jonathan M. and Smith, Robert J. and Mottram, Chris J. and Barnsley, Robert M. and Kandrashoff, Michael T. and Clubb, Kelsey I.},
   year={2011},
   month=Nov, pages={L18} }

@article{Podsiadlowski1993,
  author  = {Podsiadlowski, P. and Hsu, J. J. L. and Joss, P. C. and Ross, R. R.},
  title   = {The progenitor of supernova 1993J: A stripped supergiant in a binary system?},
  journal = {Nature},
  volume  = {364},
  pages   = {509--511},
  year    = {1993}
}

@article{Stancliffe2009,
  author  = {Stancliffe, R. J. and Eldridge, J. J.},
  title   = {Modelling the binary progenitor of supernova 1993J},
  journal = {Monthly Notices of the Royal Astronomical Society},
  volume  = {396},
  pages   = {1699--1708},
  year    = {2009}
}

@article{Claeys2011,
  author  = {Claeys, J. S. W. and de Mink, S. E. and Pols, O. R. and Eldridge, J. J. and Baes, M.},
  title   = {Binary progenitor models of Type IIb supernovae},
  journal = {Astronomy \& Astrophysics},
  volume  = {528},
  pages   = {A131},
  year    = {2011}
}

@article{Benvenuto2013,
  author  = {Benvenuto, O. G. and Bersten, M. C. and Nomoto, K.},
  title   = {A binary progenitor for the Type IIb supernova 2011dh in M51},
  journal = {The Astrophysical Journal},
  volume  = {762},
  pages   = {74},
  year    = {2013}
}

@article{Yoon2017,
  author  = {Yoon, Sung-Chul and Dessart, Luc and Clocchiatti, Alejandro},
  title   = {Type Ib and IIb supernova progenitors in interacting binary systems},
  journal = {The Astrophysical Journal},
  volume  = {840},
  pages   = {10},
  year    = {2017}
}

@ARTICLE{Woosley_2006,
       author = {{Woosley}, S.~E. and {Bloom}, J.~S.},
        title = "{The Supernova Gamma-Ray Burst Connection}",
      journal = {\araa},
         year = 2006,
        month = sep,
       volume = {44},
       number = {1},
        pages = {507-556},
          doi = {10.1146/annurev.astro.43.072103.150558},
archivePrefix = {arXiv},
       eprint = {astro-ph/0609142},
 primaryClass = {astro-ph},
       adsurl = {https://ui.adsabs.harvard.edu/abs/2006ARA&A..44..507W}
}

@article{Hsiao_2018,
   title={Carnegie Supernova Project-II: The Near-infrared Spectroscopy Program},
   volume={131},
   ISSN={1538-3873},
   url={http://dx.doi.org/10.1088/1538-3873/aae961},
   DOI={10.1088/1538-3873/aae961},
   number={995},
   journal={Publications of the Astronomical Society of the Pacific},
   publisher={IOP Publishing},
   author={Hsiao, E. Y. and Phillips, M. M. and Marion, G. H. and Kirshner, R. P. and Morrell, N. and Sand, D. J. and Burns, C. R. and Contreras, C. and Hoeflich, P. and Stritzinger, M. D. and Valenti, S. and Anderson, J. P. and Ashall, C. and Baltay, C. and Baron, E. and Banerjee, D. P. K. and Davis, S. and Diamond, T. R. and Folatelli, G. and Freedman, Wendy L. and Förster, F. and Galbany, L. and Gall, C. and González-Gaitán, S. and Goobar, A. and Hamuy, M. and Holmbo, S. and Kasliwal, M. M. and Krisciunas, K. and Kumar, S. and Lidman, C. and Lu, J. and Nugent, P. E. and Perlmutter, S. and Persson, S. E. and Piro, A. L. and Rabinowitz, D. and Roth, M. and Ryder, S. D. and Schmidt, B. P. and Shahbandeh, M. and Suntzeff, N. B. and Taddia, F. and Uddin, S. and Wang, L.},
   year={2018},
   month=Nov, pages={014002} }

@INPROCEEDINGS{McLean_1998,
       author = {{McLean}, Ian S. and {Becklin}, Eric E. and {Bendiksen}, Oddvar and {Brims}, George and {Canfield}, John and {Figer}, Donald F. and {Graham}, James R. and {Hare}, Jonah and {Lacayanga}, Fred and {Larkin}, James E. and {Larson}, Samuel B. and {Levenson}, Nancy and {Magnone}, Nick and {Teplitz}, Harry and {Wong}, Woon},
        title = "{Design and development of NIRSPEC: a near-infrared echelle spectrograph for the Keck II telescope}",
    booktitle = {Infrared Astronomical Instrumentation},
         year = 1998,
       editor = {{Fowler}, Albert M.},
       series = {Society of Photo-Optical Instrumentation Engineers (SPIE) Conference Series},
       volume = {3354},
        month = aug,
        pages = {566-578},
          doi = {10.1117/12.317283},
       adsurl = {https://ui.adsabs.harvard.edu/abs/1998SPIE.3354..566M}
}

@article{Hoogendam2025a,
    author = {{Hoogendam}, W.~B. and {Jones}, D.~O. and {Ashall}, C. and
              {Shappee}, B.~J. and {Foley}, R.~J. and {Tucker}, M.~A. and
              {Huber}, M.~E. and {Auchettl}, K. and {Desai}, D.~D. and
              {Do}, A. and {Hinkle}, J.~T. and {Romagnoli}, S. and
              {Shi}, J. and {Syncatto}, A. and others},
    title = {Seeing the Outer Edge of the Infant Type Ia Supernova 2024epr
             in the Optical and Near Infrared},
    journal = {The Open Journal of Astrophysics},
    year = {2025},
    volume = {8},
    eid = {120},
    doi = {10.33232/001c.143462},
    archivePrefix = {arXiv},
    eprint = {2502.17556},
    primaryClass = {astro-ph.HE}
}

@article{Hoogendam2025b,
    author = {{Hoogendam}, W.~B. and {Ashall}, C. and {Jones}, D.~O. and
              {Shappee}, B.~J. and {Tucker}, M.~A. and {Huber}, M.~E. and
              {Auchettl}, K. and {Desai}, D.~D. and {Hinkle}, J.~T. and
              {Kong}, M.~Y. and {Romagnoli}, S. and {Shi}, J. and
              {Syncatto}, A. and {Kilpatrick}, C.~D.},
    title = {Early and Extensive Ultraviolet through Near Infrared
             Observations of the Intermediate-Luminosity Type Iax
             Supernova 2024pxl},
    journal = {\apj},
    year = {2025},
    volume = {988},
    eid = {209},
    pages = {209},
    archivePrefix = {arXiv},
    eprint = {2505.04610},
    primaryClass = {astro-ph.HE}
}

@article{Desai_2025,
   title={Plasma instabilities dominate radioactive transients magnetic fields: the self-confinement of leptons in Type Ia and core-collapse supernovae, and kilonovae},
   volume={541},
   ISSN={1365-2966},
   url={http://dx.doi.org/10.1093/mnras/staf1117},
   DOI={10.1093/mnras/staf1117},
   number={3},
   journal={Monthly Notices of the Royal Astronomical Society},
   publisher={Oxford University Press (OUP)},
   author={Desai, Dhvanil D and Haggerty, Colby C and Shappee, Benjamin J and Tucker, Michael A and Davis, Zachary and Ashall, Chris and Chomiuk, Laura and Gootkin, Keyan and Caprioli, Damiano and Bret, Antoine and Hakobyan, Hayk},
   year={2025},
   month=July, pages={2197–2215} }

@ARTICLE{Matzner1999,
       author = {{Matzner}, Christopher D. and {McKee}, Christopher F.},
        title = "{The Expulsion of Stellar Envelopes in Core-Collapse Supernovae}",
      journal = {\apj},
         year = 1999,
        month = jan,
       volume = {510},
       number = {1},
        pages = {379-403},
          doi = {10.1086/306571},
archivePrefix = {arXiv},
       eprint = {astro-ph/9807046},
 primaryClass = {astro-ph},
       adsurl = {https://ui.adsabs.harvard.edu/abs/1999ApJ...510..379M}
}

@article{Piro2017,
    author  = {{Piro}, Anthony L. and {Muhleisen}, Marc and {Arcavi}, Iair and
               {Sand}, David J. and {Tartaglia}, Leonardo and {Valenti}, Stefano},
    title   = {Numerically Modeling the First Peak of the Type IIb SN 2016gkg},
    journal = {\apj},
    year    = {2017},
    volume  = {846},
    number  = {1},
    pages   = {94},
    doi     = {10.3847/1538-4357/aa8595},
    eid     = {94},
    archivePrefix = {arXiv},
    eprint  = {1703.00913},
    primaryClass = {astro-ph.HE}
}

@article{Piro2021,
    author  = {{Piro}, Anthony L. and {Haynie}, Annastasia and {Yao}, Yuhan},
    title   = {Shock Cooling Emission from Extended Material Revisited},
    journal = {\apj},
    year    = {2021},
    volume  = {909},
    number  = {2},
    pages   = {209},
    doi     = {10.3847/1538-4357/abe2b1},
    eid     = {209},
    archivePrefix = {arXiv},
    eprint  = {2007.08543},
    primaryClass = {astro-ph.HE}
}

@ARTICLE{Spyromilio_1988,
       author = {{Spyromilio}, J. and {Meikle}, W.~P.~S. and {Learner}, R.~C.~M. and {Allen}, D.~A.},
        title = "{Carbon monoxide in supernova 1987A}",
      journal = {\nat},
         year = 1988,
        month = jul,
       volume = {334},
       number = {6180},
        pages = {327-329},
          doi = {10.1038/334327a0},
       adsurl = {https://ui.adsabs.harvard.edu/abs/1988Natur.334..327S}
}

@article{Matsuura_2011,
   title={Herschel Detects a Massive Dust Reservoir in Supernova 1987A},
   volume={333},
   ISSN={1095-9203},
   url={http://dx.doi.org/10.1126/science.1205983},
   DOI={10.1126/science.1205983},
   number={6047},
   journal={Science},
   publisher={American Association for the Advancement of Science (AAAS)},
   author={Matsuura, M. and Dwek, E. and Meixner, M. and Otsuka, M. and Babler, B. and Barlow, M. J. and Roman-Duval, J. and Engelbracht, C. and Sandstrom, K. and Lakićević, M. and van Loon, J. Th. and Sonneborn, G. and Clayton, G. C. and Long, K. S. and Lundqvist, P. and Nozawa, T. and Gordon, K. D. and Hony, S. and Panuzzo, P. and Okumura, K. and Misselt, K. A. and Montiel, E. and Sauvage, M.},
   year={2011},
   month=Sept, pages={1258–1261} }

@article{De_Looze_2016,
   title={The dust mass in Cassiopeia A from a spatially resolved<i>Herschel</i>analysis},
   volume={465},
   ISSN={1365-2966},
   url={http://dx.doi.org/10.1093/mnras/stw2837},
   DOI={10.1093/mnras/stw2837},
   number={3},
   journal={Monthly Notices of the Royal Astronomical Society},
   publisher={Oxford University Press (OUP)},
   author={De Looze, I. and Barlow, M. J. and Swinyard, B. M. and Rho, J. and Gomez, H. L. and Matsuura, M. and Wesson, R.},
   year={2016},
   month=Nov, pages={3309–3342} }

@article{Gomez_2012,
   title={Dust in historical Galactic Type Ia supernova remnants with Herschel★: Dust in the Kepler and Tycho remnants},
   volume={420},
   ISSN={0035-8711},
   url={http://dx.doi.org/10.1111/j.1365-2966.2011.20272.x},
   DOI={10.1111/j.1365-2966.2011.20272.x},
   number={4},
   journal={Monthly Notices of the Royal Astronomical Society},
   publisher={Oxford University Press (OUP)},
   author={Gomez, H. L. and Clark, C. J. R. and Nozawa, T. and Krause, O. and Gomez, E. L. and Matsuura, M. and Barlow, M. J. and Besel, M.-A. and Dunne, L. and Gear, W. K. and Hargrave, P. and Henning, Th. and Ivison, R. J. and Sibthorpe, B. and Swinyard, B. M. and Wesson, R.},
   year={2012},
   month=Jan, pages={3557–3573} }

@article{Owen_2015,
   title={THE DUST AND GAS CONTENT OF THE CRAB NEBULA},
   volume={801},
   ISSN={1538-4357},
   url={http://dx.doi.org/10.1088/0004-637X/801/2/141},
   DOI={10.1088/0004-637x/801/2/141},
   number={2},
   journal={The Astrophysical Journal},
   publisher={American Astronomical Society},
   author={Owen, P. J. and Barlow, M. J.},
   year={2015},
   month=Mar, pages={141} }

@article{De_Looze_2019,
   title={The dust content of the Crab Nebula},
   volume={488},
   ISSN={1365-2966},
   url={http://dx.doi.org/10.1093/mnras/stz1533},
   DOI={10.1093/mnras/stz1533},
   number={1},
   journal={Monthly Notices of the Royal Astronomical Society},
   publisher={Oxford University Press (OUP)},
   author={De Looze, I and Barlow, M J and Bandiera, R and Bevan, A and Bietenholz, M F and Chawner, H and Gomez, H L and Matsuura, M and Priestley, F and Wesson, R},
   year={2019},
   month=June, pages={164–182} }

@ARTICLE{Mera_2026,
       author = {{Mera}, T. and {Hoeflich}, P. and {Burns}, C.~R. and {Ashall}, C. and {Medler}, K. and {Fereidouni}, E. and {Hoogendam}, W.~B. and {Shahbandeh}, M. and {Shiber}, S. and {Pfeffer}, C.~M. and {Baron}, E. and {Lu}, J. and {Morrell}, N. and {Hsiao}, E.~Y. and {Phillips}, M.~M.},
        title = "{Probing the 3D Structures of Supernovae through IR Signatures of CO and SiO}",
      journal = {\apj},
         year = 2026,
        month = jun,
       volume = {1003},
       number = {2},
          eid = {135},
        pages = {135},
          doi = {10.3847/1538-4357/ae66ed},
archivePrefix = {arXiv},
       eprint = {2604.18339},
 primaryClass = {astro-ph.HE},
       adsurl = {https://ui.adsabs.harvard.edu/abs/2026ApJ..1003..135M}
}

@ARTICLE{Wongwathanarat_2015,
       author = {{Wongwathanarat}, A. and {M{\"u}ller}, E. and {Janka}, H.-Th.},
        title = "{Three-dimensional simulations of core-collapse supernovae: from shock revival to shock breakout}",
      journal = {\aap},
         year = 2015,
        month = may,
       volume = {577},
          eid = {A48},
        pages = {A48},
          doi = {10.1051/0004-6361/201425025},
archivePrefix = {arXiv},
       eprint = {1409.5431},
 primaryClass = {astro-ph.HE},
       adsurl = {https://ui.adsabs.harvard.edu/abs/2015A&A...577A..48W}
}

@misc{vanbaal2024,
      title={Diagnostics of 3D explosion asymmetries of stripped-envelope supernovae by nebular line profiles}, 
      author={Bart van Baal and Anders Jerkstrand and Annop Wongwathanarat and Thomas Janka},
      year={2024},
      eprint={2404.01763},
      archivePrefix={arXiv},
      primaryClass={astro-ph.HE},
      url={https://arxiv.org/abs/2404.01763}, 
}

@article{Wooden1993,
  author  = {Wooden, D. H. and Rank, D. M. and Bregman, J. D. and Witteborn, F. C. and Tielens, A. G. G. M. and Cohen, M. and Pinto, P. A. and Axelrod, T. S.},
  title   = {Airborne Spectrophotometry of SN 1987A from 1.7 to 12.6 Microns: Time History of the Dust Continuum and Line Emission},
  journal = {The Astrophysical Journal Supplement Series},
  volume  = {88},
  pages   = {477--507},
  year    = {1993},
  doi     = {10.1086/191830}
}

@article{Kotak2009,
  author  = {Kotak, R. and Meikle, W. P. S. and Farrah, D. and Gerardy, C. L. and Foley, R. J. and Van Dyk, S. D. and Fransson, C. and Lundqvist, P. and Sollerman, J. and Fesen, R. and Filippenko, A. V. and Mattila, S. and Silverman, J. M. and Andersen, A. C. and H{\"o}flich, P. A. and Pozzo, M. and Wheeler, J. C.},
  title   = {Dust and the Type II-Plateau Supernova 2004et},
  journal = {The Astrophysical Journal},
  volume  = {704},
  number  = {1},
  pages   = {306--323},
  year    = {2009},
  doi     = {10.1088/0004-637X/704/1/306}
}

@ARTICLE{2024ApJ...974..316D,
       author = {{Dong}, Yize and {Valenti}, Stefano and {Ashall}, Chris and {Williamson}, Marc and {Sand}, David J. and {Van Dyk}, Schuyler D. and {Filippenko}, Alexei V. and {Jha}, Saurabh W. and {Lundquist}, Michael and {Modjaz}, Maryam and {Andrews}, Jennifer E. and {Jencson}, Jacob E. and {Hosseinzadeh}, Griffin and {Pearson}, Jeniveve and {Kwok}, Lindsey A. and {Boland}, Teresa and {Hsiao}, Eric Y. and {Smith}, Nathan and {Elias-Rosa}, Nancy and {Srivastav}, Shubham and {Smartt}, Stephen and {Fulton}, Michael and {Zheng}, WeiKang and {Brink}, Thomas G. and {Shahbandeh}, Melissa and {Bostroem}, K. Azalee and {Hoang}, Emily and {Janzen}, Daryl and {Mehta}, Darshana and {Meza}, Nicolas and {Shrestha}, Manisha and {Wyatt}, Samuel and {Auchettl}, Katie and {Burns}, Christopher R. and {Farah}, Joseph and {Galbany}, Llu{\'\i}s and {Padilla Gonzalez}, Estefania and {Haislip}, Joshua and {Hinkle}, Jason T. and {Howell}, D. Andrew and {De Jaeger}, Thomas and {Kouprianov}, Vladimir and {Kumar}, Sahana and {Lu}, Jing and {McCully}, Curtis and {Moran}, Shane and {Morrell}, Nidia and {Newsome}, Megan and {Pellegrino}, Craig and {Polin}, Abigail and {Reichart}, Daniel E. and {Shappee}, B.~J. and {Stritzinger}, Maximilian D. and {Terreran}, Giacomo and {Tucker}, M.~A.},
        title = "{Characterizing the Rapid Hydrogen Disappearance in SN 2022crv: Evidence of a Continuum between Type Ib and IIb Supernova Properties}",
      journal = {\apj},
         year = 2024,
        month = oct,
       volume = {974},
       number = {2},
          eid = {316},
        pages = {316},
          doi = {10.3847/1538-4357/ad710e},
archivePrefix = {arXiv},
       eprint = {2309.09433},
 primaryClass = {astro-ph.HE},
       adsurl = {https://ui.adsabs.harvard.edu/abs/2024ApJ...974..316D}
}

@ARTICLE{Pessi2025,
       author = {{Pessi}, T. and {Desai}, D.~D. and {Prieto}, J.~L. and {Kochanek}, C.~S. and {Shappee}, B.~J. and {Anderson}, J.~P. and {Beacom}, J.~F. and {Dong}, S. and {Stanek}, K.~Z. and {Thompson}, T.~A.},
        title = "{Supernova rates and luminosity functions from ASAS-SN: II. 2014─2017 core-collapse supernovae and their subtypes}",
      journal = {\aap},
         year = 2025,
        month = nov,
       volume = {703},
          eid = {A34},
        pages = {A34},
          doi = {10.1051/0004-6361/202556799},
archivePrefix = {arXiv},
       eprint = {2508.10985},
 primaryClass = {astro-ph.HE},
       adsurl = {https://ui.adsabs.harvard.edu/abs/2025A&A...703A..34P}
}
\bibliographystyle{aasjournal}
\end{document}